\documentclass[aps,prd,a4paper,preprintnumbers,amsmath,amssymb,superscriptaddress,twocolumn,floatfix,showpacs]{revtex4}
\usepackage{amssymb}
\usepackage{float,graphicx}
\newcommand{\Oo}{{\cal O}}
\newcommand{\dd}{{\rm d}}
\newcommand{\TeV}{\,{\rm TeV}}
\newcommand{\GeV}{\,{\rm GeV}}
\newcommand{\Mpc}{\,{\rm Mpc}}
\newcommand{\kpc}{\,{\rm kpc}}
\newcommand{\pc}{\,{\rm pc}}
\newcommand{\MeV}{\,{\rm MeV}}
\newcommand{\keV}{\,{\rm keV}}
\newcommand{\eV}{\,{\rm eV}}

\newcommand{\mumetre}{\,\mu{\rm m}}

\newcommand{\km}{\,{\rm km}}
\newcommand{\mpl}{M_{\rm Pl}}
\newcommand{\omegap}{\omega_{\rm pl}}
\newcommand{\etal}{\emph{et al. } }
\newcommand{\Pphi}{P_{\gamma \leftrightarrow \phi}}
\newcommand{\be}{\begin{eqnarray}}
\newcommand{\ee}{\end{eqnarray}}
\newcommand{\ba}{\left( \begin{array}{cc}}
\newcommand{\ea} {\end{array} \right)}
\newcommand{\bv}{\left( \begin{array}{c}}
\newcommand{\ev} {\end{array} \right)}
\newcommand{\hatb}[1]{\hat{\mathbf{#1}}}
\usepackage{float,graphicx}
\begin{document}
\title{Detecting Chameleons:  The Astronomical Polarization Produced by
 Chameleon-like Scalar Fields}
\author{Clare Burrage}
\email{c.burrage@damtp.cam.ac.uk}
\affiliation{Department of Applied Mathematics and Theoretical Physics,
Centre for Mathematical Sciences,  Cambridge CB3 0WA, United Kingdom}
\author{Anne-Christine Davis}
\email{a.c.davis@damtp.cam.ac.uk}
\affiliation{Department of Applied Mathematics and Theoretical Physics,
Centre for Mathematical Sciences,  Cambridge CB3 0WA, United Kingdom}
\author{Douglas J. Shaw}
\email{D.Shaw@qmul.ac.uk}
\affiliation{Queen Mary University of London, Astronomy Unit, Mile End Road, London E1 4NS, United Kingdom}
\date{11 September 2008}
\begin{abstract}
We show that a coupling between chameleon-like scalar fields and
photons induces  linear and circular polarization in the light
from astrophysical sources. In this context chameleon-like scalar
fields includes those of  the  Olive-Pospelov (OP) model, which describes a varying fine structure constant. We determine  the form of this polarization
  numerically and give   analytic expressions  in two useful limits.  By comparing
the predicted signal with current observations we are able to improve
the constraints on the chameleon-photon coupling and the coupling in the OP model by over two orders
of magnitude. It is argued that, if observed, the distinctive form of the chameleon induced circular polarization would represent a smoking gun for the presence of a chameleon. We also report a tentative statistical detection of a
chameleon-like scalar field from observations of starlight
polarization in our galaxy.
\end{abstract}
\maketitle

\section{Introduction}\label{sec:intro}
Extensions of the Standard Model of particle physics, such as string theory,
introduce many new scalar fields which are not seen in the Standard
Model. Such scalar fields are commonly invoked to explain the observed
acceleration of the universe,  as inflation \cite{Inflation} or dark energy \cite{Peebles}
fields, or to cause variations in fundamental constants \cite{Uzan}.   If new scalar fields do indeed
exist in the Universe, it is important to understand the properties of the theoretical models that describe them, e.g. the interactions of the scalar fields with themselves and with matter, which may
give rise to additional observable effects that 
could be tested and constrained by experiments.

In this article we consider the effect of scalar field theories
with a self-interaction  potential, $V(\phi)$, and couplings to matter
and light on observations of the polarization of light from
astrophysical sources.  These scalar field theories are described by  the action:
\begin{eqnarray}
S&=&\int d^4x \sqrt{-g}\left(\frac{1}{2\kappa_4^2}R-
\frac{1}{2}g^{\mu\nu}\partial_\mu\phi \partial_\nu \phi -V(\phi)
\right. \label{action} \\ && \left.-\frac{B_{F}(\phi/M_{0})}{4} F^2\right)+
\sum S_{\rm m}^{(i)}  B_{i}(\phi/M_{0})
g_{\mu\nu},\psi_m^{(i)}) \nonumber
\end{eqnarray}
where the $S_m^{(i)}$  are  the matter
actions for the matter fields $\psi^{(i)}_m$, and the functions $B_{i}(\phi/M_{0})$ and
$B_F(\phi/M_{0})$  determine the couplings of the scalar
field, $\phi$, to  the $i^{\rm th}$ matter species, $\psi_{i}$, and to
the photon field respectively.  A scalar field with such couplings to matter fields
might be expected to give rise to fifth force effects or violations of
the weak equivalence principle.  In this  article we are specifically
interested in 
a  scalar field, $\phi$, which is light in  relatively low density regions such as galaxies,
galaxy clusters and  the inter-galactic medium.  More precisely, in these regions we 
require the mass,  $m_{\phi}$, of small perturbations  about the 
 background  value of the scalar field, $\phi_{\rm b}$, to satisfy
$m_{\phi} \lesssim  10^{-11}\eV/c^2$.  Hence, any force mediated by
$\phi$ would have a  range $\lambda_{\phi} = 1 /   m_{\phi} \gtrsim
20\km$.  Additionally  we  require that
coupling between  photons and $\phi$ in these regions is relatively strong:
$$g_{\phi\gamma\gamma} = \frac{1}{M}  = \left.{\rm d}\ln B_{F}/{\rm d}
\phi\right\vert_{\phi = \phi_b}  \gtrsim 10^{-11}\,{\rm GeV}^{-1}.$$
Therefore even if the coupling to matter is much weaker than the coupling to photons, since roughly
$10^{-4}$ of the mass of nuclei  is due to electromagnetic
interactions the  $\phi$-mediated force between
individual nuclei in these  backgrounds will be at least $10^{7}$
times the strength of  gravity on scales smaller than $\lambda_{\phi}$.

One might, at first glance,  conclude that a scalar field
theory with these  properties is already strongly ruled out by
laboratory constraints,  e.g. \cite{EotWash04, EotWash06,
  Irvine, Colorado, Stanford1,  Stanford2},  on the strength of fifth
forces.  Specifically,  measurements of the displacement of a micro-machined
silicon cantilever using a fibre interferometer, reported in
 \cite{Stanford2}, require that  a Yukawa-type fifth force with
strength $10^{7}$ times that  of gravity has a range $\lambda_{\phi} < 5
\mumetre$ with 95\% confidence.   This, however, does not rule out the
models we wish to consider  as
neither the strength nor the  range of the $\phi$-mediated force are
necessarily the same in the relatively  high density environment of
the laboratory as they are in the  low density background of space.

In recent years, two classes of  models have arisen that allow a
scalar field that  is strongly interacting in low density
environments  and yet is currently undetected in laboratory tests:
the chameleon model,  \cite{chamKA, chamstrong}, and the
Olive-Pospelov (OP) model  \cite{OP}.  Both models are described in
detail in the following Section.  The mechanism by which these models
 avoid laboratory
tests can be understood by extremizing  Eq. (\ref{action}) with respect
to $\phi$ to give the  following field equation:
\begin{equation}
\square \phi = V_{{\rm eff},\phi}(\phi, T_{\rm m}, F^2/4) \label{phiField}
\end{equation}
where
\begin{equation}
V_{\rm eff}(\phi; T_{i}, F^2) = V(\phi) + \frac{B_{F}(\phi)}{4} F^2 -
\frac{\ln B_{m}(\phi)}{2}  T_{\rm m},
\label{Veff}
\end{equation}
and $T_{\rm m} = g_{\mu \nu}T_{\rm m}^{\mu \nu}$ is the trace of the
energy momentum tensor  for matter:
$$
T_{\rm m}^{\mu \nu} = \frac{2}{\sqrt{-g}} \frac{\delta S_{\rm m}}{\delta g_{\mu \nu}}.
$$
For non-relativistic matter $T_{\rm m} \approx -\rho_{\rm m}$, where
$\rho_{\rm m}$ is the energy  density of the matter.  Both the
chameleon and OP models play  the scalar field
potential, $V(\phi)$, off against the  matter couplings, $B_{F}$ and $B_{\rm m}$,
to make the vacuum expectation value (VEV), and  hence the properties of the field, depend
strongly on the local density  of matter.  For convenience we shall
refer to such a  scalar field  $\phi$ as the chameleon or chameleon field,
however our analysis  applies equally well to  both the  chameleon and
OP models.  In this analysis we posit a universal
coupling to the different matter  species i.e. $B_{i}(\phi/M_{0}) =
B_{m}(\phi/M_{0})$.   Although it is not required by
either model, we make this assumption because it simplifies the analysis whilst
having little effect our  conclusions.  The best constraints on $M_0$ come from the
requirement that corrections to particle  physics are small, which limits $M_0 \gtrsim 10^{4}\,{\rm GeV}$ \cite{chamPVLAS}.   

  A coupling between matter fields and the chameleon potentially causes
  violations of the weak equivalence principle (WEP) and other fifth force
  effects such as an effective alteration to Newton's inverse square law.  A coupling between photons and chameleons introduces
  additional observable phenomena for the chameleon field. If such a coupling has super-gravitational strength, it
can result in a non-negligible  conversion of photons to chameleons
and vice versa. The detectable effects associated with this
conversion are similar to those predicted  for axion-like-particles
(ALPs) which interact with light  \cite{chamPVLAS}.  Mixing requires the interaction between two photons and one
scalar particle, and so the effects of the mixing are most likely to be
seen when a photon, or a scalar particle is passing through an
external electromagnetic field.  The
chameleon-photon coupling induces  both birefringence and dichroism
\cite{chamPVLAS,chamPVLASlong} in a coherent photon beam passing
  through an external magnetic field.  These effects could be detected by
laboratory searches, such as the  polarization experiments PVLAS, Q\&A, and BMV
\cite{Zavattini:2005tm,Zavattini:2007ee,Chen:2006cd,Robilliard:2007bq},
that are sensitive to new hypothetical particles with a small mass and
coupling to photons.  Such experiments can constrain the coupling $M$
  in the chameleon model, indeed  the otherwise anomalous detection of
birefringence with a $5.5\,{\rm T}$ magnetic  field by PVLAS
\cite{Zavattini:2007ee}, could, at  least in principle, be explained
by the presence of a chameleon  field \cite{chamPVLASlong}. For the
most widely studied class of potentials,  the PVLAS data was found to
rule out $M \lesssim 2 \times 10^{6}\,\GeV$
\cite{chamPVLASlong}. In the OP model the mass of the scalar field  in
  the laboratory is too large to produce a  detectable effect in these experiments.

If chameleons exist and  couple to photons,  then they could, as
suggested in  \cite{chamJar, RingJar}, be trapped  and slowly
converted back into photons resulting in a long lived  chameleonic
afterglow.  A number of experiments, most notably  GammeV
\cite{GammeV}, are searching or aiming to search for this  afterglow
effect.  The GammeV chameleon search recently  announced its first
results which, for models with $m_{\phi}  < 10^{-3}\eV$ in the
interior of the 
experiment, ruled out $2.4 \times 10^{5}\GeV  < M < 3.9 \times
10^{6}\GeV$ \cite{GammeVResults}. Ultimately  GammeV may be sensitive
to $M \lesssim 10^{8}\,{\rm GeV}$  \cite{chamJar}, and an optimal
sensitivity for afterglow searches of  $M < 10^{10}\GeV$ is feasible
within the constraints of currently  available technology
\cite{chamJar}.  Indeed for any of the effects  associated with the
coupling to photons to be large enough  to be detected in the
laboratory, either now or in the foreseeable  future, one must have $M
\lesssim 10^{10}\,{\rm  GeV}$. Such laboratory constraints do not, however, apply to the OP models.

These laboratory experiments need to be performed in a very good approximation to a
vacuum otherwise the chameleon becomes too heavy to have a noticeable effect.  A complimentary approach 
to testing chameleon-photon couplings is to look for the effects of the coupling  in observations of astronomical objects.  The
densities of interstellar space are typically very low and so the
effects of the chameleon may be significant. Light from all astronomical objects  travels a significant
distance through  magnetic fields in galaxies, galaxy clusters and
possibly  in the intergalactic medium, before reaching the earth.   However astronomical magnetic
fields are typically made up of large numbers of randomly oriented
magnetic domains - a very different scenario to the well controlled
constant magnetic fields of laboratory experiments.  In contrast to laboratory tests, such astrophysical effects should be see in the OP as well as the chameleon models.

The coupling between photons and chameleons means that photon number
is not conserved, however, as the 
flux of photons emitted by astronomical objects is difficult to
determine, measurements of flux cannot be used to bound the parameters of the chameleon
model.  In the  following sections we show how the coupling between photons and chameleons  generates
polarization in the light from astronomical objects.  Therefore measurements of polarization can be used to 
constrain the parameters of the chameleon model because the intrinsic 
polarization of astronomical objects is often very well constrained.   Astronomers are interested in measuring polarization
because it can provide information both about the source of the 
 radiation and about any magnetic fields present
between the source and the earth.  Very precise astronomical
polarization measurements are therefore available, and can be used to constrain
the chameleon model.

This article is organized as follows: in \S \ref{sec:model} we
introduce and provide further details of the two classes of model to
which our analysis applies i.e the chameleon and OP models. The
coupling of a  chameleon-like-particle to photons, is essentially the
same as that  which is assumed for a scalar axion-like particle (ALP).
There is  a great deal of literature concerning constraints (both
local and astrophysical)  on ALPs. However, the density dependent mass
of a chameleon field,  allows chameleon theories to  evade the
tightest of these constraints.  In \S \ref{sec:ALPs}, we review
previous constraints on ALPs,  and consider to what extent they do, or
do not, apply to  chameleon-like models.  In \S
\ref{sec:Optics}, we consider how the existence  of a chameleon-like
field alters the polarization of light from astrophysical  objects  as
it passes through an astrophysical magnetic field, and  derive the
form of the induced polarization.  In \S \ref{sec:Magnetic} we discuss
the observed and predicted  properties of the different types of large
scale astrophysical magnetic fields.  In \S \ref{sec:obs}, we apply
the results of the previous sections,  and use astrophysical
polarization observations to  constrain the chameleon to photon 
coupling. We find that such  measurements place the tightest
constraints yet on this coupling. Applying the analysis to starlight 
polarisation in our galaxy we find a tentative statistical detection of
a chameleon-like scalar field. Finally we summarize our results
in \S \ref{sec:sum}.

\section{The Models}\label{sec:model}
\subsection{Chameleon Model}\label{sec:Model:cham}
In the chameleon model the coupling functions $B_{\rm m}$ and $B_{F}$ in Eq. (\ref{action}) are well
approximated by linear  functions of $\phi$ i.e.  $B_{F} \approx 1 +
\phi/M$ and $B_{\rm m}  \approx 1 + 2\phi/M_{0}$; for reasons of
naturalness, if $1/M \neq 0$,   $M \sim \mathcal{O}(M_0)$ is usually
assumed.  The strength of  the matter coupling is determined by
$B_{F,\phi}$ and $B_{m,\phi}$, hence in the chameleon model the
coupling strength does not  depend explicitly on the VEV of $\phi$. The model was originally
proposed by Khoury and Weltman  \cite{chamKA} with $M_{0} \sim
\mathcal{O}(\mpl)$, which results  in a gravitational strength
coupling between matter and the  chameleon field, $\phi$.  The ability
of the chameleon with this coupling to behave as dark energy was
discussed in \cite{chameleonDE}.  The
coupling to photons, $B_{F}$,  was not expressly considered in  \cite{chamKA}
although with $M \sim M_0 \sim \mpl$  the most pronounced new effect is a
virtually undetectable density dependence  in the fine structure
constant.  With a gravitational  strength coupling to matter (and
possibly also photons) the chameleon  field could be detected by
laboratory, satellite, solar system and  astrophysical tests (e.g. structure formation  \cite{green}) of
gravity.  A potentially much
wider phenomenology was opened up,  when Mota and Shaw
\cite{chamstrong} showed that coupling between  chameleon fields and
matter could be many orders of  magnitude stronger than gravity, $M_{0}
\ll \mpl$, and yet still be  compatible with all existing experimental
data. The  properties of such strongly coupled chameleon
fields can probed using experiments  designed to measure the Casimir
force \cite{chamstrong, chamcas}.  In addition, with a strong coupling and $M \sim
\mathcal{O}(M_0)$, Brax \etal~  \cite{chamPVLAS} noted that interactions between
chameleons fields and photons would  result in potentially detectable
effects similar to those predicted  for axion-like-particles (ALPs)
which interact with light.

It should be noted, that it is generally seen as `natural', from the
point of  view of
string theory, to have $M \approx M_{\rm pl}$. This relation also arises
in $f(R)$  modified gravity theories (see e.g. Ref. \cite{chamfR} and
references  therein).  It has
also been suggested, however,  that the chameleon field arises from the
compactification  of extra dimensions, \cite{morecham3}. In this case,
there is  no particular reason why the true Planck
scale (i.e. that of the whole of space time including the
extra-dimensions) should be the same as the effective
4-dimensional Planck scale defined by $M_{\rm pl}$.  Indeed having the
true Planck scale  much lower than $M_{\rm pl}$ has been suggested as
a  means of solving the Hierarchy problem (e.g. the ADD scenario
\cite{HP}). In string-theory too, there is no particular reason
why the string-scale should be the same as the effective
four-dimensional Planck scale.  It is also possible that the
chameleon might arise as a result of new physics with an
associated energy scale greater than the electroweak scale but
much less than $M_{\rm pl}$.   Therefore in this article we consider
$M$ as a free energy scale to be constrained by experiment. This said,
to date,   no one has managed to find such a chameleon theory (with
either $M \sim  M_{\rm pl}$ or otherwise) in the low-energy limit of a
more fundamental high  energy theory (e.g. supergravity).  

The chameleon model evades the strong constraints imposed by local
tests of gravity  \cite{chamKA, chamstrong} through non-linear self  interactions of the field described by the potential
$V(\phi)$, and hence  the
field may  couple with super-gravitational strength.   `Non-linear' in this case means that $V_{,\phi}$ is a
strongly non-linear  function of $\phi$, and the mass of the
scalar field,  $m_{\phi} = \sqrt{V_{,\phi \phi}(\phi)}$, therefore  depends
strongly on the VEV of  $\phi$.  The VEV of $\phi$ in a given
background is determined by  the minimum of the effective potential, (\ref{Veff}), and therefore the position of the minimum depends on
$\rho_{\rm m}$ and $F^2$.   $V(\phi)$ is chosen so that $m_{\phi}$ is
larger in high density  regions than it is in low density regions.  It
is then possible to ensure  that in a galaxy $\lambda_{\phi} \gtrsim
20\,{\rm km}$ whereas in the  laboratory $\lambda_{\phi} < 5
\mumetre$. Assuming that $\ln B_{F,\phi} > 0$  and $\ln B_{m,\phi} >
0$, the chameleon mechanism  requires:
$$
V_{,\phi} < 0,\,V_{,\phi \phi}>0,\,V_{,\phi \phi \phi} < 0.
$$

To provide some intuition for what we expect $m_{\phi}$ to be in a low
density region such as a galaxy  or galaxy cluster, we consider the
most widely studied class of  chameleon models where:
$$
V(\phi) \approx {\rm const.} +  \frac{\Lambda^{4+n}}{n\phi^{n}},
$$
with  $\phi/\Lambda \ll 1$, $n > -1$ and $n \sim O(1)$.  We
note that this includes potentials  with the form $V = {\rm const.} -
\Lambda^{4} \ln (\phi/M)$.   $\Lambda$ is constrained by experiments
to be at most a few orders  of magnitude larger than the dark energy
energy scale $\Lambda_{0} =  \rho_{\rm de}^{1/4} = \left(2.4 \pm
0.3\right) \times 10^{-3}\eV$ \cite{chamstrong,chamcas}. When the chameleon is posited as an
explanation for  dark energy, it is therefore considered natural
to take $\Lambda \approx  \Lambda_{0}$
\cite{chamKA, chamstrong}.    The minimum of the effective potential
occurs when $\phi = \phi_b$  where $-V_{,\phi}(\phi_b) =
\rho_{b}/M_{0}$. The mass of the chameleon  at this minimum is given
by $\sqrt{V_{,\phi \phi}(\phi_b)}$,  so:
$$
m_{\phi} = \Lambda \sqrt{n+1} \left(\frac{\rho_{\rm b}}{\Lambda^3
  M_{0}}\right)^{ \frac{n+2}{2(n+1)}}.
$$
Assuming  $M_{0} \approx M > 10^{8}\GeV$  (i.e. the region which is
not currently  accessible to laboratory experiments), we have $m_{\phi}  < 10^{-12}\,{\rm eV}$
 for all $-1 < n \lesssim 5.6$ in
a background, such  as a  galaxy or galaxy cluster, with  $\rho_{\rm b}
\approx 10^{-24}\,{\rm  g\,cm}^{-3}$.

\subsection{Olive-Pospelov Model}\label{sec:model:OP}
The Olive-Pospelov (OP) model \cite{OP} was proposed as a way to allow
particle masses and coupling  `constants' to depend on the local
energy density of matter.   The model could therefore provide an
explanation for the $6\sigma$  difference between the value of the
fine structure constant,  $\alpha =e^2/\hbar c$, in the laboratory  and that extrapolated from
the spectra of 128 QSO  absorption systems at redshifts $0.5<z<3$ by
Webb \etal \cite{webb}:  $\Delta \alpha/\alpha\equiv (\alpha_{\rm
  qso}-\alpha_{\rm lab}) /\alpha_{\rm lab}
=-0.57\pm 0.10\times 10^{-5}$.  There is now a great deal of tension between other potential theoretical explanations for this data see e.g. Refs. \cite{Uzan,valpha}, and the most recent local atomic clock constraints on any local time variation of $\alpha$ \cite{Rosenband, seasonal, BarrowLi}.  In the OP model, $\alpha$ is locally time-independent and hence these constraints are avoided.

The OP model could also describe a density dependent electron-proton
mass ratio $\mu =m_{\rm p}/m_{\rm e}$.  Reinhold \etal \cite{reinhold}
reported a $4\sigma$ indication of a  variation in $\mu$.  They
analysed the $H_{2}$ wavelengths of the spectra  of two absorbers at
$z\approx 2.6$ and $z \approx 3.0$ observed  using the Very Large
Telescope (VLT), finding $\Delta \mu /\mu =24.4\pm  5.9\times
10^{-6}$.  It was shown, however, in Ref. \cite{King}  that due to
wavelength calibration errors in the spectrograph  on the VLT
identified in Ref. \cite{MurphyRe}, the result of Reinhold  \etal
could no longer be trusted. The reanalysis performed by  King \etal
\cite{King}, in which data from an additional object at $z\approx 2.8$
was also included, found $\Delta \mu / \mu = (2.6 \pm 3.0) \times
10^{-6}$, which is consistent with no change.  Very recently Levshakov
\etal \cite{LevMu} have reported
evidence for a spatial
variation in $\mu$  found by measuring ammonia  emission lines in the Milky Way:  $\delta \mu/\mu = -(4-14)\times 10^{-8}$.    

All of these astronomical measurements of $\mu$ and
$\alpha$, were made in regions  where the average density
of matter, $\rho_{\rm b}$ is very low  compared to the ambient density of
matter in a laboratory.    The background density for all of
these measurements,  $\rho_{\rm b}$,  is similar to the average
density of a galaxy or galaxy cluster i.e.  $\rho_{\rm b} \sim
10^{-24}\,{\rm g\,cm}^{-3}$.  These  measurements could therefore be
an indication that some or all  of the  `constants' of
Nature depend on the ambient density  of matter.   The OP model
realises  just such a density dependent  variation in a manner that
does not conflict with local tests of  gravity.  In this model, the
coupling functions, $B_{\rm m}$ and  $B_{\rm F}$, are chosen so that
they are close to their minimum  (which occurs at $\phi_{m}$):
\begin{eqnarray}
B_{\rm F} &=& 1 +
\frac{\xi_{\rm F}}{2}\left(\frac{\phi-\phi_{m}}{M_{0}}\right)^2,  \nonumber \\ 
B_{\rm m} &=& 1 +
\frac{\xi_{\rm m}}{2}\left(\frac{\phi-\phi_{m}}{M_{0}}\right)^2. \nonumber
\end{eqnarray}
For reasons of naturalness, one would expect $\xi_{\rm F},\xi_{\rm m} \sim
\Oo(1)$ \cite{OP}. In contrast to the chameleon model, the OP
model does not require that  the potential, $V(\phi)$, contain
non-linear self interaction terms,  and in the simplest model:
\begin{eqnarray}
V(\phi) &=& \Lambda^4_{0} + \frac{\Lambda^4_{1}}{2}
\left(\frac{\phi}{M_{0}}\right)^2.  \nonumber
\end{eqnarray}
In a background with density $\rho_b$, assuming $\vert F^2 \vert \ll
\rho_{b}$, as is usually the case,  and fixing the definition of $M_0$
by setting $\xi_{\rm m} =1$, the  value of $\phi$ at the minimum of
the effective potential,  $\phi_{\rm min}$ is given by:
\begin{equation}
\frac{\phi_{\rm min}}{\phi_m} = \frac{\rho_{\rm b}}{\rho_{\rm b} + \Lambda_1^4}.
\end{equation}
In the laboratory environment, $\rho_{\rm b} \gg \Lambda_1^4$, and so
$\phi_{\rm min} \approx \phi_m$.  Additionally the effective matter
coupling is small enough to evade  experimental constraints.  In low
density regions such as galaxies  $\Lambda_1^4 \gg \rho_{\rm b}$, so
$\phi_{\rm min} \approx 0$. The change in $\alpha$
between the laboratory and a low  density region such as galaxy  is given by:
$$
\frac{\delta \alpha}{\alpha} = \frac{\alpha_{\rm low} - \alpha_{\rm
    lab}}{\alpha_{\rm lab}}  \approx -\frac{\xi_{F}}{2} \left(\frac{\phi_m}{M_0}\right)^2.
$$
To explain the Webb \etal value \cite{webb} of $\Delta \alpha /
\alpha$, one would  require  $\phi_{\rm m}/M_0 \approx 3 \xi_{\rm F}^{-1/2} \times 10^{-3}$.  

Olive and Pospelov \cite{OP} found that the current best constraints
on $M_{0}$ are $M_0 \gtrsim 15\, {\rm TeV}$ and $M_{0}\xi_{F}^{-1/2}
\gtrsim 3\,{\rm TeV}$. We define  $m_{\phi}^{\rm vac} =
\Lambda_{1}^{2}/M_{0}$ to be the mass of small  perturbations in
$\phi$ in a low density region (i.e.  $\rho_{b} \ll \Lambda_{1}^4$),
and let $\lambda_{\phi}^{\rm vac}  = 1/m_{\phi}^{\rm vac}$
specify the range of the $\phi$-mediated  force in such a region.
It was  found in \cite{OP}  that
$$
\left(\frac{\phi_m}{10^{-3} M_0}\right)^2 \left(\frac{M_0}{1\,{\rm
    TeV}}\right)^2\left(\frac{1\km}{\lambda_{ \phi}^{\rm vac}}\right)^{4} \lesssim 10^3-10^4.
$$ 
For there to be measurable differences between the particle masses and
coupling constants in the  laboratory and in regions with  $\rho
\approx 10^{-24}\,{\rm g\,cm}^{-3}$,  one must require:
$$
\left(\frac{M_0}{1\,{\rm
    TeV}}\right)\left(\frac{1\km}{\lambda_{\phi}^{\rm vac}}\right)
\gtrsim 3.3\times   10^{-7}.
$$ 

In the low density regions where $\phi \approx 0$ the effective coupling
to the photon field for small perturbations  in $\phi$ is:
\begin{eqnarray}
g_{\phi\gamma \gamma} &=& 1/M = \left.\frac{\dd \ln B_{F}}{\dd
  \phi}\right\vert_{\phi = 0} \approx  -\frac{\xi_{F} \phi_{m}}{M_0^2}, \nonumber \\
&=& -10^{-6}\GeV\, \left(\frac{\xi_{\rm F}^{1/2} \phi_{\rm
  m}}{10^{-3}M_{0}}\right)\left(\frac{1  \TeV}{\xi_{F}^{-1/2}M_{0}}\right). \nonumber
\end{eqnarray}
It is clear then that a field with the required properties,
$g_{\phi\gamma \gamma} \gtrsim 10^{-11}\, {\rm GeV}^{-1}$ and
$\lambda_{\phi}^{\rm vac} \gtrsim 2\, {\rm km}$, is perfectly
compatible with current experimental  constraints.  

We note that in  \cite{OP}, the value suggested for $\Lambda_{1}$,
which is compatible with  all current constraints, is $\Lambda_{1} \sim
\Oo(1) \eV$.  Now:
$$
m_{\phi} = 10^{-12}\eV \,\left(\frac{\Lambda_{1}}{1\,{\rm
    eV}}\right)^2 \left(\frac{1 {\rm TeV}}{M_{0}} \right),
$$
so for $M_{0} \gtrsim 15\,{\rm TeV}$, we have $m_{\phi} \lesssim 7
\times 10^{-14} \,{\rm eV}$ which  corresponds roughly to $\lambda_{\phi} \gtrsim 2800\km$.

\section{Constraints on Axion-Like-Particles} \label{sec:ALPs}
Axion-like-particles (ALPs) can either be scalar or pseudo-scalar
fields which couple to the electromagnetic field strength.  If it were
not for the chameleon mechanism (i.e. the density-dependent mass)
present in the chameleon and OP models, they would essentially
describe a standard  scalar ALP.   There are tight constraints on the
coupling, $g_{\phi \gamma \gamma}$, of ALPs to photons. In the
previous section, we discussed constraints from local experiments on
chameleon-like particles, however such particles are also constrained
by searches for ALPs. For a recent review of the astrophysical constraints on ALPs see Ref. \cite{ALPrev} and reference therein.  In all cases, these constraints only apply when the ALP mass, $m_{\phi}$, lies within a certain range e.g $m_{\rm low} < m_{\phi} < m_{\rm high}$.  The mass of a chameleon-like particle is not, however, fixed and so applying these constraints to chameleon models is non-trivial. We must take great care to identify the ambient density of the region wherein the constraint on $m_{\phi}$ is required to derive the bound on $g_{\phi \gamma \gamma}$.  

The strongest astrophysical constraints in Ref. \cite{ALPrev} come
from axion production in the cores of stars.  The application  of the
constraints of solar axion production to chameleon-like models has
previously been studied in Ref. \cite{chamPVLAS} and \cite{AxLocalG}.
The Sun may be a powerful source of ALP flux, and the predicted
effects of the loss of energy of the Sun through ALP emission allows one to constrain the coupling $g_{\phi \gamma \gamma}$.  It must be noted, that all solar ALP constraints require that the ALPs actually escape the sun. The strongest solar ALP constraints come from limits on the solar neutrino flux, and this gives:
$$
g_{\phi \gamma \gamma} \lesssim 5 \times 10^{-10}\,{\rm GeV}^{-1}.
$$
Similar constraints result from the CERN Axion Solar Telescope (CAST)
which attempts to directly detect solar axions.  However it was shown
in Ref. \cite{chamPVLAS}, that solar chameleon-like ALPs would
generally bounce off, rather than enter the CAST instrument, and so
the CAST constraints cannot be applied to chameleon models.   Similar
constraints are found from the life-time of Helium burning (HB) stars
in globular clusters: $g_{\phi \gamma \gamma} \lesssim 10^{-10}{\rm
  GeV}^{-1}$.  Solar axion constraints are derived from production of
axions in the solar core by the Primakoff process. In this region the
temperature is $T \approx 1.3\,{\rm keV}$, and the typical density is
$150\,{\rm g}\,{\rm cm}^{-3}$.  In the Helium burning stars, $T
\approx 10\,{\rm keV}$ and $\rho \approx 10^{4}\,{\rm g}\,{\rm
  cm}^{-3}$.   All of the constraints assume that $m_{\phi} \ll T$.
It was shown in Ref. \cite{AxLocalG}, that all solar axion production
bounds are evaded if $m_{\phi} \gtrsim 10\,{\rm keV}$ in the solar
core.  Similarly, the Helium burning star constraints are effectively
evaded if $m_{\phi} \gtrsim 30\,{\rm keV}$ in their cores.  For
example, if one considers a chameleon potential like $\Lambda^4
(\Lambda/\phi)$,  where $\Lambda \approx 2.3 \times 10^{-3}\,{\rm
  eV}$, one finds that with a matter coupling of $10^{10}\,{\rm GeV}$,
we have $m_{\phi} \approx 1.5 {\rm MeV}$ in the core regions of the HB
stars, and $m_{\phi} \approx 64\,{\rm keV}$ in the solar core.  With
the  different choice of potential, $\Lambda^4 \exp (\Lambda/\phi)$, one finds $m_{\phi} \approx 30\,{\rm keV}$ in the cores of Helium burning stars but $m_{\phi} \approx 3\,{\rm keV}$ in the Sun when $M = 10^{10}\,{\rm GeV}$. Thus the solar and HB star axion constraints on $g_{\phi \gamma \gamma} = 1/M$ would apply to the latter potential with $M \approx 10^{10}\,{\rm GeV}$ but be evaded by the former. If we took $M \approx 2 \times 10^{9}\,{\rm GeV}$, then both potentials would evade these constraints. Thus in the chameleon model, whether or not these astrophysical constraints are relevant depends greatly on the properties of the potential, and in particularly how it determines the behaviour of the theory at high densities. In general, these properties cannot be inferred from the low-density behaviour of the theories.  We are concerned only with the low density behaviour in this work. 

 At high densities, the scalar field in the OP model couples quadratically, rather than linearly (as an ALP would) to the QED $F^2$ term. In this way it avoids astrophysical constraints related to axion production in high density regions.  

Recently, in Ref. \cite{ALPgamma}, it was shown that polarization
measurements of $\gamma$-ray bursts could be used to constrain axion
production at the source of the burst.  Whilst later in this article
we  will consider the potential constraints on chameleon-like fields
from $\gamma$-ray burst polarization measurements, we will be
interested in constraining any  polarization that is induced by the
chameleon as the light from the $\gamma$-ray burst passes through
low-density magnetized regions of space (e.g. the interstellar and
inter-galactic mediums).  We will assume that axion production in the
immediate vicinity of  the $\gamma$-ray burst itself is negligible. In
Ref. \cite{ALPgamma}, it is assumed that, in the vicinity of a $\gamma$-ray burst, there is a magnetic field of strength $B \sim 10^{9}\,{\rm G}$ over a distance of about $L_{\rm GRB} \sim 10^{9}\,{\rm cm}$.  Given this, it is found that:
$$
g_{\phi \gamma \gamma} \lesssim 5\times 10^{-12}\,{\rm GeV}^{-1},
$$
for $8 \times 10^{-5}\,{\rm eV}< m_{\phi} < 3.5 \times 10^{-4}\,{\rm eV}$.   For larger values of $m_{\phi}$:
$$
g_{\phi \gamma \gamma} \lesssim 2.2 \times 10^{-8}\,\left(\frac{m_{\phi}}{1\,{\rm eV}}\right) \,{\rm GeV}^{-1}.
$$
It is noted that in the vicinity of the GRB, $n_{e} \approx
10^{10}\,{\rm cm}^{-3}$, corresponding to $\rho_{\rm m} \approx 2
\times 10^{-14}\,{\rm g}$.  The effective `energy density' to which
the chameleon field couples is not just  $\rho_{\rm m}$ but $\rho_{\rm
  tot} = \rho_{\rm m} + B^2/2-E^2/2$. Thus for  the GRB,  $\rho_{\rm tot} \approx B^2/2 \approx 4.4 \times 10^{-5}\,{\rm g}{\rm cm}^{-3}$.  Such a $\rho_{\rm tot}$ places one in the high-density region of the OP model, where the $\phi$ only couples to photons quadratically and hence no longer behaves as an axion. In chameleon theories, if $V(\phi) = \Lambda^4 f(\phi/\Lambda)$ where $f^{\prime}(1) \sim f^{\prime \prime}(1) \sim \mathcal{O}(1)$ and $\Lambda \approx 2.3 \times 10^{-3}\,{\rm eV}$ (as is usually assumed), one finds that $m_{\phi} \gg 10^{-3}\,{\rm eV}$ when $$\rho_{\rm tot} \gg M \Lambda^3 \approx 2.8 \times 10^{-8}\,{\rm g}{\rm cm}^{-3} \left(\frac{M}{10^{10}\,{\rm GeV}}\right). $$
Thus the strongest constraint on $g_{\phi \gamma \gamma}$ from Ref. \cite{ALPgamma} does not apply here. If we take $V = \Lambda^4 (\Lambda/\phi)$ or $V= \Lambda^4 \exp (\Lambda/\phi)$ then we predict $m_{\phi} \approx 0.8\,{\rm eV}$ or $m_{\phi} \approx 0.4\,{\rm eV}$ respectively for $M \approx 10^{10}\,{\rm GeV}$; hence $M \approx 10^{10}\,{\rm GeV}$ is allowed. Indeed, we find that the bound of Ref. \cite{ALPgamma}, would allow all such chameleon models with $M \gtrsim 10^{6}\,{\rm GeV}$.  

It should also be noted, that axion-like chameleon production from the magnetic fields of neutron stars would also be greatly suppressed. For a neutron star $B \approx 10^{12}\,{\rm G}$, which corresponds to $\rho_{\rm tot} \approx 44\,{\rm g}{\rm cm}^{-3}$ and hence a very heavy chameleon particle.  

It is clear then that astrophysical ALP constraints coming from relatively high density regions do not apply to the OP model, and the extent to which they apply to a chameleon theory depends greatly on the precise choice of potential. For at least one popular choice of potential ($V=\Lambda^4 (\Lambda/\phi)$) one of the constraints noted above applies.  Furthermore, because the chameleon field is very heavy in high density regions, we expect any initial chameleon flux from stars or objects to be greatly suppressed relative to that which one would expect for a standard ALP.

There has also been a great deal of work on conversion of photons to
very light ALPs  in relatively low density backgrounds (e.g. the
interstellar medium).  See, for example, Refs. \cite{Silkvie84,
  Raffelt88, Harai92, Carlson94, Jain02, Das05}, and for a recent
review see Ref. \cite{LightAxRev}.  In relatively low density regions,
chameleon-like particles behave  essentially like standard
axion-like-particles.  Therefore much of the analysis presented in the
aforementioned works is directly applicable. Only where a initial
axion flux from, for example, a star or quasar has been assumed will
the analysis differ.  Many of these studies have focused on
photon-axion conversion in the inter-galactic medium. Magnetic fields
with strength $B \sim 10^{-9}\,{\rm  G}$ are generally seen as
plausible in the inter-galactic medium. It is suspected that such
fields would be coherent over scales of about a megaparsec or so. We
discuss this further in \S \ref{sec:Magnetic}.    For reasonable values of the electron number  density, $n_{\rm e}$, in the inter-galactic medium, it is commonly found that $g_{\phi \gamma \gamma} \lesssim 10^{-10}\,{\rm GeV} (1{\rm n G}/B_{\rm IGM})$ or so \cite{LightAxRev}.  

Carlson and Garretson \cite{Carlson94} specifically considered the
effects of photon to ALP conversion induced by the magnetic field of
our own galaxy.  This discussion is directly relevant to our work.  In
their work they were only able to constrain $g_{\phi \gamma \gamma} <
10^{-5}\,{\rm GeV}$, however they suggested a method that would allow
couplings down to $10^{-9}\,{\rm GeV}$ to be probed.  In our work to
use a different method to constrain $g_{\phi \gamma\gamma}$ down to
$10^{-9}\,{\rm GeV}$.  Ref. \cite{Carlson94} is also interesting
because it is noted that small scale fluctuations in the
electron-density can lead to an enhancement of the photon to ALP
conversion rate.  In their work, the enhancement effect was estimated
to be very large for visible light.  We discuss this further in
Appendix \ref{app:ench}, and note that the size of the enhancement
effect found in Ref. \cite{Carlson94} was in part due to, what is now,
an old model for the electron-density fluctuations.  Using the more
recent NE2001 model \cite{NE2001}, we show in Appendix \ref{app:ench}
that the enhancement effect is expected to be no larger than
$\mathcal{O}(1)$ in the local interstellar medium.  Due to the
complexities and additional uncertainties associated with the
structure of electron-density fluctuations, particularly at parsec
scales, which determine the magnitude of any enhancement, we have
neglected the potential enhancement effect of Ref. \cite{Carlson94}
from our analysis.  As we note in Appendix \ref{app:ench}, however, we
do not expect this to greatly alter our conclusions. Similarly,   the analysis of Ref. \cite{Das05} is applicable to chameleon-like fields, however our analysis goes beyond what was presented there.

We also comment on Ref. \cite{Jain02}.  Here a supercluster magnetic
field with strength $1\,\mu{\rm G}$ coherent over a scale of $10\,{\rm
  Mpc}$ was assumed.  Additionally an enhancement effect similar to
that derived in Ref. \cite{Carlson94} was employed.  It must be noted
that the magnitude of any enhancement effect depends greatly on both
the magnitude and the spatial scale of the spectrum of
electron-density fluctuations.  The former is fairly well known for
electrons in our galaxy, whereas the latter is less well known.  In
the context of electrons in a supercluster neither is well known.
Additionally evidence for a  field strength of $B \approx 1\,\mu{\rm
  G}$ coherent over $10\,{\rm Mpc}$ was tentative at best at the time
that Ref. \cite{Jain02}, and a more recent analysis \cite{Vallee07}
suggests that if such a field does exist it is either weaker, $B \sim
0.1\,\mu{\rm G}$,  or only coherent over much smaller scales $\sim
100\,{\rm kpc}$. Even if such a field does exist, it is also not clear
precisely what distance along the line of sight  the field extends. As
such, the constraint: $g_{\phi \gamma \gamma} \lesssim 10^{-13}\,{\rm
  GeV}^{-1}$ quoted in Ref. \cite{Jain02} relies on many assumptions,
with at best only tentative observational support. Removing any one of
these assumptions, would allow for much larger couplings.  In this
work, we are primarily concerned with constraints on photon to
chameleon conversion in astrophysical regions where there is strong
evidence for magnetic fields, and the properties of such magnetic
fields are relatively well known.   We also note in Appendix
\ref{app:ench} that  making the reasonable assumption  $\left\langle (\delta n_{e})^2\right\rangle^{1/2} /\left \langle n_{e} \right \rangle \sim O(1)$ or smaller (where the $\left\langle \cdot \right\rangle$ indicate a spatial average), any enhancement of the photon-chameleon conversion rate due to electron fluctuations is expected to be sub-leading order at optical (and higher) frequencies for cluster and super-cluster scale magnetic fields.  

\section{Chameleon Field Optics} \label{sec:Optics}
In this section we consider how the presence of a chameleon field
alters the properties of light propagating through one, or many,  magnetic regions. Varying the action Eq. (\ref{action}) with respect to both $\phi$ and $A_{\mu}$ gives Eq. (\ref{phiField}) and
\begin{eqnarray}
\nabla_{\mu} \left[B_{F}(\phi/M_0) F^{\mu \nu}\right] = J^{\nu}
\end{eqnarray}
where $J^{\mu}$ is the background electromagnetic 4-current:
$\nabla_{\mu}J^{\mu} = 0$. We consider propagation of light in an
astrophysical background which contains a magnetic field of strength
$\mathbf{B}$.  The background value of $\phi$ is denoted $\phi_0(t)$.
We write the perturbation in the photon field as $a_{\mu}$ and the
perturbation in the chameleon field as $\delta \phi$.  Ignoring terms
that are $\mathcal{O}(\delta \phi a_{\mu})$ and assuming that the
proper frequency of the photons, $\omega$, is large compared to the
Hubble parameter, $H$,  we find that:
\begin{eqnarray}
-\ddot{\mathbf{a}} + \nabla^2 \mathbf{a} &=& \frac{\nabla \delta \phi \times \mathbf{B}}{M},  \label{aeqn} \\
-\ddot{\delta \phi} + \nabla^2 \delta \phi &=& \frac{\mathbf{B} \cdot (\nabla \times \mathbf{a})}{M} \label{phieqn} \\ &&+ (V_{,\phi}(\phi_0 + \delta \phi)-V_{,\phi}(\phi_0)), \nonumber 
\end{eqnarray}
where $1/M = (\ln B_{F})_{,\phi}(\phi_0)$ and
$$
\square \phi_0 = V_{{\rm eff},\phi}(\phi_m, \rho_{\rm b}, F^2_0/4 = B^2/4).
$$
and here $\rho_{\rm b}$ is the background density of matter. 
 
We assume that $\delta \phi$ is small enough that we may make the approximation:
\begin{equation}
V_{,\phi}(\phi_0 + \delta \phi)-V_{,\phi}(\phi_0) = m_{\phi}^2 \delta \phi \nonumber
\end{equation}
where $m_{\phi}^2= V_{,\phi \phi}(\phi_0)$ is the chameleon mass. If the photons are moving through a plasma with electron number density $n_{\rm e}$, they will behave as if they had an effective mass  squared $\omegap^2$, where $\omegap^2 = 4\pi \alpha n_{e}/m_e$ is the plasma frequency; $\alpha$ is the fine structure constant and $m_e$ is the electron mass.

\subsection{A Single Magnetic Domain}\label{sec:Optics:sing}
We define $\gamma_{\perp}$ and $\gamma_{\parallel}$ to be the
components of the photon field perpendicular and parallel to the
magnetic field $\mathbf{B}$, and take the photon field to be propagating in the $z$-direction. From Eq. (\ref{aeqn}) we have:
\begin{eqnarray}
-\ddot{\gamma}_{\parallel} + \frac{\partial^2 \gamma_{\parallel}}{\partial z^2} &=& \omega_{p}^2 \gamma_{\parallel},\nonumber \\
-\ddot{\gamma}_{\perp} + \frac{\partial^2 \gamma_{\perp}}{\partial z^2} &=& \omega_{p}^2 \gamma_{\perp} + \frac{\partial \phi}{\partial z}  \frac{B}{M},\nonumber \\
-\ddot{\phi} + \frac{\partial^2 \phi}{\partial z^2} &=&  m^2 \phi-\frac{B}{M} \frac{\partial}{\partial z} \gamma_{\perp},\nonumber
\end{eqnarray}
For such a system, it is well known that the probability of a photon,
with frequency $\omega$, converting to a chameleon particle (or vice
versa) whilst travelling a  distance $L$ through a region with a homogeneous magnetic field is:
\begin{equation}
\Pphi = A^2,
\end{equation}
where 
\begin{eqnarray}
A &=& \sin 2\theta \sin\left(\frac{\Delta}{\cos 2\theta}\right) \\
\Delta &=& \frac{m_{\rm eff}^2 L}{4\omega}, \\
\tan 2\theta &=&  \frac{2B\omega}{M m_{\rm eff}^2},
\end{eqnarray}
and $m_{\rm eff}^2 = m_{\phi}^2 - \omega_{\rm pl}^2 -
B^2/M^2$. Generally $\vert B^2/M^2 m_{\rm eff}^2 \vert \ll 1$ and so
the last term in $m_{\rm eff}^2$ is dropped.  Following
\cite{chamPVLASlong, chamJar, chamSN}, we find that, up to an overall
phase factor, $\gamma_{\perp}$, $\gamma_{\parallel}$ and $\phi =
i\chi$ are transformed by passing through a homogeneous magnetic
domain in the following way:
\begin{eqnarray}
\gamma_{\parallel} &\rightarrow \gamma_{\parallel} &,  \label{gP1} \\
\gamma_{\perp} &\rightarrow & e^{i\alpha}\left(\sqrt{1-A^2} \gamma_{\perp} + i e^{-i \varphi} A \chi\right), \label{gP2} \\
\chi &\rightarrow & e^{-i\beta}\left(\sqrt{1-A^2} \chi + i e^{i \varphi} A \gamma_{\perp}\right),\label{gP3}
\end{eqnarray}
where $\alpha = \varphi - \Delta$ and $\beta = \varphi + \Delta$ and
\begin{equation}
\tan \varphi = \cos 2\theta \tan\left(\frac{\Delta}{\cos 2\theta}\right). \label{varphiEq}
\end{equation}
Since we must, in realistic situations, allow the light to be partially polarized (or even unpolarized), it is insufficient to consider simply the evolution of the photon, $\gamma_{\perp}$ and $\gamma_{\parallel}$, and chameleon $\phi = i\chi$ amplitudes.  We must instead represent the properties of the photon field by its Stokes vector.  We therefore make the following definitions:
\begin{eqnarray}
I_{\gamma} &=& \left\langle \vert \gamma_{\perp} \vert^2 \right\rangle +\left\langle \vert \gamma_{\parallel} \vert^2 \right\rangle, \label{ampDef} \\
Q &=& \left\langle \vert \gamma_{\perp} \vert^2 \right\rangle -\left\langle \vert \gamma_{\parallel} \vert^2 \right\rangle, \nonumber \\
U + iV &=& 2\left\langle\bar{\gamma}_{\perp} \gamma_{\parallel}\right\rangle, \nonumber \\
J + iK &=& 2e^{i\varphi}\left\langle\bar{\gamma}_{\parallel} \chi\right\rangle, \nonumber \\
L + iM &=& 2e^{i\varphi}\left\langle\bar{\gamma}_{\perp} \chi\right\rangle. \nonumber
\end{eqnarray}
The Stokes vector for the photon field is  $S = (I_{\gamma}, Q, U,
V)^{\rm T}$, where  $V$ describes the amount of circular polarization
(CP), and $Q$ and $U$ describe the  amount of linear polarization (LP). We also define the reduced Stokes vector: $S_{\rm red} = (Q/I_{\gamma}, U/I_{\gamma}, V/I_{\gamma})^{\rm T}$. The fraction of light which is polarized is:
$$
p = \frac{I_{p}}{I_{\gamma}} = \frac{\sqrt{Q^2 + U^2 + V^2}}{I_{\gamma}},
$$
and the fractional circular polarization is:
$$
m_{c} = \frac{V}{I_{\gamma}}.
$$
We also define $q = \vert m_{\rm c} \vert$.  The fractional linear
polarization is $m_{\rm l} = \sqrt{p^2 - m_{c}^2}$. 

We normalise the photon and chameleon fluxes so that $I_{\gamma} + I_{\phi} = 1$ (this quantity is conserved), where $I_{\phi} = \vert \phi \vert^2$. We also define $X = 3I_{\gamma}- 2$. With these definitions we find that, on passing through a single homogeneous magnetic domain, the components of the Stokes vector transform as:
\begin{eqnarray}
X &\rightarrow & \left(1-\frac{3}{2}A^2\right)X - \frac{3}{2}A^2 Q \label{IgEq}\\ && + 3A\sqrt{1-A^2}(L \sin 2\varphi - M \cos 2\varphi), \nonumber \\
Q &\rightarrow &\left(1-\frac{1}{2}A^2\right)Q - \frac{1}{2}A^2 X\\ &&  + A\sqrt{1-A^2}(L \sin 2\varphi - M \cos 2\varphi), \nonumber, \\
U+iV &\rightarrow & \sqrt{1-A^2}e^{-i\alpha}(U+iV)  \\ &&- A e^{i\beta}(K+iJ) \nonumber,
\end{eqnarray}
Additionally, the $J$, $K$, $L$ and $M$ amplitudes transform as
\begin{eqnarray}
K+iJ &\rightarrow & \sqrt{1-A^2}e^{i\beta}(K+iJ)  \\ && + A e^{-i\alpha}\left(U+iV\right), \nonumber \\
L &\rightarrow & L\cos 2\varphi + M\sin 2\varphi  \\
M & \rightarrow &(1-2A^2)(M\cos 2\varphi - L\sin 2\varphi) \label{LastEq} \\ && + A\sqrt{1-A^2}(Q+X). \nonumber 
\end{eqnarray}
From these equations it is clear that the presence of a light scalar
field coupling to  photons can result in the production of
polarization.   This is because, when a chameleon (or another 
axion-like particle) is converted back into a photon, that photon is
polarized perpendicular to the magnetic field.  If we consider the
simple case where initially there is no chameleon flux so that
$I_{\gamma} =1 \Rightarrow X = 1$ and $K = J = L = M =0$, and we set  $Q
= Q_0$, $U=U_0$ and $V=V_0$ initially then,  using $A^2 = \Pphi$, it is clear that upon exiting the magnetic domain:
\begin{eqnarray}
X &=& 1 - \frac{3}{2}\Pphi(1+Q_0), \nonumber \\
Q &=& (1-\frac{1}{2}\Pphi)Q_0 - \frac{1}{2}\Pphi, \nonumber \\
U+iV &=& \sqrt{1-\Pphi}e^{-i\alpha}(U_0 + iV_0).\nonumber
\end{eqnarray}
If the initial total and circular polarization fractions are $p_{0}$ and $q_{0}$, their final values are
\begin{eqnarray}
p = \sqrt{\frac{p_{0}^2 + C_{0}}{1+C_0}}, \\
q =\sqrt{\frac{q_{0}^2 + D_{0}}{1 + C_{0}}},
\end{eqnarray}
where $C_0 = (A^4(1+Q_0)^2/4 - A^2Q_0)/(1-A^2)$ and $D_0 = (U_0^2 -
V_0^2)\sin^2 \alpha - U_0 V_0 \sin 2 \alpha$.  It is therefore
possible for both linearly and circularly polarized light to be
produced.  In a single magnetic domain, the production of the former
is due to the conversion of photons into chameleons and then back into
photons, and the latter is due to the birefringence of the medium
which is induced by the presence of the chameleon field.    If
initially $p = p_{0} = 0$, then after passing through a single domain:
$$
p = \frac{\Pphi^2}{2-\Pphi^2}
$$
We also note that if there is no initial chameleon flux or
polarization, $q_{0} = D_{0} = 0$, no CP can be produced in a single
magnetic domain.  As we shall see below, the same is \emph{not} true if there are multiple magnetic domains.

\subsection{Multiple Magnetic Domains}
In many realistic astrophysical settings, including the ones we will
be primarily concerned with in subsequent sections,  light passes
through many magnetic domains on its way from a source to an
observer. In each domain the angle, $\theta_{n}$, describing the
inclination  of the background  magnetic field to the direction of
propagation is essentially random.  Solving the full system of
evolution equations for a large number of magnetic domains involves
diagonalising an 8 by 8 matrix as well as evaluating multiple sums
involving the random angle $\theta_{n} \sim U[0,2\pi)$, and  we have
  been unable to find a general analytic solution, however, it is
  straightforward to solve the system numerically. This said,
  approximate analytical solutions exist in a number of  interesting
  and important limits.  A full presentation of the equations that
  must be solved in this set-up, and their  analytic solutions in
  these  limits is provided in Appendix \ref{appA}.  We present the results of that analysis below.  We define $N$ to be the number of magnetic domains through which the light has passed, and in all cases assume that there is no initial chameleon flux.

For fixed $m_{\rm eff}^2$ and magnetic domain length $L$, we define a critical frequency $\omega_{\rm crit}$ such that  $\Delta(\omega_{\rm crit}) = \pi/2$, and hence $\omega_{\rm crit} = m^2_{\rm eff}L/2\pi$. When $\omega \gg \omega_{\rm crit}$, $\Pphi$ is almost independent of frequency, however when $\omega \ll \omega_{\rm crit}$, $\Pphi \propto \omega^2$.  We also define $\lambda_{\rm crit} = 2\pi/\omega_{\rm crit}$ to be the critical wavelength and $\lambda_{\rm osc} = \lambda_{\rm crit}/N$.

\subsubsection{Weak Mixing Limit}\label{sec:Optics:Many:Weak}
In a great many realistic situations we have $N \alpha \ll 1$ and $N
\Pphi \ll 1$ and, as we shall show, the frequency dependence of the  production of linearly and circularly polarized light in this limit is qualitatively similar to that seen in general. In this limit we must have either $\Delta/\cos 2\theta$ and $\Delta \tan 2\theta \ll 1$, or $\tan 2\theta$ and $\Delta \tan^2 2\theta \ll 1$, and so
$$
\alpha \approx \frac{\tan^2 2\theta}{2}\left[2\Delta - \sin 2\Delta\right].
$$
In Appendix \ref{appA} we find that  when an initially unpolarized
light beam, with frequency $\omega$, from a single source  passes through $N \gg 1$ domains (and requiring $N \alpha \ll 1$ and $N \Pphi \ll 1$), the final polarization fraction, $p_{0}$, and final fractional CP, $q$, are essentially random variables and are described by the following distributions:
\begin{eqnarray}
p &=& \frac{N\Pphi}{2}\left[\sigma_{+}^2 (X_{1}^{2} + X_{2}^{2}) + \sigma_{-}^2(X_{3}^2 + X_{4}^2)\right], \nonumber \\
m_{\rm c} &=& N\Pphi \sigma_{+}\sigma_{-} \left(X_{1}X_{3} - X_{2}X_{4}\right). \nonumber
\end{eqnarray}
where at fixed $\Delta = m^2_{\rm eff} L/4\omega$, the $X_{i}$ are approximately independent identically distributed $N(0,1)$ random variables and
$$
\sigma_{\pm}^2 = \frac{1}{4}\left[1\pm \frac{\cos(2(N-1)\Delta)\sin 2N\Delta}{N \sin 2\Delta}\right].
$$
When $\lambda \ll \lambda_{\rm crit}/N = \lambda_{\rm osc}$ the
$X_{i}$ are roughly independent of $\Delta$, but when $\lambda \gg
\lambda_{\rm osc}$ there is a strong $\Delta$, and hence wavelength
dependence.  The above expressions describe the total and circular polarization fractions
for a monochromatic light beam from a single source. If one has observations of many objects all at the same frequency the average value of $p$, denoted $\bar{p}$, and r.m.s. average of $m_{\rm c}$, denoted $\bar{q}$, are  more useful quantities for comparing with observations.   For the distributions above
\begin{eqnarray}
\bar{p}&=& \frac{1}{2}N\Pphi, \label{fpbar1}\\
\bar{q} &\approx& \sqrt{2}\sigma_{+}\sigma_{-}N\Pphi,\label{fcbar1}
\end{eqnarray}
 where $N$ is now the average number of  magnetic regions.  When $N\Delta \gg 1$, $\sigma_{+}\sigma_{-} = 1/4$ and when $N\Delta \ll 1$, $\sigma_{+}\sigma_{-} = N\Delta / \sqrt{3}$.  
 
When some initial polarization is present ($p =p_{0}$ and $q=q_{0}$
say, so that the initial linear polarization is $m_{l0} = \sqrt{p_{0}^2
  - q_{0}^2}$), we find a different behaviour:  When $N\Pphi(1-p_{0}^2)/2p_{0} \gg 1$ (but keeping $N\Pphi \ll 1$) the average final polarization fractions, $\bar{p}$ and $\bar{q}$ are still given, to $\mathcal{O}(N\alpha^2,\, N\Pphi)$, by Eqs. (\ref{fpbar1}) and (\ref{fcbar1}) respectively.  When $N \Pphi (1-p_{0}^2)/2p_{0} \ll 1$ we find instead that to $\mathcal{O}(N\alpha^2,\, N\Pphi)$:
\begin{eqnarray}
\bar{p} &=& p_{0}, \label{fpbar2}
\end{eqnarray}
and
\begin{eqnarray}
\bar{q}^2 &=& q_{0}^2(1-\alpha^2 N) + \frac{\alpha^2 N m_{l0}^2}{2}\label{fcbar2} \\ &&+ 2N^2\Pphi^2 N^2 \sigma_{+}^2 \sigma_{-}^2 + N^2\Pphi^2 \sigma_{2}^2 m_{l0}^2, \nonumber
\end{eqnarray}
where this expression is only accurate to leading order in $\bar{q}-q_{0}$.

We shall see that for realistic astrophysical magnetic fields the critical wavelength $\lambda_{\rm crit}$ generally corresponds to UV or X-ray light.  As such most polarimetry measurements of astrophysical objects will have been made at wavelengths $\gg \lambda_{\rm osc} = \lambda_{\rm crit}/N$.  For such wavelengths, the analytical solutions found in Appendix \ref{appA} show that the reduced Stokes parameters, $Q/I_{\gamma}$, $U/I_{\gamma}$ and $V/I_{\gamma}$, exhibit a strong and oscillatory frequency dependence.  This behaviour is very important when one wishes to make comparisons with observations. Polarimeters always have some finite wavelength ($\lambda = 2\pi/\omega$) resolution, $\delta \lambda$. This is usually referred to as the spectral resolution.
Thus a measurement of the reduced Stokes parameters at some wavelength
$\lambda_{0}$, will actually measure an average of their values in the
window $\lambda \in (\lambda_{0}-\delta \lambda/2,
\lambda_{0}+\delta\lambda/2)$.  If one averages the reduced Stokes
parameters over wavelength bins much larger than $\lambda_{\rm osc}$
much of the  information about a chameleonic contribution will be lost.  Specifically if we  assume $\delta \lambda / \lambda \ll 1$ and $\delta \lambda \gg \lambda_{\rm osc}$ then, to $\mathcal{O}(N\alpha^2,\, N\Pphi)$, $p = \hat{p}$ and $m_c = \hat{m}_{c}$ where: 
\begin{eqnarray}
\hat{p}(\delta_{v},\omega) &=& \bar{p}_{0}, \\
\hat{m}_{c}(\delta_{v},\omega) &=& m_{c0}\left[1-\frac{\alpha^2 N}{4}\left(Y_{1}^2 + Y_{2}^2\right)\right] \\ && - \frac{\alpha \sqrt{N}}{\sqrt{2}} m_{l0} Y_{1} \nonumber,
\end{eqnarray}
where for $N \gg 1$, $Y_{1}$ and $Y_{2}$ are independent identically distributed $N(0,1)$ random
variables. Thus, in this case, constraints on the parameters of the
scalar field theory could only be derived by measuring \emph{both} the
total polarization fraction and the CP fraction. When $\lambda \gg
\lambda_{\rm osc}$, if the spectral resolution is too poor or the data
is placed into too wide wavelength bins, the measured polarization
fraction, $\hat{p}$, carries little or no information about the
properties of $\phi$.  For optimal sensitivity to chameleonic effects,
the spectral resolution of the polarimeter and the size of the
wavelength bins must satisfy $\delta \lambda \lesssim \lambda_{\rm
  osc}$.  We discuss in \S \ref{sec:Optics:Sig} below how when the
spectral resolution is sufficiently good, the strong wavelength
dependence at wavelengths $\lambda \gtrsim \lambda_{\rm osc}$ can be exploited to extract strong constraints on chameleon-like theories from observations of a single object.

\subsubsection{Maximal Mixing Regime}
When $N \gg 1$, if the chameleon to photon coupling is strong enough and $N\Delta \ll 1$, i.e. if $\lambda \ll \lambda_{\rm osc}$, maximal mixing will occur. In this limit the equations that must be solved simplify greatly. Further details of the calculations are given in Appendix \ref{appA}.  The strong mixing limit is appropriate when $N\Pphi \gg 1$, $N\Delta \ll 1$ and $N \gg 1$.  When $N \Delta \ll 1$ there is little production of circular polarization, and so the main effect is the production of linear polarization.  Additionally since at high frequencies we expect $q_{0} \ll p_{0}$ for astrophysical objects we set the initial circular polarization fraction to zero, and the final CP fraction, $q$, remains $\ll p$.
\begin{figure}[htb!]
\includegraphics*[width=6.5cm]{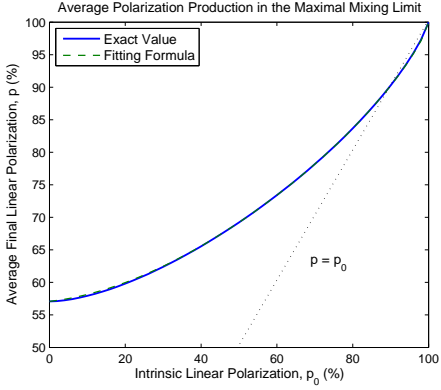}
\caption[]{Dependence of the mean linear polarization, $\bar{p}$, in the maximal mixing limit on the intrinsic polarization ($p_{0}$).  The solid line is the exact value of $\bar{p}$ whereas the dashed line is the value calculated from the fitting formula Eq. (\ref{fpbar3}).  The thin dotted line shows $\bar{p} = p_{0}$ as would be the case when chameleon photon mixing is weak or non-existent.  We can see that for $p_{0} \lesssim 90\%$ maximal chameleon-photon mixing increases the average linear polarization, whereas for $p_{0} \gtrsim 90\%$ it  slightly decreases it.}
\label{maxPol}
\end{figure}
We find that the final (linear) polarization fraction, $p = m_{\rm l}$, in this limit, does not explicitly depend on $\Pphi$ or any other properties of the chameleon field and that it is given by the following distribution:
\begin{eqnarray}
&p &= F(X^2,\cos 2\psi; p_{0})\\ &=&  \sqrt{1- \frac{4(1-p_{0}^2) X^2}{\left[(1+X^2) - p_{0}(1-X^2)\cos 2\psi \right]^2}}. \nonumber
\end{eqnarray}
where $\psi$ and $X$ are independent uniform random variables: $\psi
\sim U[0, \pi)$ and $X \sim U[0,1)$. We note that when $N\Delta \ll
    1$, $f_{\rm p}$ does not depend on frequency.   If we average over
    observations of many sources (each with $N \Delta \ll 1$ and $N\Pphi \gg 1$) then we would measure the average polarization fraction $\bar{p}$.  In the simplest case where $p_{0} = 0$ we have:
$$
p= \frac{1-X^2}{1+X^2}.
$$
and so
$$
\bar{p} = \int_{0}^{1}\,{\rm d}X\,\frac{1-X^2}{1+X^2} = \frac{\pi}{2}-1 \approx 0.57.
$$
More generally
$$
\bar{p}(p_{0}) = \frac{1}{\pi}\int_{0}^{\pi}\,{\rm d}\psi\,\int_{0}^{1}\,{\rm d}X\, F(X^2, \cos 2\psi;p_{0}).
$$
$\bar{p}(p_{0}) $ is a monotonically increasing function of $p_{0}$ and increases from $0.57$ to $1$ as $p_{0}$ goes from $0$ to $1$.  It is clear that in the strong mixing limit, the presence of  a chameleon-like field coupling to the photon can induce a significant amount of linear polarization.  
We find that  the following formula 
\begin{eqnarray}
\bar{p}_{\rm fix}(p_{0}) &=& \frac{\pi p_{0}}{48}\sqrt{1-p_{0}^2}(1-2p_{0}^2) \label{fpbar3} \\ &&+ \left(\frac{\pi}{2}-2\right)(1-p_{0}^2) + 1. \nonumber
\end{eqnarray}
fits $\bar{p}(p_{0})$  extremely well.  
We plot $\bar{p}_{\rm fix}$ against the exact value of $\bar{p}$ in FIG. \ref{maxPol}.  The solid line is the exact value and the dashed line shows $\bar{p}_{\rm fix}$.   Also shown on this plot is the line $\bar{p} = p_{0}$.  For $p_{0} \lesssim 90\%$, the average polarization after maximal chameleon mixing is larger than the intrinsic polarization $p_{0}$, whereas for $p_{0} \gtrsim 90\%$ it is slightly less.
\begin{figure}[htb!]
\includegraphics*[width=6.5cm]{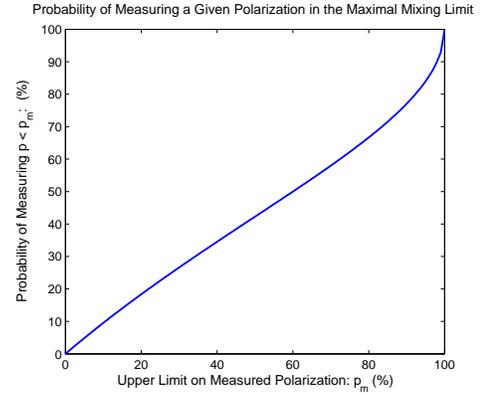}
\caption[]{Probability of measuring the linear polarization degree ($p$) less than some $p_{\rm m}$ for a random object if chameleon-photon mixing is maximal.}
\label{maxProb}
\end{figure}
If one were to measure $\bar{p} < 57\%$ for a large number of astrophysical objects then at least one of the following must be true: $N\Pphi \ll 1$ or $\lambda \gtrsim \lambda_{\rm osc}$.  When $\lambda \ll \lambda_{\rm osc}$, the chameleon induced polarization is largely independent of frequency, and so the spectral resolution of the polarimeter is not as important as it is in the weak mixing regime.  The probability of measuring the total linear polarization of a random object to be less than some $p_{\rm m}$, when $\lambda \ll \lambda_{\rm osc}$ and mixing is maximal ($N\Pphi \gg 1$), is shown in FIG. \ref{maxProb}; we have assumed no knowledge of the initial intrinsic polarization and hence marginalized over a uniform prior for it.

\subsubsection{General Behaviour}
When, as is often the case, one excepts little or no intrinsic circular polarization of the light beam i.e. $m_{{\rm c}0} = 0$, we are able to combine the results presented in the previous two subsections to provide a fitting formula for the general form of the mean value of $p$ after the light beam has passed through $N \gg 1$ magnetic domains.  We find
\begin{equation}
\bar{p}(N) \approx  \sqrt{p_{0}^2 + (\bar{p}^{\rm fix}(p_{0}) - p_{0}^2)\left[1-\left(1-\frac{b^2 \Pphi^2}{4}\right)^{N^2}\right]} \nonumber
\end{equation}
where
$$b^2 = \frac{\sqrt{1-p_{0}^2}}{\bar{p}^{\rm fix}(0)}. $$

In the maximal mixing limit $\bar{p} = \bar{p}^{\rm fix}(p_{0})$.  In the weak mixing limit, when $N\Pphi \ll 1$, we  have:
$$
\bar{p}(N) \approx \sqrt{p_{0}^2 + \frac{b^2 (\bar{p}^{\rm fix}(p_{0}) - p_{0}^2)N^2 \Pphi^2}{4}} 
$$
So if $p_{0}^2$ is small, we have $\bar{p}(N) \approx N\Pphi/2$, as required. If instead $p_{0}$ is larger, $\bar{p}(N) = p_{0} + O(N^2 \Pphi^2)$. This provides a very good fit to the simulated data in all cases.

\subsection{Optical Signatures of Chameleon Fields}\label{sec:Optics:Sig}
We presented above the results of a mathematical analysis of how the
presence of a light scalar field coupling  to matter would alter the polarization properties of light passing through a magnetic field, the details of which can be found in Appendix \ref{appA}. By combining the results of this analysis with numerical simulations, we now  detail the main physical signatures that a chameleon field would imprint on the polarization properties of light from astrophysical sources.  Above we found that there were two main effects:
\begin{itemize}
\item the production of polarization,
\item the production of circular polarization.
\end{itemize}
Each of these effects depends on frequency in a characteristic manner that is well illustrated by considering the weak mixing limit of \S \ref{sec:Optics:Many:Weak} above with no initial polarization ($p_{0} = 0$). In this limit $N \gg 1$ but $NP_{\gamma \leftrightarrow \phi} \ll 1$ for all $\Delta$ and $N\alpha^2 \ll 1$. This requires $BL/2M \ll 1$; $P_{\rm max} = \lim_{\Delta \rightarrow 0} \Pphi \approx (BL/2M)^2 \ll 1$ is the maximum value of $\Pphi$. 
\begin{figure}[htb!]
\includegraphics*[width=6.5cm]{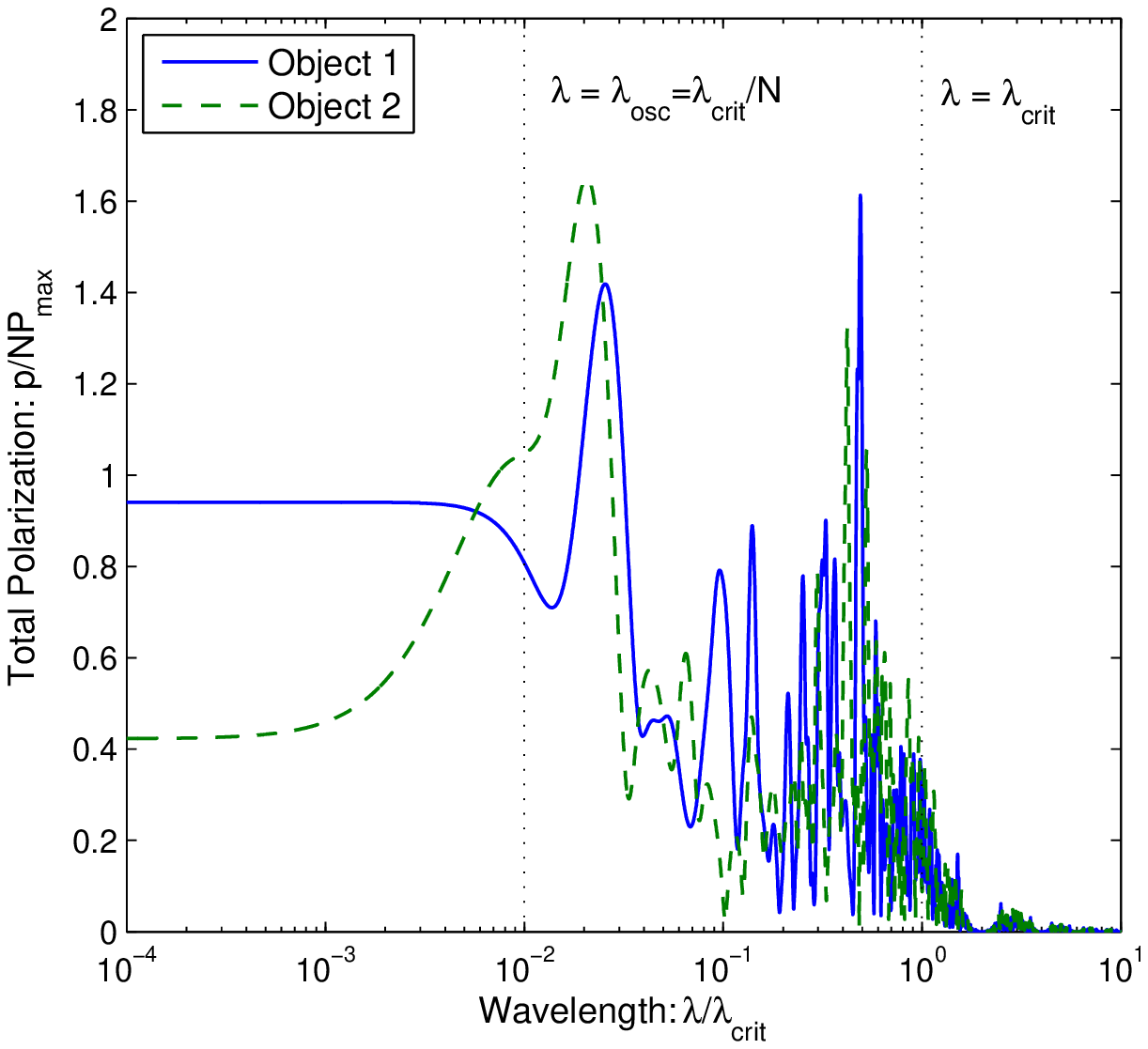}
\includegraphics*[width=6.5cm]{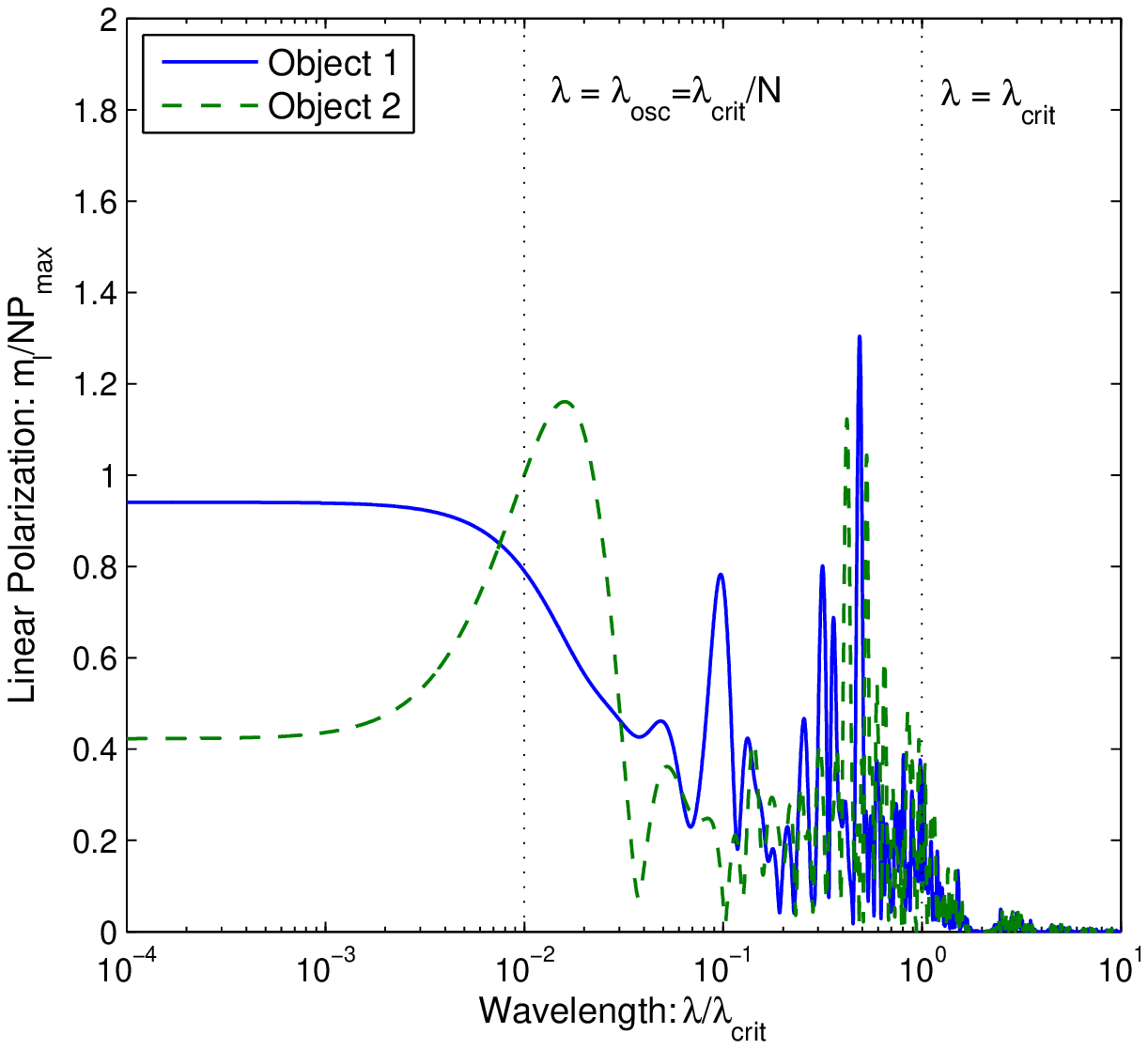}
\includegraphics*[width=6.5cm]{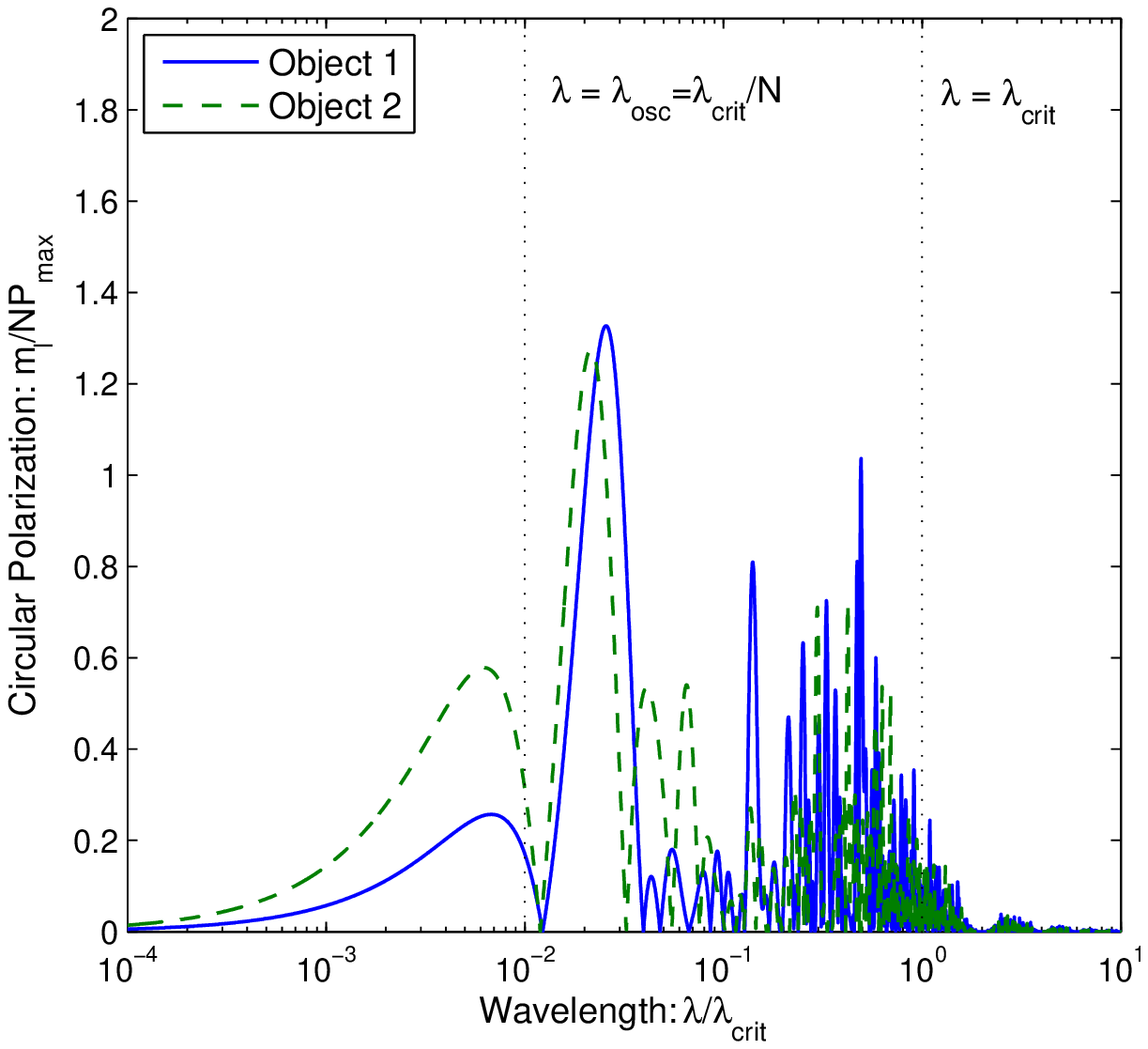}
\caption[]{Dependence of the total polarization degree, $p$, the linear polarization degree, $m_{l}$ and the circular polarization degree $q$ on wavelength for two hypothetical objects with $N = 100$ and $NP_{\rm max} \ll 1$.  Here $\lambda_{\rm crit} = 4\pi^2/\vert m_{\rm eff}^2\vert L$ where $L$ is the coherence length of the magnetic field regions.  The total path length of the light through the magnetic field is given by $L_{\rm path} = NL$.  We define $\lambda_{\rm osc} = \lambda_{\rm crit}/N$. We have assumed that initially $p=0$ and that there is no initial chameleon flux.}
\label{fig1}
\end{figure}

We consider the weak mixing limit, assuming that the polarimeter has wavelength resolution $\lesssim \lambda_{\rm osc} = \lambda_{\rm crit}/N = 4\pi^2/\vert m_{\rm eff}^2 \vert L_{\rm path}$ where $m_{\rm eff}^2 = m_{\phi}^2 - \omegap$ and $L_{\rm path}$ is the total path length of the light through the magnetic field.  In this limit, when there is no initial polarization, both the induced degree of polarization: $p$, and circular polarization $q = \vert m_{\rm c} \vert$ are proportional to $N\Pphi$.    FIG. \ref{fig1} shows  possible simulated forms for the rescaled total polarization degree, $p/NP_{\rm max}$, linear polarization $m_{\rm l}/NP_{\rm max}$ and CP  $q/NP_{\rm max}$ for two different hypothetical objects, where for example $N \approx 100$ in both cases.  We can clearly see from this that production of linear polarization is greatest for $\lambda \lesssim \lambda_{\rm crit}$ and CP polarization production is peaked in the region $\lambda_{\rm osc} \lesssim \lambda \lesssim \lambda_{\rm crit}$.  As expected, we can also see that both the induced linear and circular polarization degrees are highly frequency dependent for $\lambda \gtrsim \lambda_{\rm osc}$. 
\begin{figure}[htb!]
\includegraphics*[width=6.5cm]{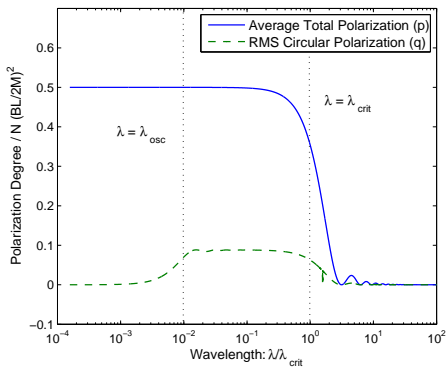}
\caption[]{Dependence of the total average polarization degree, $\bar{p}$, and r.m.s. circular polarization degree, $\bar{q}$, when averaged over many sources, each with $N = 100$ and $N P_{\rm max} \approx N(BL/2M)^2 \ll 1$.  Here $\lambda_{\rm crit} = 4\pi^2/\vert m_{\rm eff}^2\vert L$ where $L$ is the coherence length of the magnetic field regions.  The total path length of the light through the magnetic field is given by $L_{\rm path} = NL$.  We define $\lambda_{\rm osc} = \lambda_{\rm crit}/N$. Initially we have assumed that $p=0$ and that there is no initial chameleon flux.}
\label{fig2}
\end{figure}

Assuming $\delta \lambda \lesssim \lambda_{\rm osc}$, averaging $p$,
and $q$ over many sources each at roughly the same distance, and hence
with roughly the same $N$, gives $\bar{p}$ and $\bar{q}$. The forms of
$\bar{p}$, and $\bar{q}$ are shown in FIG. \ref{fig2}.  We see that
both quantities grow strongly when $\lambda \approx \lambda_{\rm
  crit}$ and that $\bar{q}$ is peaked between $\lambda_{\rm osc}$ and
$\lambda_{\rm crit}$.  The  form of the averaged CP degree,
$\bar{q}$, is very distinctive. The height of the peak, as well as the
maximum value of $\bar{p}$ are determined by $N (BL/2M)^2$, whereas
the position of the  peak and its width are fixed by $\lambda_{\rm osc}$ and $\lambda_{\rm crit}$. If such a peak should be resolved, one would in principle be able to determine both $N(BL/2M)^2$, $m_{\rm eff}$ and the coherence length, $L$, of the magnetic field regions. Measurements of circular polarization for $\lambda_{\rm osc} \lesssim \lambda \lesssim \lambda_{\rm crit}$ could therefore provide a powerful tool with which to constrain chameleon theories; we discuss this further in \S \ref{sec:CP}. Qualitatively similar behaviour is seen for the chameleonic induced polarizations in the more general case where $NP_{\rm max}$ can take any value and the intrinsic (i.e. non-chameleonic) polarization is not restricted  to vanish. When $N P_{\rm max} \gg 1$ is allowed, the chameleonic production of linear polarization is peaked for $\lambda \lesssim \lambda_{\rm max}$ where:
\begin{equation}
\frac{\lambda_{\rm max}}{\lambda_{\rm crit}} = {\rm max} \left(1, \frac{BL}{\pi \sqrt{N} M}\right).
\end{equation}
CP production is in general peaked between $\lambda_{\rm osc}$ and $\lambda_{\rm max}$.

    In practice, however, as we shall discuss further in \S \ref{sec:obs} below, it is rare for current polarimeters to have $\delta \lambda \ll \lambda_{\rm osc}$; although measurements do exist with $\delta \lambda \sim \mathcal{O}(\lambda_{\rm osc})$.  As well as requiring $\delta \lambda \lesssim \lambda_{\rm osc}$, to measure $\bar{p}$, $\bar{m}_{l}$ and $\bar{q}$ one must also have measurements of many sources where the light from each source is expected to have passed through roughly the same number of magnetic regions, each with roughly the same properties, as the light from any other source.  This requirement introduces a fair amount of uncertainty and will ultimately limit ones ability to accurately constrain the averaged quantities.  Another problem is that even when the intrinsic polarization is small ($p_{0} \ll 1$), if $NP_{\phi} \ll 1$, the form of $\bar{p}$, and $\bar{q}$ are highly dependent on $p_{0}$.  Unless one can measure or accurately predict the intrinsic polarization, this again limits ones ability to accurately constrain chameleon theories.

Many astrophysical polarization measurements are made at wavelengths
$\lambda \gg \lambda_{\rm osc}$ where the chameleon induced
contribution to the Stokes parameters exhibits a highly oscillatory
wavelength dependence. When $p_{0} \ll 1$ and provided $\delta \lambda
\sim \mathcal{O}(\lambda_{\rm osc})$ or smaller,  we can exploit this
property to extract strong constraints about the properties of any
chameleon-photon interaction from observations of a single object
without any detailed prior knowledge of $p_{0}$.   We do this by
defining a smoothing scale $\delta \lambda_{\rm smooth}$ which is
picked to be $\gg \lambda_{\rm osc}$ but smaller than the wavelength
scale over which the intrinsic polarizations, $p_{0}$, $m_{{\rm
    l}\,0}$ and $q_{0}$, are expected to vary strongly. By removing
the smoothing signal from the measured signal we should recover a
superposition of any induced chameleonic signal and the
noise. Assuming that the noise is either random  or that it does not
have a wavelength structure that mimics that of the induced chameleon
signal  we can extract constraints on $M$.  Further details of how $M$ can be constrained in this manner are given in Appendix \ref{app:pol}.  Using this method, it is possible to extract strong constraints on $M$ using data from only a single source. 

\section{Large Scale Astrophysical Magnetic Fields}\label{sec:Magnetic}
The largest scale magnetic fields that are known to exist are those associated 
with galaxies and galaxy clusters.  In both cases the mean field strength has 
been measured to be roughly a few micro Gauss.  It is also thought likely that 
a weak, $B<10^{-9}{\rm G}$, magnetic field permeates the inter-galactic 
medium (IGM).  We discuss the observed properties of first two fields as well as the 
hypothesised properties of the latter below.  The electron density,
$n_{\rm e}$, determines the plasma frequency, $\omegap$, which plays a critical r\^{o}le in 
determining the effective chameleon mass, $m_{\rm eff}^2$, and hence also 
critical frequency, $\omega_{\rm crit}$,  above which polarization production
is peaked.  Therefore we also quote the observed or estimated values of 
$n_{\rm e}$ for each of the three regions.

\subsection{Galactic magnetic fields}

Galactic magnetic fields, particularly those of our own galaxy, could
produce detectable polarization effects if chameleon-like  fields
interact strongly enough with photons. Galactic magnetic fields  have
been observed to be a superposition of a regular magnetic field,
$\mathbf{B}_{\rm reg}$, and a random magnetic field, $\mathbf{B}_{\rm
  rand}$ (see  \cite{galRev} and references therein).  The regular
component of the magnetic field has a coherence length, $L_{\rm reg}
\sim {\rm few}\,{\rm kpc}$ i.e. about the scale of the galaxy
\cite{galRev}.  The component of the regular part of the magnetic
field along the line of sight to distant objects such as pulsars and
extragalactic radio sources has  been measured using Faraday rotation.  These measurements are performed using electromagnetic waves whose frequency is well below $\omega_{\rm crit}$. The interpretation of such measurements would therefore be largely unaltered by the presence of a chameleon field or similar light scalar field. The average regular magnetic field in own galaxy is locally (within about $2\,{\rm kpc}$ of the Sun): $B_{\rm reg} = 1.8 \pm 0.4 \,\mu {\rm G}$ \cite{Han94, Indrani98}, rising to about $4.4\pm 0.9 \mu{\rm  G}$ in the more central Norma arm \cite{Han02}.  The magnetic field is aligned with the disk of the galaxy, and is coherent out to a galactic radius of about $5\kpc$, field reversals then occur at $R = 5\kpc$, $6\kpc$ and $7.5\kpc$ \cite{galRev, Sun08}.

The random magnetic field, $\mathbf{B}_{\rm rand}$, is often slightly
larger than the regular magnetic field.   The largest scale of the
turbulent field was determined from pulsar rotation measures (RMs) as
$L_{\rm rand} = 55\,{\rm pc}$ by Rand and Kulkarni \cite{Rand89}, with
a turbulent field strength about about $5\,\mu{\rm G}$.  A similar
study by Ohno and Shibata  \cite{Ohno93} found $L_{\rm rand} =
10\,–\,100\,{\rm pc}$ with a random field strength of
$4\,-\,6\,{\rm \mu G}$. $L_{\rm rand}$ has also been estimated by the
depolarization of light by turbulent fields at centimetre radio wavelengths, and by Faraday dispersion at decimetre radio wavelengths \cite{Sun08}. Both methods give results consistent with $L_{\rm rand} \approx 20\,{\rm pc}$.  

Recently Sun \etal~ \cite{Sun08} combined radio telescope and WMAP measurements of diffuse polarized radio emission from the Milky Way with Faraday rotation measurements to obtain an overall model of the Milky Way's magnetic field.  They found that on average  $B_{\rm reg} = 2\,\mu$G with field reversals occurring over kiloparsec scales, and $B_{\rm rand} = 3\,\mu$G with $L_{\rm rand} = 20\,{\rm pc}$.  The average electron density was taken by Sun \etal~ to be $n_{\rm e} = 0.03\,{\rm cm}^{-3}$.   Generally  light observed from objects within our own galaxy will have passed through $N \sim O(1)$ regions of the regular magnetic field, but $N \gg 1$ different coherent regions of the random magnetic field.  Taking $L = L_{\rm rand} = 20\,{\rm pc}$ and $B = 3\mu$G for the random magnetic field and $n_{\rm e} = 0.03\,{\rm cm}^{-3}$ we have:

\begin{equation}
\left(\frac{\vert B \vert L}{2M}\right)_{\rm rand} = 0.92 \times 10^{-2} \left(\frac{10^{10}\GeV}{M}\right),
\end{equation}
and $\omegap = 6.4 \times 10^{-12}\eV$ so
\begin{equation}
\omega_{\rm crit}^{\rm (rand)} = \frac{\vert m^2_{\rm eff} \vert L}{2\pi} =  20.4\eV\,\left(\frac{\vert m^2_{\rm eff} \vert}{\omegap^2}\right).
\end{equation}
When $m_{\phi} \ll 6.4 \times 10^{-12}\eV$ and hence $\vert m_{\rm
  eff}^2\vert = \omegap^2$,  $\lambda_{\rm crit}^{\rm (rand)} =
2\pi/\omega_{\rm crit} = 608$ \AA.  For an object in our galaxy at a
distance $d$ we take $N \approx d/20\,{\rm pc}$.  Therefore if, as is typical, $d \sim 1\,{\rm kpc}$ we have $N \approx 50$.

Taking $B = 2\,\mu$G for the regular magnetic field and $L = 2\,{\rm kpc}$ we have:
\begin{equation}
\left(\frac{\vert B \vert L}{2M}\right)_{\rm reg} = 0.612 \left(\frac{10^{10}\GeV}{M}\right),
\end{equation}
and
\begin{equation}
\omega_{\rm crit}^{\rm (reg)} = \frac{\vert m^2_{\rm eff} \vert L}{2\pi} =  2.04\keV\,\left(\frac{\vert m^2_{\rm eff} \vert}{\omegap^2}\right),
\end{equation}
so when $\vert m_{\rm eff}^2\vert =  \omegap^2$, $\lambda_{\rm crit}^{\rm (reg)} = 2\pi/\omega_{\rm crit} = 6.08$ \AA. We note that $\vert m_{\rm eff}^2 \vert \leq \omegap^2$ for $m_{\phi} \lesssim 1.3 \times 10^{-11}\eV$. 

In the weak mixing regime, i.e. when the chameleon induced polarization is small, we find that the total induced polarization is a sum of that which would be separately induced by the random and regular magnetic fields. 

For the dark energy inspired chameleon model discussed in \S \ref{sec:Model:cham}, we have $m_{\phi} \lesssim \sqrt{2}\omegap$ in the galaxy, and hence $\vert m_{\rm eff}^2\vert \leq \omegap^2$, for all $M_{0} > 3.9 \times 10^{6}\GeV$ when $n \lesssim 3.3$.

\subsection{Intracluster magnetic fields}
In galaxy clusters electron densities of $n_{\rm e} \approx
10^{-3}\,{\rm cm}^{-3}$ are typical, as are magnetic field strengths
of a few $\mu{\rm G}$, rising to tens of $\mu$G at the centre of
cooling core clusters. These magnetic fields are coherent over length
scales of about $L \approx 10\,-\,100\,{\rm kpc}$ \cite{clustRev}.
Galaxy clusters typically extend over  a length scale of $L_{\rm clust} \sim 1\Mpc$. A light beam traversing a galaxy cluster would therefore pass through roughly $N = L_{\rm clust}/L \approx 100\,-\,1000$ magnetic regions. 

In a study of data from 53 radio sources located in and behind Abell
clusters and a control sample of 99 sources Kim \emph{et al.}
\cite{Kim91} found the mean core electron density of a cluster to be
$n_{\rm e} = 3.5 \pm 2.7 \times 10^{-3}\,{\rm cm}^{-3}$, where the
radius of the core is $r_{\rm core} = 0.65 \pm 0.41 h^{-1}\,{\rm
  Mpc}$, and found cluster magnetic field strengths of $O(1)\mu$G with coherence length $\sim 10\kpc$.  A study of 18 radio sources close in angular position to
the Coma Cluster by Kim \etal \cite{Kim90} found the following result for the strength of magnetic fields in the intracluster medium (ICM):
\begin{equation}
\left\langle \vert B \vert \right\rangle_{\rm ICM} = 2.5 h_{75}^{1/2} \left(\frac{L}{10\kpc}\right)^{-1/2},  \label{clustmag}
\end{equation}
where $h_{75}$ is defined in terms of the Hubble parameter today: $H_{0} = 75h_{75}\,{\rm km}\,{\rm s}^{-1}\,{\rm Mpc}^{-1}$. A subsequent study, again of the Coma cluster, by Feretti  \etal \cite{Feretti} found tangled magnetic fields with length scales of about $1\kpc$, so $B_{\rm ICM} \approx 7.9 h_{75}^{1/2}$.  A study of 16 low redshift ($z < 0.1$) ``normal'' galaxy clusters by Clarke, Kronberg and B\"{o}hringer \cite{Clarke} found that the ICM of these clusters was permeated by a slightly larger magnetic field:
\begin{equation}
\left\langle \vert B \vert \right\rangle_{\rm ICM} = (5-10) h_{75}^{1/2} \left(\frac{L}{10\kpc}\right)^{-1/2}. \nonumber
\end{equation}

Based on the studies of Kim \etal \cite{Kim91,Kim90}, we take the follow representative values for the parameters which describe magnetic fields in the ICM:
$$
L = 1\kpc, \qquad B = 7.9 h_{75}^{1/2}, \qquad n_{\rm e} = 3.5 \times 10^{-3}\,{\rm cm}^{-3}.
$$
We define $L_{\rm path}$ to be the path length a given light beam
traverses through a cluster, and take as a representative value
$L_{\rm clust} = 1\,{\rm Mpc}$.  The number of magnetic regions, $N$,
is given by $N = L_{\rm path}/L$, and  hence we take 
$$N = 1000.$$

With these values we have $\omegap = 2.2 \times 10^{-12}\eV$ and 
\begin{equation}
\left(\frac{\vert B \vert L}{2M}\right)_{ICM} = 1.2 \left(\frac{10^{10}\GeV}{M}\right),
\end{equation}
and 
\begin{equation}
\omega_{\rm crit}^{\rm (ICM)} = \frac{\vert m^2_{\rm eff} \vert L}{2\pi} =  120\eV\,\left(\frac{\vert m^2_{\rm eff} \vert}{\omegap^2}\right).
\end{equation}
When $\vert m^{2}_{\rm eff} \vert  = \omegap^2$, $\lambda_{\rm crit}^{\rm (ICM)} = 2\pi/\omega_{\rm crit} = 104$ \AA.  We note that $\vert m_{\rm eff}^2 \vert \leq \omegap^2$ when $m_{\phi} \lesssim 4.4 \times 10^{-12}\eV$.  

In galaxy clusters, we have $m_{\phi} \lesssim \sqrt{2}\omegap$ for the chameleon model introduced in \S \ref{sec:Model:cham} for all $M_{0} > 3.9 \times 10^{6}\GeV$ when $n \lesssim 3.5$.

\subsection{Intergalactic magnetic fields}
Although a number of different mechanisms have been suggested that
would produce large scale magnetic fields in the intergalactic medium
(IGM), at the present time very little is known about whether such
fields actually exist, let alone their typical strengths. A coherent
magnetic field on the current horizon scale would produce an
anisotropic expansion. CMB and Faraday rotation constraints on such a
scenario limit $B \lesssim  10^{-9}$~G
\cite{Barrow97,Blasi99}. Faraday rotation also constrains smaller
scale magnetic fields.  For a $50\Mpc$ coherence length one has $B
\lesssim 6 \times 10^{-9}$~G, and $B \lesssim 10^{-8}$~G for Mpc scale
coherence lengths \cite{Blasi99}. The CMB has also been shown to
constrain fields with a coherence length between $400\,{\rm pc}$ and
$0.6\Mpc$ to be $< 3 \times 10^{-8}$~G \cite{Jedamzik00}. Motivated by
the need to explain the origin of galactic magnetic fields it is
thought that IGM magnetic fields with coherence lengths of a few Mpc
are likely (see  \cite{IGMrev} and references therein).   Most of the
proposed theoretical mechanisms for generating such fields would,
however, only produce them with strengths well below the current
observational upper bounds \cite{IGMrev}.  These seed fields are then
amplified by some dynamo mechanism during galaxy formation to the
$\sim \mu$G galactic magnetic  fields observed.  

Typical electron densities in the IGM are $n_{e} \approx 2.5 \times 10^{-7}\,{\rm cm}^{-3}$ giving $\omegap = 1.8 \times 10^{-14}\,{\rm eV}$ and so
\begin{eqnarray}
\frac{BL}{2M} &=& 0.153 \left(\frac{10^{10}\,{\rm GeV}}{M}\right)\left(\frac{B}{10^{-9}\,{\rm G}}\right)\left(\frac{L}{1\,{\rm Mpc}}\right), 
\end{eqnarray}
and 
$$
\omega_{\rm crit} = 3.4\eV \,\left(\frac{\vert m_{\rm eff}^2\vert}{\omegap^2}\right)\left(\frac{L}{1\Mpc}\right),
$$
hence for $\vert m_{\rm eff}^2 \vert= \omegap^2$ and $L = 1\Mpc$, we
have $\lambda_{\rm crit} \approx 3647$\AA. For the dark energy
chameleon potentials discussed in \S \ref{sec:Model:cham}, the mass of
the chameleon due to the density of the IGM is $< \sqrt{2}\omegap$
i.e. $\vert m_{\rm eff}^2 \vert \leq \omegap^2$ for $n \lesssim 4.5$
if $M_{0} \gtrsim 3.9 \times 10^{6}\GeV$ and the chameleon couples
only to baryons. If the chameleon couples to dark matter with equal
strength  then the same is true but only for $n \lesssim 3.5$.

\section{Current Polarization Constraints on Chameleon-like Models}
\label{sec:obs}
In this Section we  review a number of astronomical polarization
observations and deduce how they constrain the properties of any chameleon-like field.

We noted in \S \ref{sec:Optics:Many:Weak} that at wavelengths $\lambda
\lesssim \lambda_{\rm osc} \equiv \lambda_{\rm crit}/N$, any chameleon
induced polarization signal is a highly oscillatory function of
wavelength, with oscillation length $\approx \lambda_{\rm osc}$.
This is particularly important as many astrophysical polarization
measurements are made at optical frequencies for which $\lambda <
\lambda_{\rm osc}$, and either the Stokes parameters are put into
wavelength bins with width $\gg \lambda_{\rm osc}$, or the spectral
resolution of the polarimeter is so poor that it effectively averages
over a range of wavelengths which is $\gg \lambda_{\rm osc}$.  In either
situation, any signal of chameleon mixing will be washed out, and no
constraints on the chameleon model are possible. If there is  no initial polarization  the polarization fraction depends on $\lambda$ via $\Pphi$ and  $\sigma^2_{\pm}$, these are all highly oscillatory functions of $\lambda$ except when  $\lambda < \lambda_{\rm osc}$.   We note that in some cases the spectral resolution of the polarimeter is good enough to resolve any chameleon induced polarization, but the published data only quotes the Stokes parameters in bins much wider than $\delta \lambda$. In these cases, the published data cannot bound chameleon-like theories but constraints should follow from a reanalysis of the raw data. By way of an example, we have performed such a reanalysis for observations of three stars in our galaxy, however in general such a reanalysis is beyond the scope of this article, and is intended to form the basis of a future work.

\subsection{Starlight Polarization}\label{starpol}
Polarization is not usually produced by the thermal emission of stars.  In \cite{Fosalba:2001wr} a statistical analysis of the largest available compilation of galactic starlight data \cite{Heiles} was
performed.   The data is statistically significant  for sources out
to distances of $6\kpc$, and the average polarization of light from stars at such
distances is  $2\%$. 

This data, provided in the polarization catalogue \cite{Heiles}, is in very wide wavelength bins that generally cover the whole range of optical frequencies, i.e. the bin width is $\delta \lambda \approx 2000 - 8000$\AA.  For comparison, the oscillation length, $\lambda_{\rm osc} = \lambda_{\rm crit}/N$, in the galaxy for such stars is $\lambda_{\rm osc} \approx 2 \,-\, 12$\AA\ for both the random and regular components of the magnetic field.  Thus $\delta \lambda \gg \lambda_{\rm osc}$ and the data provided in  \cite{Heiles} as well as the subsequent analysis of  \cite{Fosalba:2001wr} does not provide useful constraints on chameleon-like theories.

Existing starlight polarization measurements can, however, constrain
chameleon-like theories.     UV polarization of starlight was measured
for 121 objects by the Wisconsin Ultraviolet Photo-Polarimeter
Experiment (WUPPE), which flew on the ASTRO-1 and ASTRO-2 NASA space
shuttle missions, and had a nominal spectral resolution of $6$\AA
\cite{WUPPE}.  The data from these observations is available from the
Multimission Archive at STScI (MAST) \cite{MASTweb}.   A  full
reanalysis of the data for all 121 objects is beyond the scope of this
work, however we have derived preliminary confidence limits on $BL/2M$
in the galaxy using data from three objects: HD2905, HD37903 and
HD34078.   These objects were picked as they all lie at distances
between $500\,{\rm pc}$ and $1000\,{\rm pc}$, which is not so close
that a chameleonic signal would be too small to detect, and not so far
away that numerical calculations involved in extracting the confidence
limits on $BL/2M$ are too time consuming.  Other than that, the choice
of objects is entirely arbitrary. A detailed account of the method and
resulting confidence limits derived from the polarization measurements
of these objects is given in Appendix \ref{app:pol}.  Intriguingly, we
found that the data from all three objects preferred a non-zero value
of $BL/2M$ at a prima facie statistically significant level.  This   analysis shows that there is some structure in the polarization
data which is consistent with the signal that we predict would be
induced by a chameleon field.  It would be premature, however, to claim
this as an actual detection before a similar analysis has been
conducted for more objects, and before a thorough analysis of all
systematics which could be sources this signal has been undertaken.  The most conservative, in the sense that they are the widest and are expected to be the most robust, confidence intervals were found using the bootstrap-t method (see Appendix \ref{app:pol} for further details).  At 95\% confidence we found, taking $L \approx 20\,{\rm pc}$ and $\lambda_{\rm crit} = 608$\AA:
\begin{eqnarray}
\left(\frac{\vert B\vert L}{2M}\right)_{\rm rand} &=& \left(4.68_{-1.70}^{+1.44}\right) \times 10^{-2} \,{\rm (HD2905)}, \\
\left(\frac{\vert B\vert L}{2M}\right)_{\rm rand} &=& \left(7.59_{-1.42}^{+1.63}\right) \times 10^{-2} \,{\rm (HD37903)}, \\
\left(\frac{\vert B\vert L}{2M}\right)_{\rm rand} &=& \left(8.58^{+2.15}_{-1.85}\right) \times 10^{-2} \,{\rm (HD34078)}. 
\end{eqnarray}
If we assume that the same value of $\vert B\vert L/2M$ is appropriate for each object, by combining the polarization data for all three stars we find that the estimate of $BL/2M$ is  approximately normally distributed with mean $6.27 \times 10^{-2}$ and variance $\sigma^2$; $\sigma = 0.58 \times 10^{-2}$.  Hence we find the following approximate confidences
\begin{eqnarray}
\left(\frac{\vert B\vert L}{2M}\right)_{\rm rand} = (6.27 \pm 1.14) \times 10^{-2}\,(95\%), \\
\left(\frac{\vert B\vert L}{2M}\right)_{\rm rand} = (6.27 \pm 1.91) \times 10^{-2}\,(99.9\%).
\end{eqnarray}
From this preliminary analysis, it therefore
appears as if the  polarization data of the three objects considered is
consistent with a value of $BL/2M$ which deviates from $0$ by more
than $10\sigma$.

 Although this analysis is only preliminary, it does appear as if
 there is a reasonably  significant, and robust, statistical preference towards the existence of a chameleon-like field in the starlight polarization data of the three objects we have considered here. This is a highly surprising result, and as such it would be
  premature to claim it as a detection.  Whilst it is well beyond
  the scope of this particular article, a thorough analysis of
  possible backgrounds and  sources of systematic error which could
  mimic the signal from a chameleon field would have to be undertaken
  before any such claim could be made with true confidence.  In
  particular, since all the data analysed comes from a  single
  experiment (WUPPE) it possible that the `detection' of a non-zero
  value for $BL/2M$ is actually due to effects intrinsic to the
  instrument.  In order to quantify the magnitude of such instrumental
  effects it would be necessary to study similar data from other
  polarimeters.  We have only considered three of the well over  one hundred objects measured by WUPPE. When more objects have been analyzed it should be possible to better estimate the effect of systematic error in the determination of $BL/2M$ by considering the spread in the values of $BL/2M$ determined for each object.
What we can say with confidence is that there is some structure in the
polarization of three objects considered which is not consistent   with either random error or that predicted to be induced by interstellar dust.  Furthermore this structure exhibits non-trivial oscillatory frequency correlations which at least in part mimic that predicted by the chameleon model.  
At the present time we cannot rule out possible systematic effects having a relative magnitude of $O(1)$.  The presence of such effects would raise both the extracted upper and lower bounds bounds on $BL/2M$.  Whilst this means that any non-zero lower bound on $BL/2M$ can only be seen as tentative at best, the upper bounds on $BL/2M$ should be robust. We therefore believe it to be better to see the data as providing the following 95\% and 99.9\% confidence upper bounds on $BL/2M$:
\begin{eqnarray}
\left(\frac{BL}{2M}\right)_{\rm rand} < 7.2 \times 10^{-2}\,(95\%), \\
\left(\frac{BL}{2M}\right)_{\rm rand} < 8.1\times 10^{-2}\,(99.9\%).
\end{eqnarray}

We also consider observations of the UV polarization of two stars made
with the Faint Object Spectrograph (FOS) of the Hubble Space Telescope
(HST) and reported in Ref. \cite{somerville}.  In this case, we have
only undertaken a preliminary analysis of the data, postponing a  full
reanalysis to a later work.    Observations were made for $1279 $\AA\
$< \lambda < 3300$\AA, and the HST FOS has a nominal spectral
resolution of $2-4$\AA\ in this range.  The data published in
Ref. \cite{somerville} was binned to give ten data points in each
frequency region they considered, although the precise width of the
bins is not stated.  The shortest wavelength region was $1279 - 1603$
\AA.  Assuming that each of the ten bins in this region had equal
width, the bin width is $32.4$\AA.  The two stars, HD7252 and
HD161056, are respectively  $824\pc$ and $295\pc$ from the earth.
This gives an oscillation wavelength, $\lambda_{\rm osc} =
\lambda_{\rm crit}/N$,  no smaller than $15$\AA\ and $41$\AA\  for
the random and regular components of the galactic magnetic field
respectively,  when $m_{\phi} <  9 \times 10^{-12}\eV$.  A significant
amount of a  chameleonic signal should therefore survive the rebinning
process in the $1279-1603$\AA\ wavelength grating; this may not be the
case for the  lower frequency gratings. In the $1279-1603$\AA\
grating, the polarization angle of HD7252 was found to be independent
of frequency with a standard deviation of about $5$ degrees.  It is
noted in Ref. \cite{somerville} however that the systematic
uncertainty in the polarization angle could be $10$ degrees or so.
This corresponds to the component of the reduce Stokes vector,
$P_{\perp}$ say, that is perpendicular to the mean polarization
detection  in the region $1279-1603$\AA\, satisfying $\vert P_{\perp}
\vert < 0.2\%$.  Assuming $m_{\phi} < 9 \times 10^{-12}\eV$ in the
galaxy and that $L_{\rm rand} \sim \mathcal{O}(20\pc)$ (the precise
value of $L_{\rm rand}$ does not greatly alter the resulting
constraint) so that $\lambda_{\rm crit} = 608$\AA, and using the
method outlined Appendix \ref{App:BUp}, we find the following $95$\%
and $99$\% confidence  limits
\begin{eqnarray}
\left(\frac{BL}{2M}\right)_{\rm gal} &<& 8.9 \times 10^{-2} \, (95\%), \nonumber \\ 
\left(\frac{BL}{2M}\right)_{\rm gal} &<& 12.7 \times 10^{-2} \, (99.9\%). \nonumber
\end{eqnarray}
These constraints are consistent with, but weaker than, those found from the WUPPE data.

\subsection{The Crab Nebula}
The polarization of X-ray light from the Crab nebular was reported in
\cite{Weisskopf}.  The measured linear polarization fraction was $p=18\pm 4\%$ at a
frequency of $\omega=5.2\keV$ and $p=16 \pm 2 \%$ at a
frequency of $2.6\keV$. This confirmed the hypothesis of
synchrotron X-ray emission.  The Crab nebula is at a distance of 2kpc
from the solar system so photons from the Crab nebular pass through
$\mathcal{O}(1)$ regular magnetic domains and $\mathcal{O}(100)$
random magnetic domains to reach the earth.  When $\vert m_{eff}^2\vert \approx \omegap^2  = 6.4 \times 10^{-12}\eV$, we have $\lambda_{\rm osc} \approx 6$\AA; $N\omega_{\rm crit} \approx 2\,{\rm keV}$.  Thus both measurements are in the $\omega \gtrsim N\omega_{\rm crit}$ region,  where  $p$ is almost independent of frequency.  The spectral resolution of these measurements is $\delta \lambda \lesssim 0.7$\AA $\ll \lambda_{\rm crit}/N$.

The linear polarization fraction,  $p$ is given by a probability
distribution even if we are in the maximal mixing regime and so the
amount of information one can extract from a single measurement is
limited. We found that the average polarization fraction for a set of
objects in the maximal mixing limit is $\geq 0.57$.  However, if
$p_{0} \ll 0.16$ initially, one would still expect to measure $p
\lesssim 0.16-0.18$ for a given object about $17\%$ of the time.  Even
the possibility of maximal mixing at X-ray frequencies cannot therefore be ruled out by the Crab Nebula data.

\subsection{Type Ia supernovae}

In \cite{Wang96,Wang97} supernova polarimetry data published before
1996 was studied.   The degree of polarization of light from type Ia supernovae  was less than $0.2-0.3\%$.  In
\cite{Wang:2003ge} high-quality spectro-polarimetry data was reported for
SNIa 2001e1.  It was found that the maximum linear polarization of the
light from the supernovae was $p \approx 0.2-0.3\%$.  The supernova
was observed at frequencies $\omega\approx
1.4 \eV - 3.8 \eV$ and the spectral
resolution of the polarimeter was $\delta\lambda\approx
12.7$\AA.  The Stokes parameters were later re-binned into $\delta
\lambda=15$\AA bins.  The supernova lies at a redshift of
$z\approx 5\times 10^{-3}$ corresponding to a distance of roughly $20\Mpc$. 

If, as light travels from the supernova to the earth, mixing with
chameleons occurs mostly in the intergalactic medium (as opposed to in
galaxies or clusters) then the PVLAS bound rules out maximal mixing.
The critical frequency for weak mixing in the intergalactic medium is
$\omega_{\rm crit}^{\rm (IGM)} \approx  3.4(L_{\rm IGM}/\Mpc)\eV$,
where $L$ is the coherence length of the IGM magnetic field. Since
$NL_{\rm IGM} = 20\Mpc$ we have $N\omega_{\rm crit}^{\rm IGM} =
68\eV$; $\lambda_{\rm osc} = 182$\AA.  Hence SNIa 2001e1 was observed
at frequencies $\omega\ll N\omega_{\rm crit}^{\rm IGM}$.  Any
chameleon induced polarization fraction would therefore be a highly
oscillatory function of the frequency.  For this chameleon signal to
survive the binning process and be detected one must ensure that the
polarimeter's spectral resolution and width of the wavelength bins
satisfy $\delta \lambda < 182$\AA.   In this case $\delta \lambda =
15$\AA and so the data can indeed be used to constrain chameleon-like
theories.  At the wavelengths observed, the chameleonic signal would
look like random noise that grows with frequency.  A full analysis of
the raw data reported in  \cite{Wang:2003ge} is beyond the scope of
this work. However, a preliminary analysis of the scatter in the
component of the Stokes vector perpendicular to the mean direction of
polarization, $P_{\perp}$, in the frequency range $4181 - 8631$\AA\
provides strong constraints.  At five different epochs, it was found
that $P_{\perp}$ was consistent with zero to within about $\pm 0.3
\%$.  We take $\vert P_{\perp} \vert < 0.3\%$ and extract approximate
upper confidence limits on the chameleon to photon coupling using the
method detailed in Appendix \ref{App:BUp}.  When $m_{\phi} < 2.5
\times 10^{-14}\eV$ in the IGM, and $L_{\rm IGM} \sim
\mathcal{O}(1\Mpc)$, we find the following $95$\% and $99.9$\% confidence limits
\begin{eqnarray}
\left(\frac{\vert B\vert L}{2M}\right)_{\rm IGM} < 5.2 \times 10^{-2}\, (95\%), \nonumber \\ 
\left(\frac{\vert B\vert L}{2M}\right)_{\rm IGM} < 7.2 \times 10^{-2}\, (99.9\%).\nonumber
\end{eqnarray}

If the intergalactic magnetic field is sufficiently small (i.e. $B \lesssim 10^{-11}\,{\rm G}$) mixing between light from the
supernova and chameleons will occur mostly in galaxies and galaxy
clusters.   SNIa 2001e1 is located in the nearly edge-on spiral galaxy
NGC 1448, however the line of sight does not intersect with either the
core or the disk of the host galaxy \cite{Wang:2003ge}. Additionally,
at only $20\Mpc$ away, light from SNIa 2001e1 does not pass through
any significant intra-cluster magnetic fields.  Our solar system
currently lies close to the midpoint of the galactic plane, and models
of the galactic magnetic field and electron density suggest that it
has a scale height above the midpoint galactic plane of about a
kiloparsec.  At the very least then, light from SNIa2001 will have
passed through roughly $1\kpc$ of the random galactic magnetic field.
For the random galactic magnetic field we have $\lambda_{\rm
  crit}^{\rm rand}/N \approx 12$\AA, where $N \approx 50$ when $\vert
m_{\rm eff}^2 \vert \approx \omegap^2 \approx 6.4 \times
10^{-12}\,{\rm eV}$.  Therefore  $\delta \lambda \sim \lambda_{\rm
  crit}^{\rm rand}/N \approx 12$\AA, and a  chameleon signal could be detected.  Since $\delta \lambda \sim \lambda_{\rm crit}^{\rm rand}/N$, the chameleon induced polarization would look like random noise.  Again a preliminary analysis of the data of  \cite{Wang:2003ge}, gives the follow 95\% and 99\% confidence limit, where we have assumed $m_{\phi} < \sqrt{2}\omegap \approx 9 \times 10^{-12} \eV$ in the galaxy and that the coherence length of the random component of the galaxy magnetic field is $\mathcal{O}(20\pc)$ so that $\lambda_{\rm crit} \approx 608$\AA:
\begin{eqnarray}
\left(\frac{\vert B\vert L}{2M}\right)_{\rm rand} &<& 0.14\, (95\%), \nonumber \\ 
 \left(\frac{\vert B\vert L}{2M}\right)_{\rm rand} &<& 0.18 \, (99.9\%).\nonumber
\end{eqnarray}

\subsection{High Redshift Quasars}
The optical and UV polarization  of some high redshift quasars have
been measured  \cite{Impey:1994qh,Shields} often using the HST FOS.
Below frequencies $\omega\sim 1 \eV$  the quasars have a polarization
of about 1\% but there is an interesting rise in the polarization
above frequencies $\omega\approx 2.5 \eV$. At electron-Volt
frequencies mixing between photons and chameleons is expected to be
highly frequency dependent. The HST FOS has a nominal spectral
resolution of $2-4$\AA, which is in principle good enough to resolve
the expected  chameleon signal if $m_{\phi} \ll 6.4 \times
10^{-12}\eV$ in galaxies or galaxy clusters. The data in
\cite{Impey:1994qh,Shields} is then rebinned with bin widths of
$\delta \lambda = 32 - 270$\AA.  Extracting the most stringent
constraints on chameleon theories would require a full reanalysis of
original data.  This is beyond the scope of this article.  However, by
analysing the data of Impey \etal \cite{Impey:1994qh} for object PG
1222+228 at $z \approx 2$ as rebinned and presented in
Ref. \cite{Shields} we can extract useful constraints.  Specifically,
we focus on the spread of the  Stokes parameter that is perpendicular to
the mean polarization angle.  The light from this QSO will have
travelled at least  $\approx 1\kpc$. We assume that the coherence length of the random component of our galaxy's magnetic field, $L_{\rm rand}$, is $\mathcal{O}(20\kpc)$. Making the conservative assumption that the total path length through our galaxy's magnetic field is $1\kpc$, when $m_{\phi} < \sqrt{2}\omegap \approx 9 \times 10^{-12}\eV$ in the galaxy, we find the following 95\% and 99\% confidence limits:
\begin{eqnarray}
\left(\frac{\vert B\vert L}{2M}\right)_{\rm rand} &<& 0.6\, (95\%), \nonumber \\ 
\left(\frac{\vert B\vert L}{2M}\right)_{\rm rand}  &<& 1.1\, (99.9\%).\nonumber
\end{eqnarray}
We expect that a full reanalysis of the original data would raise this limit greatly as currently the bounds are considerably weakened by the relatively large size of the wavelength bins (compared to $\lambda_{\rm osc}$).

If there is a sufficiently strong intergalactic magnetic field then this would also produce chameleon-photon mixing.  We make the conservative assumption that IGM magnetic fields only go out as far as $z=1$, so that the propagation distance through the IGM  magnetic field is about $2.5\,{\rm Gpc}$.  Assuming the IGM magnetic field is coherence over roughly megaparsec scales, we find the following confidence limits when $m_{\phi} < \sqrt{2}\omegap \approx 2.5 \times 10^{-14}\eV$ in the IGM:
\begin{eqnarray}
\left(\frac{\vert B\vert L}{2M}\right)_{\rm IGM} &<& 1.4 \times 10^{-2} \, (95\%), \nonumber \\ 
\left(\frac{\vert B\vert L}{2M}\right)_{\rm IGM} &<& 2.1 \times 10^{-2}\, (99.9\%). \nonumber
\end{eqnarray}
These constraints are particularly strong because the quasar is so far away, and as such the light from it travels through many different coherent regions, $\sim \mathcal{O}(2500)$, of any IGM magnetic field. This counter balances the loss of information due to the relatively large width of the wavelength bins.   A full reanalysis of the raw data would likely raise these bounds on $M$.
\subsection{Gamma Ray Bursts}

 Measurements of linearly polarized gamma rays have been made for four GRBs and these are summarised in Table \ref{tabGRB}.
 \begin{table}[ht]
 \caption{ \label{tabGRB} GRB Polarization Measurements}
 \begin{tabular}{l l c c}
 \hline
 \hline
GRB930131 & \cite{Willis} & $0.35 < p < 1$ & $3\keV < \omega < 100\keV$\\
GRB960924 &\cite{Willis} & $0.5 < p < 1$ & $3 \keV < \omega < 100\keV$\\
GRB041219a &\cite{McGlynn} & $0.56 < p < 1$ & $100\keV < \omega < 350\keV$\\
GRB021206 &\cite{Coburn} & $0.6 < p < 1$ & $0.15 \MeV < \omega < 2\MeV$\\
\hline
\end{tabular}
\end{table}
The last observation has been challenged \cite{Rutledge}. 
GRBs are the only objects we consider that are believed to be highly
polarized initially.  Theory predicts the emission of  highly linearly polarized light with
$0.6< m_{\rm l} <0.8$ due to synchrotron emission.  This hypothesis was  confirmed by
observations of polarization in the GRB afterglow \cite{Lazzati}.  

Mixing at gamma ray frequencies is maximal in all galaxies and clusters
regions if $M\ll 10^{9}\GeV$: maximal in galaxies if $M \lesssim \,{\rm few}\, \times 10^{9}\GeV$ and in the ICM if
$M \lesssim 4 \times 10^{11}\GeV$.  If $B = 10^{-9}\,{\rm G}$ in
the IGM, then maximal mixing  would occur if  $M \lesssim 5 \times 10^{10}\GeV$; however if $m_{\phi} \ll 2.2 \times 10^{-12}\eV$ in the IGM then this scenario is strongly ruled out by the bounds obtained above.   If the mixing is maximal the mean observed linear
polarization at high frequencies should be $\bar{p} \geq 0.57$; consistent with current GRB
observations.  It is not possible to make a more precise prediction
than this without knowing more accurately the initial polarization of
the GRB.  A better understanding of the central engine of the GRB and    better polarimetry for GRBs 
would allow strong constraints to be placed on the chameleon model. If future observations constrain
$\bar{p} <0.57$  maximal mixing in the chameleon model would be ruled out, and strong constraints on $M$ would follow.  If $\bar{p} >0.8$ is observed such a high degree of polarization cannot be  explained by the synchrotron mechanism and a chameleonic explanation would be favoured. 

If $M$ is very large the mixing between chameleons and light from
GRBs would be weak.  Then it becomes difficult to put bounds on the
chameleon model both because of the limitations of polarimeters  and, if there is no intergalactic magnetic field, the difficulty of estimating how many magnetic domains have been traversed.

\subsection{CMB Polarization}
The upcoming Planck satellite will measure the polarization of the CMB
to a high degree of accuracy.  However it is extremely hard to
estimate how many magnetic domains radiation from the CMB would have
passed through, particularly as if there is an intergalactic magnetic
field it is not known whether this field is primordial.  Neglecting the intergalactic magnetic field
it might be possible to use galaxy and cluster surveys to estimate how
many magnetic domains the radiation had passed through.  However
because the frequency of CMB radiation is so low mixing with the
chameleons will be weak and highly oscillatory and the amplitude of
these oscillations is damped as $\omega^2$.  A weak and highly oscillatory chameleon signal would be very hard to detect.  

\subsection{Summarised Constraints}
The tightest constraints on the chameleon to matter coupling come from
the WUPPE starlight polarization data, in the context of photon to
chameleon conversion in the galaxy, and from HST FOS measurements of
the polarization of high redshift quasars in the context of conversion
in the intergalactic medium.  Our preliminary analysis of starlight
polarization data appears to provide a non-zero lower bound on $1/M$,
however for the purposes of this discussion we only consider the upper
bounds on $1/M$ here.  For the IGM we took the coherence length, $L$,
to be $1 \,{\rm Mpc}$ and for the galaxy we assumed, $L = 20\,{\rm
  pc}$, however the precise values of these quantities do not greatly effect the upper bounds on $BL/2M$.  Taking these typical values for $L$ and $B \approx 3\,\mu{\rm G}$ for the strength of the random component of the galactic magnetic field, we find at $95\%$ confidence:
\begin{eqnarray}
M &>& 1.3 \times 10^{9}\GeV, \\
M &>& 1.1 \times 10^{11}\GeV \left(\frac{B_{\rm IGM}}{10^{-9}\,{\rm G}}\right).
\end{eqnarray}
At $99.9\%$ confidence we find similarly
\begin{eqnarray}
M &>& 1.1 \times 10^{9}\GeV, \\
M &>& 7.3 \times 10^{10}\GeV \left(\frac{B_{\rm IGM}}{10^{-9}\,{\rm G}}\right).
\end{eqnarray}
In both cases the upper constraint applies if $m_{\phi} \lesssim 1.3
\times 10^{-11}\eV$ in the galaxy and the lower one if $m_{\phi}
\lesssim  2.5 \times 10^{-14}\eV$ in the IGM.  Since $B_{\rm IGM}$ is
currently unmeasured, the strongest constraint on $M$ is comes from
the starlight polarization measurements, the interpretation of which
relies only on knowledge of the galactic magnetic field. If $B_{\rm
  IGM} \gtrsim 10^{-11}\,{\rm G}$, however, then the constraints
coming from high redshift quasars currently provide the tightest lower
bounds on $M$.  In terms of chameleon theories, these constraints
represent an improvement of almost 2.5 order of magnitude  on the
previous best lower bounds on $M$ coming from laboratory tests,
specifically $M > 3.9 \times 10^{6}\,\GeV$ at $99.9\%$ confidence from
the GammeV experiment \cite{GammeV}.   

GammeV and other similar laboratory tests do not constrain the OP
model.    Provided $m_{\phi} \lesssim 1.3 \times 10^{-11}\eV$ in the galaxy,
and it was shown in  \S \ref{sec:model:OP} that this is expected to be
the case, the starlight polarization constraint on the OP model translates to:
\begin{equation}
\xi_{\rm F}^{-1/2} M_{0} > 1.6 \times 10^{3} \TeV \, \left\vert\frac{\delta \alpha}{10^{-6}\alpha}\right\vert ^{1/2},
\label{OPconstraint}
\end{equation}
at $99.9\%$ confidence where $\delta \alpha/\alpha$ is the fractional
difference between $\alpha$ in the laboratory and $\alpha$ in a
background such as the galaxy.  For comparison, the previous best
constraints were $M_0 > 15\TeV$ and $\xi_{F}^{-1/2} M_{0} > 3\TeV$.
If $\vert \delta \alpha /\alpha \vert \sim O(10^{-6})$ as suggested by
the analysis of Webb \etal\, \cite{webb}, then this represents an
improvement of two to three  orders of magnitude. We note that if a subsequent analysis were to confirm the lower bound on $1/M$ found from starlight polarization measurements, then both these measurements and the Webb \etal value of $\delta \alpha/\alpha$ could be explained by an OP model with 
$$
\xi_{\rm F}^{-1/2} M_{0} \sim 3\, -\, 8 \times 10^{3}\TeV.
$$
and $\Lambda_{1} \sim \left(\mathcal{O}(10^{-2}) -
\mathcal{O}(10^{2})\right) \eV$; $\phi_{m} \sim (10\,-\,20)\TeV$.

\section{Circular Polarization:  A smoking gun?}\label{sec:CP}
The total polarization due to chameleon-photon mixing grows as the square of the frequency of the light until it reaches a critical frequency at which the mixing becomes maximal.  It has not been possible to detect this frequency pattern in current linear polarization data.  Objects whose initial polarization is well constrained have not been observed 
over a wide enough frequency range or to the required accuracy to see any such signal.  Certain GRBs have
been observed over a very large range of frequencies as they evolve,
but because their initial linear polarization is not known accurately,
and generally does not satisfy $p_{0}\ll 1$, it is difficult to search for a chameleon signal in this data.
\begin{figure}[htb!]
\includegraphics*[width=6.5cm]{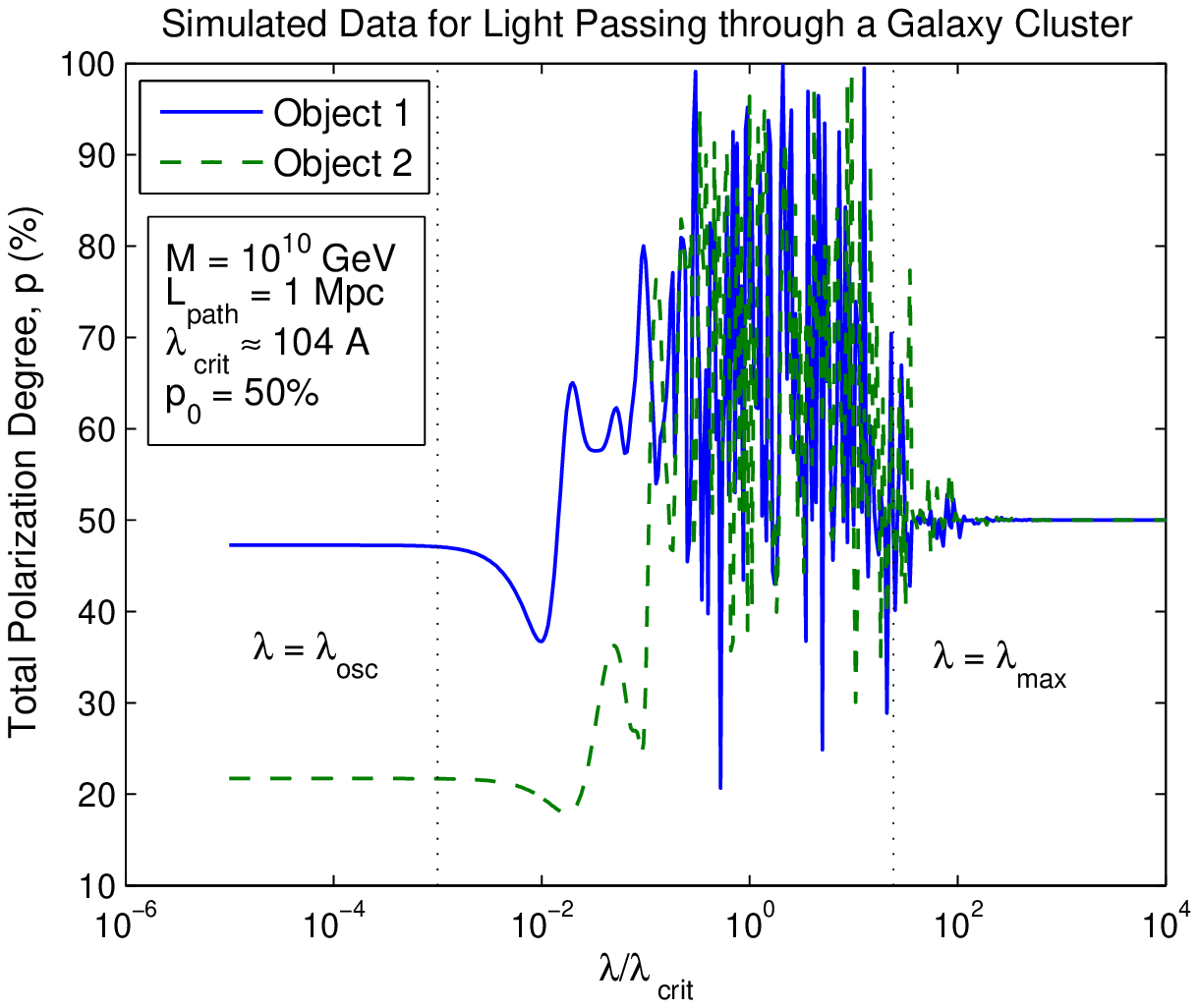}
\includegraphics*[width=6.5cm]{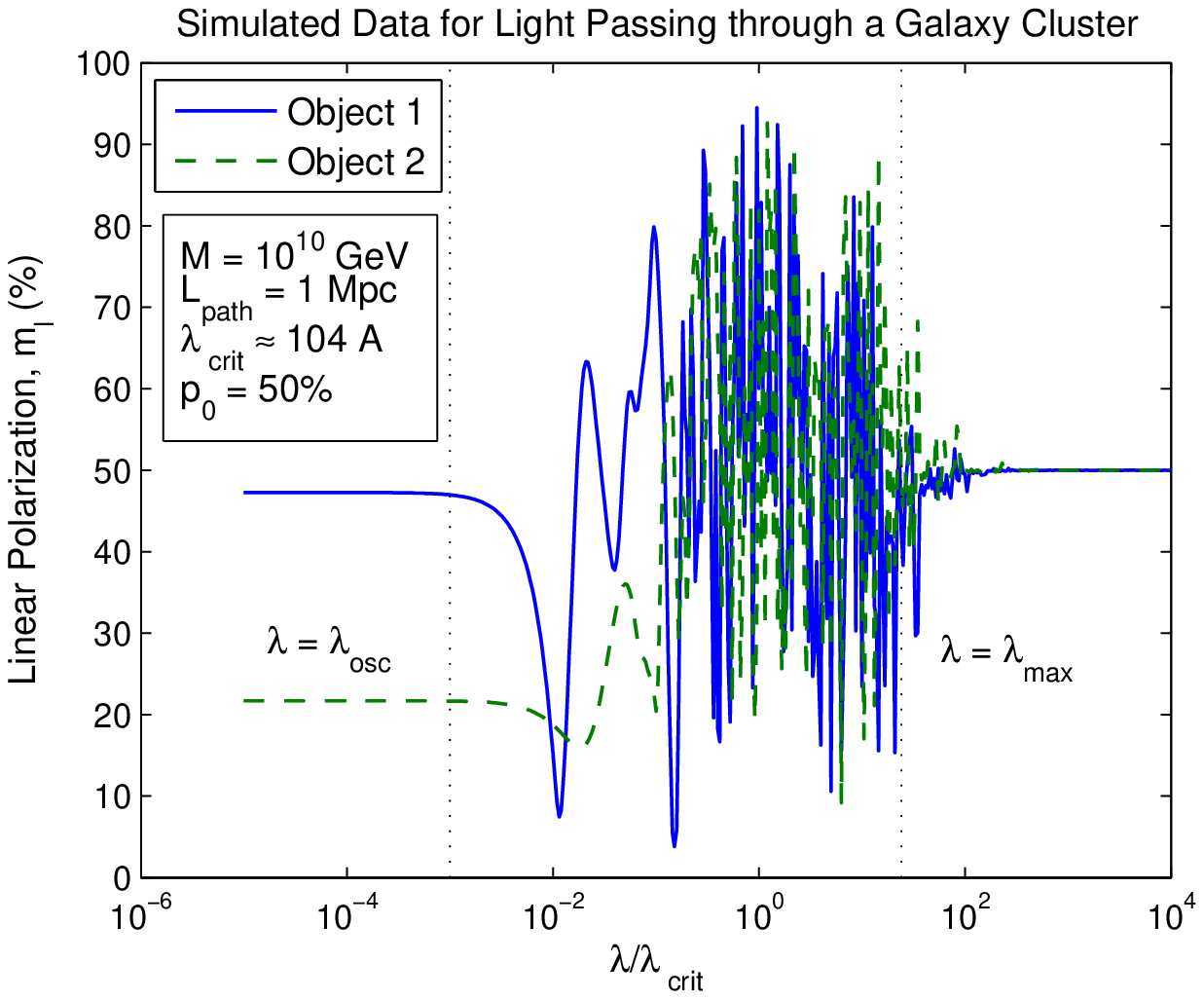}
\includegraphics*[width=6.5cm]{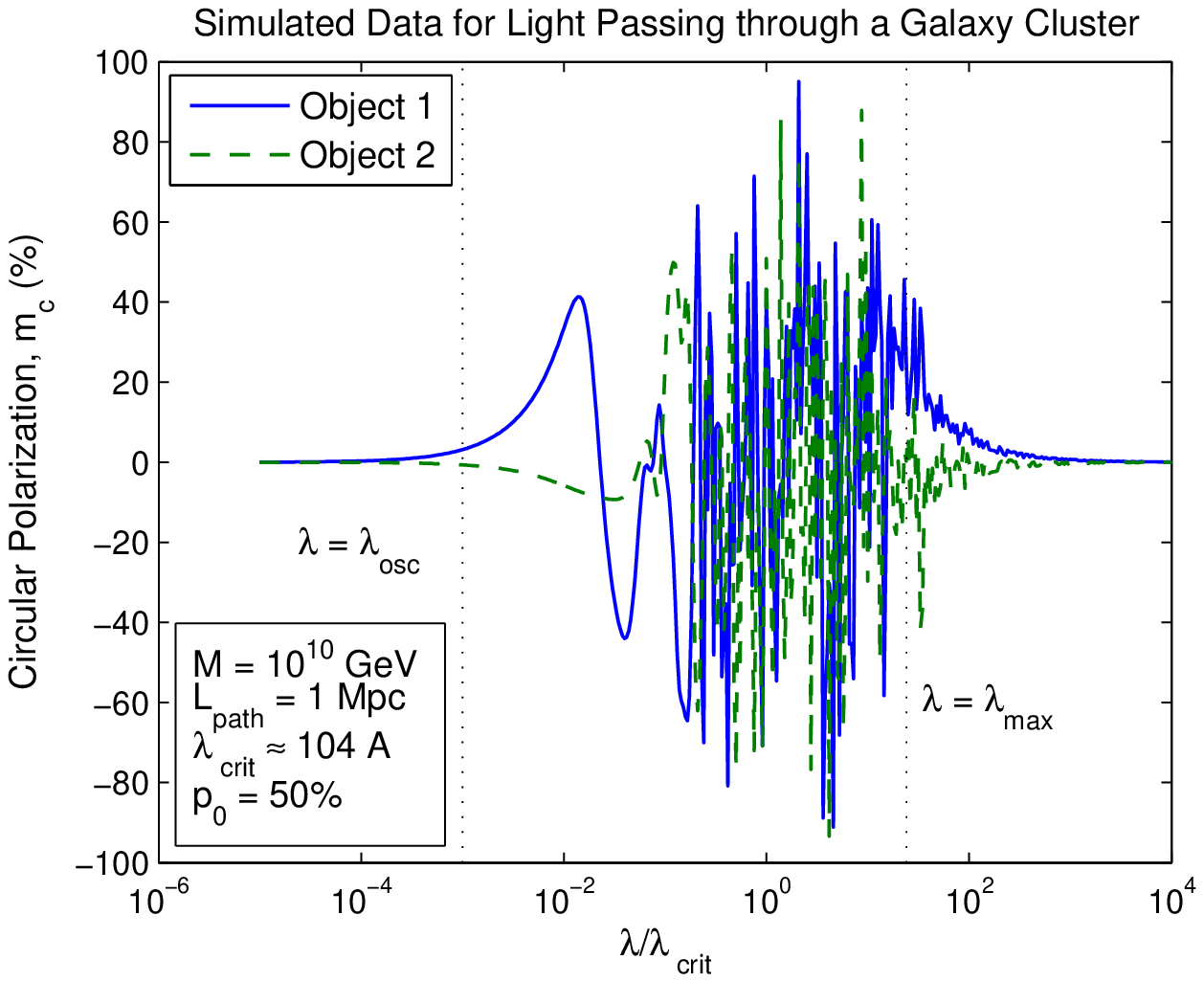}
\caption[]{Simulated data for two objects whose light has passed through roughly $1\Mpc$ of the magnetic field of a typical galaxy cluster. We have assumed $m_{\phi} \ll 2.2 \times 10^{-12}\eV$ and $M = 10^{10} \GeV$; which corresponds to strong mixing for wavelengths of $\lambda_{\rm max} \approx 24 \lambda_{\rm crit}$.  We have assumed that both objects have little or no intrinsic circular polarization, and are $50\%$ linearly polarized prior to chameleon mixing. Qualitatively  similar behaviour is seen for different values of the intrinsic linear polarization, $m_{l0}$, and in particular the behaviour CP fraction does not depend greatly on $m_{l0}$. }
\label{clustdata}
\end{figure}

The production of circular polarization by chameleon photon mixing has
a much more interesting signature. One does not usually expect significant amounts of intrinsic circular polarization (CP) for astrophysical objects.  We noted above in \S \ref{sec:Optics:Sig} that chameleonic CP production is peaked over a potentially large range of wavelengths (when $N \gg 1$) i.e. $\lambda_{\rm osc}  = \lambda_{\rm crit}/N < \lambda < \lambda_{\rm max}$,
where
$$
\lambda_{\rm max} = \lambda_{\rm crit}\, {\rm max}\left(1, \frac{\sqrt{N} BL}{\pi M} \right).
$$
Importantly, in this band, the chameleon induced circular polarization is the same order of
 magnitude as the chameleon produced linear polarization, and both exhibit
 a highly oscillatory frequency dependence in this region.  Outside of
 this wavelength band, the chameleon contribution to the circular
 polarization is much smaller than to the linear polarization.  If
 mixing is maximal, $q \sim O(1)$ is expected. Neither the magnitude,
 the shape, nor the oscillatory frequency dependence  of the chameleon
 induced circular polarization peak is likely to caused by any other
 process.  The  observation of this peak  could be considered a smoking gun for
 chameleon-photon mixing, and if such a structure could be ruled out
 then strong constraints on chameleon like theories would follow.   In
 particular if $O(1)$, highly frequency dependent values of $q$ are
 not seen in the region $\lambda_{\rm osc}  = \lambda_{\rm crit}/N <
 \lambda < \lambda_{\rm crit}$, maximal mixing could be ruled out,
 immediately limiting $M \gtrsim 10^{10} - 10^{11}\GeV$.  Strong
 constraints would result   if the CP  of a distant object whose light
 was known to pass through the magnetic field of a galaxy cluster
 could be constrained in the region $\lambda_{\rm osc} < \lambda <
 \lambda_{\rm crit}$.  To ensure the maximal sensitivity to
 chameleonic effects however the spectral resolution would have to be
 $\approx \lambda_{\rm osc}$ or smaller, which for a cluster would
 require $\delta \lambda \lesssim 0.1$\AA. Assuming light travels
 roughly $1\kpc$ through the galaxy, $1\Mpc$ through a galaxy cluster
 and about $2.5\,{\rm Gpc}$ through the IGM the typical expected
 values of $\lambda_{\rm osc}$ and $\lambda_{\rm crit}$ are shown below in  Table \ref{tab2}. We have assumed  $m_{\phi} \ll 6.4 \times 10^{-12}\eV$ in the galaxy, $\ll  2.2 \times 10^{-12}\eV$ in the ICM and $\ll 1.8 \times 10^{-14}\eV$ in the IGM.

 \begin{table}[ht]
 \caption{ \label{tab2} Position of $CP$ Peak}
 \begin{tabular}{l c c}
 \hline
 \hline
 Environment & $\lambda_{\rm osc}$ & $\lambda_{\rm crit}$ \\
 \hline
Galaxy  & $12$\AA & $608$\AA \\
ICM &$0.1$\AA & $104$\AA \\
IGM & $1.5$\AA & $3600$\AA \\
\hline
\end{tabular}
\end{table}

\begin{figure}[htb!]
\includegraphics*[width=6.5cm]{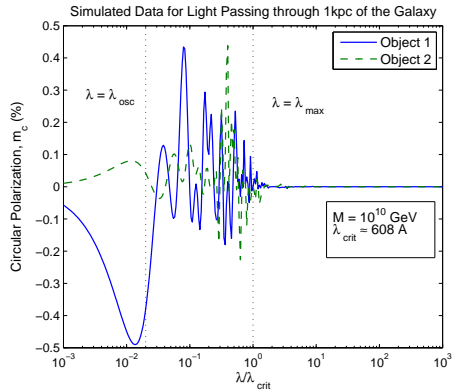}
\caption[]{Simulated data for two objects whose light has passed through roughly $1\kpc$ of our galaxies magnetic field. We have assumed $m_{\phi} \ll 6.4 \times 10^{-12}\eV$ and $M = 10^{10} \GeV$.  We have assumed that both objects have little or no intrinsic circular polarization. Potentially detectable levels of CP are seen between $\lambda_{\rm osc} \approx  12$\AA and $\lambda_{\rm crit} \approx 608$\AA. }
\label{galplot}
\end{figure}

FIG. \ref{clustdata} shows simulated data for two objects (e.g. GRBs),
with $50\%$ initial linear polarization and no intrinsic circular
polarization, whose light has passed through about $1\Mpc$ of the
magnetic field of a galaxy cluster.  The wavelength, $\lambda$, in
this plot should be interpreted as $\lambda_{\rm m}/(1+z_{\rm clust})$
where $\lambda_{\rm m}$ is the measured wavelength and $z_{\rm clust}$
is the redshift of the cluster. We have assumed, as is generally the
case, that $m_{\phi}\ll 2.2\times 10^{-12}\eV$. We also have taken $M
= 10^{10}\GeV$, which corresponds to strong mixing for $\lambda
\lesssim \lambda_{\rm max}$; in this case $\lambda_{\rm crit} \approx
104$\AA.  We can see that chameleonic production of polarization
begins for $\lambda \lesssim \lambda_{\rm max}$ and here $\lambda_{\rm
  max} \approx 24 \lambda_{\rm crit} \approx 2500$\AA\ (i.e.  in the
middle UV part of the spectrum).  Very similar behaviour is seen  for
different choices of the intrinsic polarization. Between $\lambda_{\rm
  osc}$ and $\lambda_{\rm max}$ both the linear polarization, $m_{\rm
  l}$, and the circular polarization, $m_{\rm c}$, are, as expected,
highly frequency dependent and as we expect from the strong mixing
scenario  when $\lambda_{\rm osc} < \lambda < \lambda_{\rm max}$ the
magnitude of both $m_{c}$ and $m_{l}$ oscillates between $0\%$ and
$100\%$.  For $\lambda < \lambda_{\rm osc} $, $m_{\rm l}/100\%$
settles to some, essentially random value between $0\%$ and $100\%$,
and $m_{c} \rightarrow 0$. If there is little or no intrinsic circular
polarization, the behaviour of $m_{c}$ does not depend greatly on the
value of the intrinsic linear polarization.  If such measurements
could be made it should be straightforward to either detect or rule
out   values of $M$ for which strong mixing in clusters is expected to occur; $M \lesssim 10^{11}\GeV$.    All light that reaches us from distant objects will have passed through at least a part ($\sim O(1)\kpc$) of our own galaxy's magnetic field.  In FIG. \ref{galplot}, we  show sample circular polarization data for two objects (with little or no intrinsic circular polarization) whose light has passed through $1\kpc$ of our own galaxy's magnetic field (corresponding to about $50$ regions of the random field, and one of the regular field). Most of the CP production is due to the random field.  We take $M= 10^{10}\GeV$ and $m_{\phi} \ll 6.4 \times 10^{-12} \eV$.  In this case we are in the weak mixing limit, and $\lambda_{\rm max} = \lambda_{\rm crit} = 608$\AA.  We can see that in the region $\lambda_{\rm osc} < \lambda < \lambda_{\rm crit}$,  potentially detectable levels of CP (between $0.1\%$ and $0.5\%$) are typical.  

If $m_{\phi} < 9 \times 10{-12}\eV$, measurements of CP between $\lambda \sim O(1)$\AA\ and $\lambda \sim O(1000)$\AA\ for astrophysical objects should allow one to detect or rule out theories with $M \lesssim 10^{10}\GeV$.

Thus far, circular polarization has been measured for a number of different astronomical sources; for certain stars observed in the near infrared in \cite{Serkowski72,Serkowski74}, for zodiacal light  in \cite{Wolstencroft72}, for the 
Orion molecular cloud in \cite{Chrysostomou}, for some relativistic jet sources at radio wavelengths  in \cite{Macquart}.  However for all of these observations mixing between photons and chameleons is weak, and the frequency resolution of the observations is not good enough to detect a chameleon signal. Additionally all such observations have been made at wavelengths outside the expected $\lambda_{\rm osc} - \lambda_{\rm max}$ position of any chameleonic CP peak.

\section{Summary}\label{sec:sum}

Theories of physics beyond the standard model typically predict the
existence of new scalar fields.  If these scalar fields do exist it is
important to understand both their self interactions and their
interactions with the other fields present in the model in order to test
and constrain the theory.  In this article
we have studied the results of a coupling between scalar fields and
photons on observations of astrophysical objects.  Specifically we
have studied the scalar fields of the chameleon and Olive-Pospelov
models, which are strongly interacting in low density environments yet
currently undetected in the laboratory.  For simplicity we refer to
both types of scalar field as chameleons.

If the chameleon field couples to photons then in the presence of a
background magnetic field  the chameleon mixes with the component
of the photon polarized orthogonally to the direction of the magnetic
field.  We have studied the effect of this mixing on light beams
passing through a large number of randomly oriented homogeneous
magnetic domains, in order to predict the effects of chameleon-photon
mixing on observations of light from astrophysical objects.  Typically both linear and circular
polarization are induced in the light beam by mixing in such an
environment.

We found analytic solutions to the equations describing the mixing in
two important limits.  In the weak mixing limit the
polarization fractions induced by chameleon photon mixing are highly
wavelength dependent. If the light is not polarized at the source
the averaged values of the total and circular polarization scale as
$N\Pphi$, that is as the product of the number of domains traversed and the probability of mixing in any one domain.  This limit is generally appropriate when one is considering the chameleon induced polarizations at wavelengths longer than roughly 1000\AA.  In the maximal mixing limit, which applies when the chameleon-photon coupling is strong and the wavelength is sufficiently short, little or no
circular polarization is produced by the mixing, but the production of linear polarization is at its strongest.  The distribution of the total polarization fraction after mixing in a large number of domains is
independent of the parameters of the chameleon model, and instead
depends only on the initial polarization of the light.  The average
value of the total polarization fraction is always greater than
$(\pi/2)-1 \approx 0.57$ in the maximal mixing limit.

Numerical simulations confirm the analytic analysis.  In particular
they clearly demonstrate the existence of two wavelength scales $\lambda_{\rm osc} =
\lambda_{\rm crit}/N = 4\pi^2/\vert m_{\rm eff}^2 \vert L_{\rm path}$, and $\lambda_{\rm max} = \lambda_{\rm crit}{\rm max} \left(1, \frac{BL}{\pi \sqrt{N} M}\right)$
which determine the shape of the polarization signal.  Here $B$ and $L$ are the strength and domain size of the magnetic field. $g_{\phi \gamma \gamma} = M^{-1}$ is the coupling between two photons and the scalar field. The linear
polarization is greatest for $\lambda \lesssim \lambda_{\rm max}$,
and the circular polarization is peaked for $\lambda_{\rm
  osc}\lesssim\lambda\lesssim  \lambda_{\rm max}$.  Both polarization
fractions are highly frequency dependent for $\lambda \gtrsim \lambda_{\rm
  osc}$.  This highly oscillatory behaviour means that observations of
polarization at these wavelengths must be performed with a sufficiently good
spectral resolution if any chameleon induced signal is to be resolved.

We have considered a wide variety of astrophysical observations and have used these to constrain the
parameters of the chameleon model.  From observations of starlight polarization
in our galaxy we show that at the 99\% confidence level $M>1.1\times10^9\mbox{ GeV}$,
which is an improvement of over two orders of magnitude on the previous
best constraints on $M$.  The equivalent constraint on
the Olive-Pospelov model is given in (\ref{OPconstraint}).  Both constraints could, however, be evaded if the potential of the scalar field, $V(\phi)$, is chosen so that the field is sufficiently heavy in regions with the density of our galaxy: $m_{\phi} \gg 10^{-11}\eV$.  Constraints from objects outside the galaxy are limited by our lack of
knowledge about a possible intergalactic magnetic field, $B_{\rm IGM}$. If, however, $B_{\rm IGM} \approx 10^{-9}\,{\rm G}$ and is coherent over roughly ${\rm Mpc}$ scales, then the lower bounds on $M$ and the OP model coupling scale are raised by roughly two orders of magnitude.

The circular polarization signal predicted from chameleon-photon
mixing in a large number of randomly oriented magnetic domains was shown to be very
distinctive.  Its frequency dependence is unlikely to have been caused
by any other physical process, particularly as astrophysical objects
do not normally produce significant amounts of circular polarization.
To date, no observations of astrophysical circular polarization have
yet been made with sufficient accuracy to allow us to search for a
chameleon signal. Nonetheless, we have shown how future observations of circular
polarization in the wavelength range $O(1)-O(1000)$\AA\ would be a smoking gun
for chameleon-photon coupling. 

We have also reported a seemingly strong statistical preference in observations of starlight polarization in our galaxy for the presence of a chameleon-like field. Precisely, at the 99\% confidence level, we find
\begin{equation}
\left(\frac{|B|L}{2M}\right)_{\rm rand}=(6.27\pm 1.91)\times 10^{-2}
\end{equation}
where $B$ and $L$ are the strength and domain size of the random component of the galactic magnetic field.  Formally, the central value deviates from zero (the value for
a theory without a chameleon) by more than $10\sigma$.  It must be stressed, however, that this
is only a preliminary analysis and we have only performed it for three out of a possible 121 objects. Before a detection could be claimed with any confidence, a full study of the possible backgrounds and systematics for
these observations that could bias one towards larger values of $1/M$ would have to be undertaken.  Based on initial numerical simulations of data, it does, however, seem unlikely that polarization due to interstellar dust would produce such a strong signal.  

In summary: astrophysical polarization measurements currently provide the strongest constraints on any coupling between photons and the scalar field, for many  chameleon and chameleon-like theories such as the Olive-Pospelov model, improving on previous constraints by more than two orders of magnitude.  Furthermore, future measurements of linear and, in particular, circular polarization at short wavelengths (i.e. $\lesssim 2000$\AA)  could provide one of the best tools in the continuing search for such scalar fields. 

\vspace{0.5cm}

\noindent{\bf Acknowledgements:} CB, ACD and DJS are supported by
STFC. We are grateful to J. D. Barrow, Ph. Brax, C. van de Bruck,
D. F. Mota, C. Spyrou and W. Sommerville for helpful conversations. 

\appendix

\section{Fluctuating Electron Density}\label{app:ench}
In Ref. \cite{Carlson94} it was shown that fluctuations in the
electron density, $n_{\rm e}$, and hence the plasma frequency,
$\omega_{\rm pl}$, could lead to a significant enhancement of the
photon to axion-like-particle (ALP) conversion rate when $m_{\phi} \ll
\omega_{\rm pl}$ and $\vert \Delta \vert \approx \omega_{\rm pl}^2 L
/4 \omega \gg 1$.  In this appendix we reproduce this analysis and
show that  conclusions about the magnitude of such an enhancement effect are modified in the light of more recent models of the electron-density in our galaxy (specifically the NE2001 model) than those used in Ref. \cite{Carlson94}.  We also extend the analysis to allow for fluctuations in the magnetic field $\mathbf{B}$.

We write $\omega_{\rm pl}^2(z) = \bar{\omega}_{\rm pl}^2(1+\delta_{n}(z))$ where $\delta_{n}(z) = \delta n_{e}(z)/n_{e}$. We also have $B = \vert \mathbf{B}_{\perp}\vert = \bar{B}(1+\delta_{b}(z))$. We are concerned with the limit $m^2_{\phi} \ll \omega_{\rm pl}^2$. Now in a single region of magnetic field the equations describing the evolution of the photon and the chameleon are:
\be
-\ddot{\gamma}_{\parallel} + \gamma_{\parallel,zz} = \omega_{\rm pl}^2(z) \gamma_{\parallel}, \nonumber \\ 
-\ddot{\gamma}_{\perp} + \gamma_{\perp,zz} = \omega_{\rm pl}^2(z) \gamma_{\perp} + \phi_{,z} \frac{B}{M}, \nonumber \\
-\ddot{\phi} + \phi_{,zz} = - \frac{B}{M}\gamma_{\perp,z}. \nonumber
\ee
We assume that $\omega \gg \omega_{\rm pl}$ and $\delta_{n}^{\prime}/\delta_{n} \ll \omega$.  We write the solution for $\gamma_{\parallel}$ thus:
$$
\gamma_{\parallel} = \gamma_{0}e^{i\omega(z-t)-ia(z)}.
$$
where we assume $\omega_{\rm pl} \ll \omega$ and so $\vert a_{,z} \vert \ll \omega$.  
We similarly write:
\be
\gamma_{\perp}(z) &=& \tilde{\gamma}(z) e^{i\omega(z-t)-ia(z)},\nonumber 
\\
\phi(z) &=& \tilde{\phi}(z) e^{i\omega(z-t)-ia(z)}. \nonumber
\ee
We then have:
\be
\tilde{\gamma}_{,z} &\approx & \frac{B}{M}\tilde{\phi}, \nonumber \\
\tilde{\phi}_{,z} &\approx & -\frac{B}{M}\tilde{\gamma} + \frac{i\omega_{\rm pl}^2}{2\omega} \tilde{\phi},\nonumber
\ee
We define $m_{\rm eff}^2 = m_{\phi}^2 - \bar{\omega}_{\rm pl}^2(1+\bar{\delta}_{n}) \approx  - \bar{\omega}_{\rm pl}^2(1+\bar{\delta}_{n})$ where
$$z\bar{\delta}_{n}(z) = \int_{0}^{z}\delta_{n}(z^{\prime})\,{\rm d}z^{\prime}.$$
We then let $x = m_{\rm eff}^2 z / 4\omega$, so $z= L$ implies $x =\Delta$ and remember that:
\be
\tan 2\theta = \frac{2\bar{B}\omega}{M\bar{m}_{\rm eff}^2}. \nonumber
\ee
where $\bar{m}_{\rm eff}^2 \approx -\bar{\omega}_{\rm pl}^2$.
Thus:
\be
\mathbf{u}_{,x} \equiv \bv \tilde{\gamma} \\ \tilde{\phi} \ev_{,x} = iM(x) \bv \tilde{\gamma} \\ \tilde{\phi} \ev,
\ee
where
\be
M(x) &=& \ba 0 & -i D(x)\tan 2\theta  \\ i D(x)\tan 2\theta  & -2\ea \\ &=& (\sigma_{3}-\mathbb{I}) + \sigma_{2} D(x)\tan 2\theta. \nonumber
\ee
where
$$
D(x) = 1+ \tilde{\delta}(x) \equiv \frac{1+\delta_{b}(z)}{1+\delta_{n}(z)}.
$$
In the above the $\sigma_{i}$ are the Pauli matrices.
We write $M(x) = M_0 + \sigma_2\tilde{\delta}(x)\tan 2\theta$ where $M_0 = (\sigma_{3}-\mathbb{I}) + \sigma_{2} \tan 2\theta$ and so:
\be
e^{i M_0 x} &=& e^{-ix}\left[\cos\left(\frac{x}{\cos 2\theta}\right) \right. \nonumber \\ &&\left.+ i(\sigma_3\cos 2\theta + \sigma_2 \sin 2\theta)\sin \left(\frac{x}{\cos 2\theta}\right)\right], \nonumber \\
&=& e^{-ix}\ba \sqrt{1-A^2(x)}e^{i\varphi(x)} & A(x) \\ -A(x) & \sqrt{1-A^2}e^{-i\varphi(x)} \ea, \nonumber
\ee
where 
\be
A(x) &=& \sin 2\theta \sin\left(\frac{x}{\cos 2\theta}\right), \nonumber \\
\tan \varphi(x) &=& \cos 2\theta   \tan \left(\frac{x}{\cos 2\theta}\right). \nonumber
\ee

We note that
\be
e^{-iM_{0}x} \sigma_{2} e^{i M_{0}x} = \left(\cos 2\theta \mathbf{n}_{1} + \sin 2\theta \mathbf{n}_{2}\right)\cdot\sigma
\ee
where
\be
\mathbf{n}_{1}(x) &=& \bv -\sin \left(\frac{2x}{\cos 2\theta}\right) \\ \cos 2\theta \cos \left(\frac{2x}{\cos 2\theta}\right) \\ -\sin 2\theta \cos \left(\frac{2x}{\cos 2\theta}\right)\ev, \nonumber \\
\mathbf{n}_{2}(x) &=& \bv 0 \\ \sin 2\theta \\ \cos 2\theta \ev. \nonumber
\ee
We then define $\mathbf{v} = e^{-i C(x)} e^{-iM_0 x} \mathbf{u}$  where
$$
C(x) = (\mathbf{n}_2\cdot \sigma) \tan 2\theta \int_{0}^{x} \tilde{\delta}(x^{\prime}) {\rm d}x^{\prime} \equiv \frac{\bar{\alpha}(x)x (\mathbf{n}_2\cdot \sigma)}{\cos 2\theta}.
$$
We then have:
\be
\mathbf{v}_{,x}  = i\sin 2\theta \tilde{\delta}(x)\left(\mathbf{n}(x) \cdot \sigma\right) \mathbf{v}. \label{vEqn}
\ee
where
$$
\mathbf{n} = \bv -\sin \left(\frac{2(1+\bar{\alpha})x}{\cos 2\theta}\right) \\ \cos 2\theta \cos \left(\frac{2(1+\bar{\alpha})x}{\cos 2\theta}\right) \\ -\sin 2\theta \cos \left(\frac{2(1+\bar{\alpha})x}{\cos 2\theta}\right).\ev
$$
When it is acceptable to do so we may solve Eq. (\ref{vEqn})
perturbatively, a sufficient condition is:
$$
\Vert \Delta \sin 2\theta  \tilde{\delta}(x)\Vert \ll 1,
$$
as $\mathbf{n}^2(x) = 1$. This may be satisfied if either $2\theta \ll 1$, $\Delta \ll 1$ or $\Vert \tilde{\delta}\Vert \ll 1$. To sub-leading order we have:
\be
\mathbf{v} &\approx&  \mathbf{N}(x)\mathbf{v}_{0} \equiv \left[\mathbb{I}+i\sin 2\theta\left( \mathbf{c}(x;\theta)\cdot \sigma\right) \right. \\ &&\left. - \frac{1}{2}\sin^2 2\theta\left(\mathbf{c}^2+i\mathbf{d}\cdot \sigma\right) \right]\mathbf{v}_0, \nonumber
\ee
where
\be
\mathbf{c}(x;\theta) &=& \int^{x}_{0} {\rm d}s\,\tilde{\delta}(s) \mathbf{n}(s;\theta), \\
\mathbf{d}(x;\theta) &=& \int^{x}_{0} {\rm d}s\,\tilde{\delta}(s) \left(\mathbf{n}(s;\theta)\times \mathbf{c}(s;\theta)\right).
\ee
We evaluate $\mathbf{N}(x)$ in the weak-mixing limit, $2\theta \ll 1$.  To do this we expand the diagonal terms in $\mathbf{N}(x)$ to order  $(2\theta)^2$ and the off-diagonal ones to order $2\theta$. We find
\be
\mathbf{N}  \approx \ba  1 - 2\theta^2(\Vert g \Vert^2 + i \tau) & 2\theta g^{\ast} \\ -2\theta g &  1 - 2\theta^2(\Vert g \Vert^2 - i \tau)\ea,
\ee
where:
\be
\tau &=& 2{\rm Re}(g)+h, \nonumber \\
g &=&  \int_{0}^{x} {\rm d}s\,\tilde{\delta}(s) e^{i\left(\frac{2(1+\bar{\alpha}(s)) s}{\cos 2\theta}\right)}, \nonumber \\
h &=& \mathbf{d}_{3}(x). \nonumber
\ee
Now at the end of a magnetic domain with length $L$, $z=L$ and $\mathbf{u} = \mathbf{u}(L)$ we have:
\be
\mathbf{u}(L) \approx  && e^{-i \Delta} \left[\cos\left(\frac{(1+\bar{\alpha}(L))\Delta}{\cos 2\theta}\right)\mathbb{I}  \right. \\ && \left.+ i\sin \left(\frac{(1+\bar{\alpha}(L))\Delta}{\cos 2\theta}\right)\right. \nonumber \\ && \left.\left(\sigma_{3}\cos 2\theta  + \sigma_{2}\sin 2\theta\right)\right]\mathbf{N}(L) \mathbf{u}(0) \nonumber
\ee
where $\mathbf{u}(0)$ is the initial value of $\mathbf{u}$.  In the weak mixing-limit, it follows that the probability of converting a photon to a chameleon is:
\be
\Pphi \approx 4\theta^2 \Vert \sin \left(\frac{(1+\bar{\alpha}(\Delta))\Delta}{\cos 2\theta}\right) - g(x=\Delta) \Vert^2.
\ee
The second term inside the $\Vert \cdot \Vert^2$ represents the
enhancement term from electron density fluctuations.  It is
straightforward to check that if $2\theta \ll 1$ we expect $g/ \Delta
\lesssim O(1)$ if, as expected, $\Vert \tilde{\delta}\Vert^2 \lesssim
O(1)$.  Therefore when $\Delta \ll 1$, i.e. at high frequencies, we do
not expect the new term to produce a large enhancement. This was also
noted by Carlson and Garretson in Ref. \cite{Carlson94}. We therefore
focus on the limit $\Delta \gg 1$. In this limit $\sin^2
((1+\alpha)\Delta/\cos 2\theta) \sim 1/2$ on average.   We denote the relative magnitude of the enhancement in photon to chameleon conversion by $E$ and
$$
E = 2\Vert g \Vert^2.
$$
When $E \ll 1$, the enhancement is negligible, and when $E \gg 1$ the enhancement is strong and represents a significant effect. In the weak-mixing limit in which we work $\Vert\alpha  \Vert \ll 1$, and so:
\be
E &\approx& 2\Vert \int_{0}^{\Delta}{\rm d}x\,\tilde{\delta}(x) e^{ix}\Vert^2 \\ &=& 2\int_{0}^{\Delta}\int_{0}^{\Delta}{\rm d}x\,{\rm d}y\,\tilde{\delta}(x) \tilde{\delta}^{\ast}(y) e^{i(x-y)}. \nonumber
\ee

We define the Fourier transform $\tilde{\delta}_{k}(\mathbf{k})$ of $\tilde{\delta}(x)$ thus
$$\tilde{\delta}(x) = \tilde{\delta}(\Delta z/L) = \int {\rm d}^{3}k \tilde{\delta}_k(\mathbf{k})e^{i k_{z} L x/\Delta}. $$
where $k_{z}$ is the component of $\mathbf{k}$ in the $\hatb{z}$ direction.  The power spectrum $P(k)$ is given the by expectation of $\tilde{\delta}_{k}(\mathbf{k})\tilde{\delta}_{k}^{\ast}(\mathbf{q})$ thus:
$$
\left\langle \tilde{\delta}_k(\mathbf{k})\tilde{\delta}^{\ast}_{k}(\mathbf{q}) \right\rangle = P(k) \delta^{(3)}(\mathbf{k}-\mathbf{q}).
$$
Thus:
\be
\left\langle E \right\rangle = 2\Delta^2 \int {\rm d}^3 k\, P(k){\rm sinc}^2\left(\frac{k_{z}L + \Delta}{2}\right),
\ee
where ${\rm sinc}(x) = \sin x /x$.  Electron-density and magnitude field fluctuations are often modeled by a Power spectrum with inner scale $l_0$ and outer scale $L_{0}$, and a power law behaviour between these two scales i.e.:
\be
P(k) = \frac{C^2}{\left[L_{0}^{-2} + k^2\right]^{\alpha/2}}e^{-\frac{k^2 l_{0}^2}{2}}.
\ee 
When, as is the case for visible light, $\Delta l_{0}/L_{0} \ll 1$, the role of $l_{0}$ in the estimate for $E$ is negligible, and we may approximate by setting $l_0 = 0$.  We then find that:
\be
E \equiv 2\left\langle \left\Vert \int_{0}^{\Delta} {\rm d}s\,\hat{\delta}(s) e^{is} \right \Vert^2\right\rangle  \approx \frac{8\pi^2\Delta^2 C^2 \Lambda_{0}^{\alpha-2}}{(\alpha -2)L},
\ee
where $\Lambda_0^{-2} = L_{0}^{-2} +\Delta^2 L^{-2}$. It can be similarly checked that with this form of $P(k)$:
\be
\left\langle \tilde{\delta}^2 \right\rangle \approx \pi^{3/2} C^2 L_{0}^{\alpha-3}\frac{\Gamma\left(\frac{\alpha}{2}-\frac{3}{2}\right)}{\Gamma\left(\frac{\alpha}{2}\right)}.
\ee
and so
\be
E \approx 3q_{\alpha} \left(\frac{L_{0}\Delta^2}{L}\right) \left\langle \tilde{\delta}^2 \right\rangle   \left(\frac{1}{1+ \Delta^2 L_0^2/L^2}\right)^{\frac{\alpha-2}{2}}, \label{Eeeqn}
\ee
where
$$
q_{\alpha} = \frac{8\pi^{1/2}\Gamma\left(\frac{\alpha}{2}\right)}{3(\alpha-2)\Gamma\left(\frac{\alpha}{2}-\frac{3}{2}\right)}.
$$
For $\alpha = 11/3$, which corresponds to a Kolmogorov power spectrum, $q_{\alpha} \approx 0.9958$.

We remember that $\tilde{\delta} = (\delta_b(z)-\delta_n(z))/(1+\delta_n(z))$ where $\delta_b$ is the magnetic field fluctuation and $\delta_n$ is the electron density fluctuation. The power spectrums of both fluctuations are generally taken to be described by a Kolomogorov power spectrum with some inner and outer scale.  We note that if, as is often assumed, the inner and outer scales of the magnetic and electron density fluctuations are the same, and if the two fluctuations are uncorrelated when $\delta_n \ll 1$, $\tilde{\delta}$ will also have a Kolmogorov type power spectrum.  We also note that correlations between $\delta_b$ and $\delta_n$ could potentially greatly decrease the power in $\tilde{\delta}$. Specifically if $\delta_b \approx \delta_n$ then $\tilde{\delta} \ll \delta_b, \delta_n$.  The structure of electron density fluctuations in our galaxy is much better understood than the structure of magnetic field fluctuations.  For simplicity, and to make an order of magnitude estimate of $E$ we take $\delta_b = 0$ and assume $\delta_n \ll 1$ so that $\tilde{\delta} \approx -\delta_n$.  The power spectra of $\delta_n$ and $\tilde{\delta}$ are  then equivalent.  For electron density fluctuations, estimates of the inner scale, $l_0$, place it around $10^7 -10^9,{\rm m}$.  It can be checked that for $\omega \gtrsim 10^{-7}\,{\rm eV}$ and $L \approx 50,{\rm pc}$, $\Delta l_0/L \ll 1$ as assumed above.  

In the NE2001 model \cite{NE2001} for galactic electron density fluctuations, the fluctuation parameter, $F_n$, is defined thus:
$$
F_{n} \approx \left\langle \delta_{n}^2 \right\rangle \left(\frac{1\,{\rm pc}}{L_{0}}\right)^{2/3}
$$
and the electron density fluctuations have a Kolmogorov spectrum with $\alpha = 11/3$.  We also estimated previously that:
$$
\Delta \approx 16 \left(\frac{2\,{\rm eV}}{\omega}\right),
$$
and so
\be
E \approx 15.4 F_n \left(\frac{L_{0}}{1\,{\rm pc}}\right)^{5/3} \beta^{-5/6}
\ee
where
$$
\beta = L_{0}^2 \Lambda_{0}^{-2} \approx 1 + 0.1 \left(\frac{L_{0}}{1\,{\rm pc}}\right)^2 \left(\frac{2\,{\rm eV}}{\omega}\right)^2.
$$
The fluctuation parameter varies widely across the galaxy.  On average
in the disk $F_n \approx 0.2$ however in the local interstellar medium
(out to about a ${\rm kpc}$ from the Sun) $F_n \approx 0.01-0.1$.  The
stellar objects we analyzed in \S \ref{starpol} are located in the
local ISM where $F_n$ is smaller, however even if we take the slightly
larger value of $F_n \approx 0.2$ appropriate for the disk on average, we find that for visible light $\omega \sim 2\eV$:
$$
E \sim 3 \left(\frac{L_{0}}{1\,{\rm pc}}\right)^{5/3}\beta^{-5/6}.
$$

Carlson and Garretson \cite{Carlson94} took the outer scale of
turbulence to be $L_{0} \approx 10-100\,{\rm pc}$, which results in
$E$ becoming independent of $L_0$ and $E \approx 19 - 20$. However,
more recent estimates \cite{outscale} suggest a much smaller value for
$L_{0}$ than previously expected, specifically an $L_{0}$ that is no
more than a few parsecs. In HII regions (clouds of gas and plasma in
which star formation is taking place) $L_{0} \approx 0.01\,{\rm pc}$ and the pulsar measurements \cite{Pulsar} give $L_{0} \approx 0.03\,{\rm pc}$.  The precise value of $E$ therefore depends fairly strongly on the value of outer scale for the Kolomogorov spectrum $L_{0}$, which is uncertain.  This is because the enhancement term is predominately sourced by electron fluctuations on scales of  $l \sim L/\Delta \approx 4\omega/\Vert m_{\rm eff}^2\Vert$. For visible light, $l \sim O(1){\rm pc}$.  The structure of galactic electron density fluctuations is not, however, well understood on such scales and almost all measurements of such fluctuations relate to lower scales.  This means it is difficult to make an accurate estimate of the enhancement factor.  However given $L_{0} \lesssim {\rm few} \,{\rm pc}$, $\omega \sim 2,{\rm eV}$, we estimate $0.03 \lesssim I \lesssim  10$ based on the different estimates for $L_0$.  A correlation between electron and magnetic fluctuations could significantly lower this estimate.  Hence we have estimated $E$ to be $\mathcal{O}(1)$ or smaller in the visible part of the electromagnetic spectrum. Importantly, even if the conversion rate is enhanced, the oscillatory nature of the chameleon induced polarization remains. Thus a slightly enhanced conversion probability is not expected to significantly alter the form of the signal for which we have searched. Given the great ambiguity in the precise magnitude of the enhancement term and because we estimate it to be no greater than factor of about $10$, we have chosen to neglect it in our analysis. 

We now consider the magnitude of any enhancement effect due to electron density fluctuations in galaxy superclusters, such as that considered by Jain \etal in Ref. \cite{Jain02}.  Since very little is known about electron density fluctuations in galaxy clusters and superclusters, Jain \etal assumed a simple scaling relation where all unknown dimensionful quantities scale with $n_{e}$.  They did not however include the role of an outer scale of fluctuations, $L_0$, instead assuming that $P(k)$ was everywhere a power law.  The outer scale of fluctuations is important as it is required for the total magnitude of fluctuations $\left\langle \delta^2_n \right\rangle$ to be finite.  We assume the same scaling for dimensionful quantities as that used by Jain \etal.   We therefore assume that the length scale $L_0 \propto \bar{n}_{e}^{-1/3}$, but that $\left\langle \delta^2_n \right\rangle$ is approximately the same in a galaxy cluster as it is in the galaxy. 

If $L_0 = 1{\rm pc}$ in the galaxy where $n_{e} \approx 0.03\,{\rm cm}^{-3}$ then in the galaxy supercluster considered by Jain \etal where $n_e \approx 10^{-6}\,{\rm cm}^{-3}$, one would expect $L_0 \sim 31\,{\rm pc}$.   In the same region $\omega_{\rm pl}^2 \approx 3.7 \times 10^{-14}\,{\rm eV}$ and an appropriate value for $L$, the length of the magnitude domain, is suggested in Ref. \cite{Vallee07} to be $100\,{\rm kpc}$.  This gives:
\be
\Delta \approx 2.7 \left(\frac{2\eV}{\omega}\right),
\ee
Thus using Eq. \ref{Eeeqn} for $\alpha = 11/3$ and assuming $\left \langle \tilde{\delta}^2 \right\rangle \lesssim 1$, we find that for a galaxy supercluster the enhancement factor for visible light is estimated to be:
\be
E \lesssim \mathcal{O}(10^{-2}) \left(\frac{L_0}{31\pc}\right), \nonumber
\ee
and so any enhancement due to electron density fluctuations in this region is estimated to be sub-leading order.  Jain \etal found the opposite result but ignored the role of the outer scale, $L_0$, which limits the overall magnitude of fluctuations.

\section{Chameleon Optics for Multiple Magnetic Domains}\label{appA}

In many realistic astrophysical settings, light beams pass through
many magnetized domains, and in each domain the angle of the magnetic
field relative to the direction of propagation is essentially
random. In this appendix we present, in detail, the equations which
describe this multiple domain problem and their solutions  in a number
of important limits. In \S \ref{sec:Optics:sing} we presented the
equations that  describe how the chameleon and photon fields evolve as
they pass through a single magnetic domain.  In that section we split
the photon field into components polarized parallel and perpendicular
to the direction of the magnetic field, and used this as a basis to
define the Stokes vector, $(I_{\gamma},Q, U, V)^{T}$ for the photon
field as well as four associated amplitudes, $J$, $K$, $L$ and $M$,
which describe correlations between the chameleon field and components
of the photon fields (see Eq. (\ref{ampDef}) for the definition of
these quantities).   To deal with the multiple domain case we must first fix a basis for the photon field that is independent of the direction of $\mathbf{B}$. Doing this we take the two components of the photon field to be $\gamma_{1}$ and $\gamma_{2}$, and redefine
\begin{eqnarray}
I_{\gamma} &=& \left\langle \vert \gamma_{1} \vert^2 \right\rangle +\left\langle \vert \gamma_{2}  \vert^2 \right\rangle, \nonumber \\
Q &=& \left\langle \vert \gamma_{2} \vert^2 \right\rangle -\left\langle \vert \gamma_{1} \vert^2 \right\rangle, \nonumber \\
U + i V &=& 2\left\langle\bar{\gamma}_{2} \gamma_{1}\right\rangle, \nonumber \\
J + iK &=& 2e^{i\varphi}\left\langle\bar{\gamma}_{1} \chi\right\rangle, \nonumber \\
L + iM &=& 2e^{i\varphi}\left\langle\bar{\gamma}_{2} \chi\right\rangle, \nonumber
\end{eqnarray}
and as in \S \ref{sec:Optics:sing} we define $X = 3I_{\gamma} -2$. $I_{\gamma}$ and $V$ are independent of the choice of basis. We define $\theta_{n}$ so that in the $n^{\rm th}$ magnetic domain:
\begin{eqnarray}
\gamma_{\parallel} &=& \cos\theta_n \gamma_{1} - \sin \theta_n \gamma_{2}, \nonumber\\
\gamma_{\perp} &=& \cos\theta_n \gamma_{2} + \sin \theta_n \gamma_{1}. \nonumber 
\end{eqnarray}
and define
\begin{eqnarray}
Q^{\prime} &=& Q\cos 2\theta_n  + U\sin 2\theta_n,\nonumber \\
U^{\prime} &=& -Q\sin 2\theta_n  + U\cos 2\theta_n,\nonumber \\
J^{\prime} &=& J \cos \theta_{n} - L\sin \theta_n,\nonumber \\
L^{\prime} &=& J \sin \theta_n + L \cos \theta_n, \nonumber\\
K^{\prime} &=& K \cos \theta_{n} - M\sin \theta_n, \nonumber \\
M^{\prime}_{n} &=& K \sin \theta_n + M \cos \theta_n.\nonumber 
\end{eqnarray}
The evolution of the primed quantities  as well as $X$ and $V$ in the $n^{\rm th}$ region are then described by Eqs. (\ref{IgEq}) - (\ref{LastEq}) with $Q$ being replaced by $Q^{\prime}$, $U$ by $U^{\prime}$ and so on.

Solving the full system of equations for  $N \gg 1$ domains involves
diagonalising an 8 by 8 matrix as well as evaluating  multiple sums
involving the random angles $\theta_{n}$ for $n = 0$ to $N-1$, and we
have been unable to find an analytic general solution. It is
straightforward to solve the system numerically, but analytical
solutions are often more useful for understanding the behaviour.
Fortunately, it is possible to make a great deal of analytical
progress in the weak-mixing limit where $N \alpha  \ll 1$ and $N
P_{\gamma \leftrightarrow \phi} \ll 1$, where $N$ is the number of magnetic
domains, as well as  in the strong mixing limit where $N\Delta \ll 1$ and $N\Pphi \gg 1$. 

\subsection{Weak Mixing Limit}
When $N \alpha \ll 1$ and $N P_{\gamma \leftrightarrow \phi}$ we must have either $\Delta/\cos 2\theta,\,\Delta \tan 2 \Delta \ll 1$ or $\tan 2\theta,\,\Delta \tan^2 2\theta \ll 1$.  In these limits $\varphi \approx \Delta$, $\beta \approx 2\Delta$ and
\begin{eqnarray}
\alpha = \varphi - \Delta \approx \frac{\tan^2 2\theta}{4} \left[2\Delta - \sin 2\Delta\right].
\end{eqnarray}
We assume that there is no initial chameleon flux, so that initially
$X=X_{0} = 1$ and $J=K = L = M = 0$.  Without loss of generality we
pick our coordinate basis so that $Q = 0$ initially and $U = U_0$ and
$V=V_0$.  By requiring that $N P_{\gamma \leftrightarrow \phi} \ll 1$
and $N \alpha \ll 1$, we are assuming the perturbations, $\delta X$,
$\delta Q$,  $\delta U$ and $\delta V$, are small compared to  the
quantities $X$, $Q$, $U$ and $V$, and that $J$, $K$, $L$ and $M$ are
$\ll 1$.  We define $J = Aj$ and make similar definitions for $k$, $m$
and $n$.  We compute the perturbed quantities to $\mathcal{O}(NA^2)$ and
$\mathcal{O}(N\alpha^2)$. We define $\delta X_{n}$ to be the value of
$\delta X$ after having passed through the $n^{\rm th}$ region, and make similar definitions for the other quantities.   Expanding to first order in the perturbations we find the following simplified recurrence relations
\begin{eqnarray}
\delta X_{n+1} &=& \delta X_{n} -\frac{3A^2}{2} - \frac{3A^2}{2}\bar{U}_{0}\sin 2\theta_n  \\ &&+ 3A^2 \left(l_{n} \cos \theta_n +j_{n} \sin \theta_n\right) \sin 2\Delta \nonumber \\ && - 3A^2\left(m_{n} \cos \theta_n +k_{n} \sin \theta_n\right)  \cos 2\Delta, \nonumber, \\
\delta Q_{n+1} &=& \delta Q_{n} - \frac{A^2}{2} \cos 2\theta_n   \\ &&+ A^2 \left(l_{n} \cos \theta_n -j_{n} \sin \theta_n\right) \sin 2\Delta \nonumber \\ && - A^2\left(m_{n} \cos \theta_n -k_{n} \sin \theta_n\right)  \cos 2\Delta, \nonumber \\
&&- \alpha V_{n} \sin 2\theta_{n}  + \frac{\alpha^2}{4}\sin 4\theta_n U_{0} \nonumber \\
\delta U_{n+1} &=&  \delta U_{n} - \frac{A^2}{2} \sin 2\theta_n - \frac{A^2}{2}U_{0}  \\
&&+ A^2\left(l_{n} \sin \theta_n  + j_{n} \cos \theta_n\right) \sin 2\Delta \nonumber \\ &&- A^2\left(m_{n} \sin \theta_n + k_{n} \cos \theta_n\right)  \cos 2\Delta \nonumber\\
&&+ \alpha V_{n} \cos 2 \theta_{n} -\frac{\alpha^2}{2} U_{0} \cos^2 2\theta_n \nonumber
\end{eqnarray}
\begin{eqnarray}
\delta V_{n+1} &&= \delta V_{n}-\frac{A^2}{2}V_{0} \\
&&+ A^2\left(l_{n} \sin \theta_n  - j_{n} \cos \theta_n\right) \cos 2\Delta  \nonumber \\ && + A^2\left(m_{n} \sin \theta_n - k_{n} \cos \theta_n\right)  \sin 2\Delta \nonumber \\
&&-\frac{1}{2}\alpha^2 V_{0} - \alpha U_{n} \cos 2\theta_n, \nonumber \\
&&+ \alpha Q_{n} \sin 2\theta_n\nonumber
\end{eqnarray}
and
\begin{eqnarray}
k_{n+1} + ij_{n+1} &=& e^{2i\Delta}\left(k_{n} + ij_{n}\right) + Y_{n}, \label{kjeqn} \\
m_{n+1} + il_{n+1} &=& e^{2i\Delta}\left(m_{n} + il_{n}\right) + Z_{n}, \label{mleqn}\\
Y_{n} &=& (U_{0}+iV_0) \cos \theta_{n} + \sin \theta_{n}, \nonumber\\
Z_{n} &=& (U_{0}-iV_0) \sin \theta_{n} + \cos \theta_{n}.\nonumber
\end{eqnarray}
Eqs. (\ref{kjeqn}) and (\ref{mleqn}) are solved thus
\begin{eqnarray}
k_{n} + ij_{n} &=& \sum_{r=0}^{n-1} e^{2i\Delta(n-1-r)} Y_{r},\\\nonumber
m_{n} + il_{n} &=& \sum_{r=0}^{n-1} e^{2i\Delta(n-1-r)} Z_{r}. \nonumber
 \end{eqnarray}
Assuming $N \gg 1$, we then arrive at following solutions for the perturbations to the components of the Stokes vector to $O(NA^2,\, N\alpha^2)$:
\begin{eqnarray}
\delta X_{N} &=& - \frac{3N A^2 N}{2} -3 N A^2  \vartheta^{(c-)}_{N}(2\Delta) \\ 
&&-3N A^2  U_0 \vartheta^{(s+)}_{N}(2\Delta) - 3N A^2 V_{0} \varrho^{(s-)}_{N}(2\Delta) \nonumber \\
\delta Q_{N} &=& -N A^2 \vartheta^{(c+)}_{N}(2\Delta) - N A^2 U_{0} \vartheta^{(s-)}_{N}(2\Delta) \\
&&-NA^2  V_0 \varrho^{(s+)}_{N}(2\Delta) - \sqrt{N}\alpha V_{0} \kappa_{N}^{s} \nonumber\\
&&+N \alpha^2 U_{0} \mu_{N}^{sc}, \nonumber \\
\delta U_{N} &=& - \frac{(2NA^2 + N\alpha^2)}{4} U_{0} \\ 
&&-  N A^2  \vartheta^{(s+)}_{N}(2\Delta) - N A^2 U_{0} \vartheta^{(c-)}_{N}(2\Delta)\nonumber \\
&&+ N A^2 V_0 \varrho^{(c+)}_{N}(2\Delta) + \sqrt{N}\alpha  V_{0} \kappa_{N}^{c}\nonumber \\
&&-N\alpha^2 U_{0} \mu_{N}^{cc}, \nonumber \\
\delta V_{N} &=& - \frac{(NA^2+N\alpha^2)}{2} V_{0} - NA^2  \varrho^{(s-)}_{N}(2\Delta) \\
&&-NA^2 U_{0} \varrho^{(c+)}_{N}(2\Delta) \nonumber \\
&&-NA^2 \bar{V}_{0} \vartheta^{(c-)}_{N}(2\Delta)  -\sqrt{N}\alpha  U_{0} \kappa_{N}^{c} \nonumber \\
&&-N\alpha^2 V_{0} \left(\mu_{N}^{cc} + \mu_{N}^{ss}\right)\nonumber
\end{eqnarray}
where
\begin{eqnarray}
\vartheta_{N}^{c\pm}(2\Delta) = \frac{1}{N}\sum_{n=0}^{N-1}\sum_{r=0}^{n-1} \cos (2\Delta(n-r))\cos(\theta_r\pm\theta_n), \nonumber \\
\vartheta_{N}^{s\pm}(2\Delta) = \frac{1}{N}\sum_{n=0}^{N-1}\sum_{r=0}^{n-1} \cos (2\Delta(n-r))\sin(\theta_r\pm\theta_n),
\nonumber \\
\varrho_{N}^{c\pm}(2\Delta) = \frac{1}{N}\sum_{n=0}^{N-1}\sum_{r=0}^{n-1} \sin (2\Delta(n-r))\cos(\theta_r\pm\theta_n), \nonumber \\
\varrho_{N}^{s\pm}(2\Delta) = \frac{1}{N}\sum_{n=0}^{N-1}\sum_{r=0}^{n-1} \sin (2\Delta(n-r))\sin(\theta_r\pm\theta_n). \nonumber
\end{eqnarray}
and
\begin{eqnarray}
\mu_{N}^{cc} &=& \frac{1}{N}\sum_{n=0}^{N-1}\sum_{r=0}^{n-1} \cos 2\theta_{n}   \cos 2\theta_{r}, \nonumber \\
\mu_{N}^{sc} &=& \frac{1}{N}\sum_{n=0}^{N-1}\sum_{r=0}^{n-1} \sin 2\theta_{n}   \cos 2\theta_{r}, \nonumber \\
\mu_{N}^{ss} &=& \frac{1}{N}\sum_{n=0}^{N-1}\sum_{r=0}^{n-1} \sin 2\theta_{n}   \sin 2\theta_{r}, \nonumber \\
\kappa_{N}^{c} &=& \frac{1}{\sqrt{N}} \sum_{n=0}^{N-1}\cos 2\theta_{n}, \nonumber \\
\kappa_{N}^{s} &=&\frac{1}{\sqrt{N}} \sum_{n=0}^{N-1}\sin 2\theta_{n}.
\end{eqnarray}
Each of these nine quantities vanishes when averaged over all possible values of $\theta_{n}$. When $f(n,p) = f(p,n)$ we have
\begin{eqnarray}
\frac{1}{N}\sum_{n=0}^{N-1}\sum_{p=0}^{n-1} f(n,p) = &&\frac{1}{2N} \sum_{n=0}^{N-1} \sum_{p=0}^{N-1} f(n,p) \\ &&- \frac{1}{2N} \sum_{n=0}^{N-1} f(n,n), \nonumber
\end{eqnarray}
and so
\begin{eqnarray}
\vartheta_{N}^{c\pm} &=& \frac{1}{2}\left(X_{cc}^2 \mp X_{cs}^2 + X_{sc}^2 \mp X_{ss}^2 - \frac{1}{2} \pm \frac{1}{2}\right), \label{Xeqn1}\\
\vartheta_{N}^{s+} &=& X_{cc}X_{cs} + X_{sc}X_{ss}, \label{Xeqn2} \\
\varrho_{N}^{s-} &=& X_{sc}X_{cs} - X_{ss}X_{cc}. \label{Xeqn3}
\end{eqnarray}
where 
\begin{eqnarray}
X_{cc} &=& \frac{1}{\sqrt{N}}\sum_{n=0}^{N-1}\cos 2n \Delta \cos \theta_{n},\nonumber \\
X_{cs} &=& \frac{1}{\sqrt{N}}\sum_{n=0}^{N-1}\cos 2n \Delta \sin \theta_{n},\nonumber \\
X_{sc} &=& \frac{1}{\sqrt{N}}\sum_{n=0}^{N-1}\sin 2n \Delta \cos \theta_{n}\nonumber \\
X_{ss} &=& \frac{1}{\sqrt{N}}\sum_{n=0}^{N-1}\sin 2n \Delta \sin \theta_{n}.\nonumber
\end{eqnarray}
In the large $N$ limit (and at fixed $\Delta$) each of these  four quantities are independent,  normally distributed random variables:  $X_{cc},\, X_{cs} \sim N(0, \sigma_{+}^2)$ and $X_{sc},\,X_{ss} \sim N(0, \sigma_{-}^2)$, where:
\begin{eqnarray}
\sigma_{\pm}^2 = \frac{1}{4}\left[1\pm \frac{\cos(2(N-1)\Delta)\sin 2N\Delta}{N \sin 2\Delta}\right].
\end{eqnarray}
Additionally, $\vartheta^{s-}_{N}$, $\varrho^{s+}_{N}$,
$\varrho^{c\pm}_{N}$ are, for fixed $\Delta$ and  in the large $N$ limit, well approximated by independent normally distributed random variables, with  $\vartheta^{s-}_{N} \sim N(0, \sigma_{1}^2)$ and the rest are $n(0, \sigma_{2}^2)$ where for $N \gg 1$
\begin{eqnarray}
\sigma_{1}^2 &=& \frac{1}{8}\left[1+ \frac{\sin^2 2N\Delta}{N^2 \sin^2 2\Delta}\right] \nonumber \\
\sigma_{2}^2 &=& \frac{1}{8}\left[1- \frac{\sin^2 2N\Delta}{N^2 \sin^2 2\Delta}\right]. \nonumber
\end{eqnarray}

We choose a basis so that initially $U_{0} \geq 0$ and define $m_{l0} = U_{0}$, $m_{c0} = V_{0}$, $q_{0}  = \vert m_{c0} \vert$ and $p_{0} = \sqrt{U_{0}^2 + V_{0}^2}$.  Keeping terms to order $O(N\Pphi p_{0})$ and $O(N^2 \Pphi^2)$, we find that:
\begin{eqnarray}
p^2(N) = && p_{0}^2 + 2N\Pphi(1-p_{0}^2)\left[m_{l0} \vartheta_{N}^{s+} + m_{c0} \varrho_{N}^{s-}\right] \nonumber\\ &&+ N^2 \Pphi^2 (1-p_{0}^2)^2\left(\frac{1}{2} + \vartheta_{N}^{c-}\right)^2. 
\end{eqnarray}
where we have used Eqs.  (\ref{Xeqn1}-\ref{Xeqn3}) to provide the identity:
\begin{eqnarray}
\sqrt{ \vartheta_{N}^{(c+)\,2} + \vartheta_{N}^{(s+)\,2} + \varrho_{N}^{(s-)\,2}} &&= \frac{1}{2}\left(X_{cc}^2 + X_{cs}^2\right. \nonumber\\  \left. + X_{sc}^2 + X_{ss}^2\right) &&= \vartheta_{N}^{c-} +\frac{1}{2}.
\end{eqnarray}
If $p_{0} \sim O(1)$, the last term in this expression is the same order as terms that have been omitted so it too should be dropped.  Similarly for the fractional circular polarization we have to $O(N\Pphi)$ and $O(N\alpha^2)$:
\begin{eqnarray}
m_{c}(N) &=& m_{c0}-\frac{N\alpha^2}{2}m_{c0}\left[\kappa_{N}^{c\,2} + \kappa_{N}^{s\,2}\right]  \label{mcN}\\ && - \sqrt{N}\alpha m_{l0} \kappa_{N}^{c} \nonumber \\
&& -N\Pphi (1-m_{c0}^2)\varrho_{N}^{s-}(2\Delta) \nonumber \\ && + N\Pphi m_{l0}m_{c0}\vartheta_{N}^{s+}(2\Delta)\nonumber \\ &&- N\Pphi m_{l0} \varrho_{N}^{c+}(2\Delta).\nonumber
\end{eqnarray}

If there is no initial polarization ($p_{0} = 0$), or more generally if $NP_{\gamma \leftrightarrow \phi}(1-p_{0}^2)/p_{0} \gg 1$, the final polarization fraction is given by
\begin{eqnarray}
p(N) &=& N\Pphi \left[\frac{1}{2} + \vartheta_{N}^{c-}(2\Delta)\right].
\end{eqnarray}
We may therefore write:
$$
p(N) = \frac{1}{2}N\Pphi \left(\sigma_{+}^2(X_{1}^2 + X_{2}^2) + \sigma_{-}^2(X_{3}^2 + X_{4}^2)\right),
$$
where the $X_{i}$ are independent identically distributed $N(0,1)$ random variables. When $p_{0}=0$ the circular polarization  simplifies:
$$
m_{c}(N) = N \Pphi \sigma_{+}\sigma_{-} \left(X_{1}X_{3} - X_{2}X_{4}\right).
$$

Where $p_{0} \neq 0$ and $NP_{\gamma \leftrightarrow \phi}(1-p_{0}^2)/p_{0} \ll 1$ we have to $O(NP_{\gamma \leftrightarrow \phi}(1-p_{0}^2)/p_{0})$:
\begin{eqnarray}
p(N) &=& p_{0} + \frac{N\Pphi (1-p_{0}^2)m_{l0}}{p_{0}}\left(\sigma_{+}^2 X_{1}X_{2} \right. \nonumber \\ && \left. + \sigma_{-}^2 X_{3}X_{4}\right) \nonumber \\
&&+ \frac{N\Pphi (1-p_{0}^2)m_{c0}}{p_{0}} \sigma_{+}\sigma_{-} \left(X_{1}X_{3}-X_{2}X_{4}\right) \nonumber.
\end{eqnarray}
The circular polarization is  given by Eq. (\ref{mcN}) in this case.

\subsection{Strong Mixing Limit}
We now consider the strong mixing limit.  This is the limit in which
$N\Delta \ll 1$ so that $P_{\gamma \leftrightarrow \phi}$ takes it
largest value, and   the mixing between the chameleon and photons is strong, $N P_{\gamma \leftrightarrow \phi} \gg 1$.   In this limit $\alpha,\, \beta,\, \Delta,\, \varphi \ll 1$ and so Eq. (\ref{IgEq}-\ref{LastEq}) simplify to:
\begin{eqnarray}
X &\rightarrow & \left(1-\frac{3}{2}A^2\right)X - \frac{3}{2}A^2 Q \nonumber \\ && - 3A\sqrt{1-A^2} M, \nonumber \\
Q &\rightarrow &\left(1-\frac{1}{2}A^2\right)Q - \frac{1}{2}A^2 X\\ &&  - A\sqrt{1-A^2}M, \nonumber, \\
U &\rightarrow& \sqrt{1-A^2}U - AK, \nonumber \\
M & \rightarrow &(1-2A^2)M + A\sqrt{1-A^2}(Q+X). \nonumber \\
K &\rightarrow &\sqrt{1-A^2}K + AU, \nonumber
\end{eqnarray}
and
\begin{eqnarray}
V &\rightarrow & \sqrt{1-A^2} V  - A J \nonumber, \\
J &\rightarrow & \sqrt{1-A^2} J  + A V, \nonumber \\
L &\rightarrow & L. \nonumber
\end{eqnarray}
The differently oriented magnetic fields in each domain mix $Q$ with $U$, $M$ with $K$ and $J$ with $L$.  It is clear then that the evolution of $V$, $J$ and $L$ are completely decoupled from that of $X$, $Q$, $U$, $M$ and $K$.  We are concerned with the limiting value of total polarization fraction, $p$.  Additionally since we expect the initial circular polarization fraction to be small, $q_{0} = \vert m_{c0} \vert \ll p_{0}$, we set $V = 0$. We also require that initially the chameleon flux is zero ($M=K = L = J = 0$ initially).  It is clear then from the above equations that $V$ remains zero.  From simulations we see that in the strong mixing limit the final mean polarization fraction takes a specific value, which depends on $p_{0}$.  Remarkably we can calculate both the limiting value and the final distribution of $p$ analytically without actually explicitly solving the above equations.

We assume that initially the photon is in a state with polarization
fraction $p_{0} = (1-a)/(1+a)$.  Without loss of generality we pick
coordinates so that $U=0$ initially and write the initial Stokes vector of the photon state thus:
\begin{eqnarray}
S_{0} = \left(\begin{array}{cc} I_{\gamma} \\ Q \\ U \\ V \end{array}\right) = \left(\begin{array}{cc} (1+a) \\ (1-a) \\ 0 \\ 0 \end{array}\right)
\end{eqnarray}
We can always consider such a partially polarized photon state to be a linear superposition of two fully polarized photon states (labelled $(+)$ and $(-)$), i.e. 
$$
S_{0} = S_{+}(0) + aS_{-}(0)
$$
where (dropping the $V$ component as it vanishes):
\begin{eqnarray}
S_{\pm}(0) = \left(\begin{array}{cc} 1 \\ \pm 1 \\ 0 \end{array}\right).
\end{eqnarray}
Since both $S_+(0)$ and $S_-(0)$ represent fully polarized photon
states, they can also be described in terms of a vector whose components are the photon and chameleon amplitudes, $c_{1} = \gamma_{1}$, $c_{2} = \gamma_{2}$ and $c_{\phi} = \chi = i\phi$. We define this vector to be $v_{+}$ for $S_{+}$ and $v_{-}$ for $S_{-}$, so that
\begin{eqnarray}
v_{+} = \left(\begin{array}{cc} c_{\phi} \\ c_1 \\ c_2 \end{array}\right)_{+} &=& \left(\begin{array}{cc} 0 \\ 1 \\ 0 \end{array}\right). \\
v_{-} = \left(\begin{array}{cc} c_{\phi} \\ c_1 \\ c_2 \end{array}\right)_{-} &=& \left(\begin{array}{cc} 0 \\ 0 \\ 1 \end{array}\right). 
\end{eqnarray}
We also define $v_{\rm tot} = v_{+} + v_{-}$ and note that this too is a fully polarized state.

The evolution of a fully polarized state through a single magnetic domain is given by Eqs. (\ref{gP1}-\ref{gP2}).  We note that these equations conserve the total flux $I_{\gamma} + I_{\phi} = A_{0}^2$, where $I_{\gamma} = \vert c_{1} \vert^2 +\vert c_{2} \vert^2$ and $I_{\phi} = \vert c_{\phi} \vert^2$.  For  $v_{\pm}$, the total flux is $1$ and for $v_{\rm tot}$ it is $2$.

After having passed through many randomly orientated magnetic domains,
if $NP_{\gamma \leftrightarrow \phi} \gg 1$, the mixing between the
chameleon and photon fields, and between  different components of the
photon field, will be strong. This means that on average the initial
flux should be evenly distributed among each of $c_{1}$, $c_{2}$ and $c_{\phi}$ and so
$$
\left(\begin{array}{cc} c_{\phi} \\ c_1 \\ c_2 \end{array}\right)_{N} = A_{0}\left(\begin{array}{cc} x \\ \sqrt{1-x^2} \cos \theta \\ \sqrt{1-x^2} \sin \theta \end{array}\right)
$$
where each of $c_{\phi}$, $c_{1}$, $c_{2}$ are uniformly distributed random variables on $A_{0}[-1,1)$.  This implies that $x \sim U[-1,1)$ and $\theta \sim U[0,2\pi)$.  

Now $v_{+}$, $v_{-}$ and $v_{tot}$ are all fully polarized states. If,
after having passed through many regions, $v_{+} \rightarrow v_{+}(\infty)$ and $v_{-} \rightarrow v_{-}(\infty)$ where
\begin{eqnarray}
v_{+}(\infty) &=& \left(\begin{array}{cc} x \\ \sqrt{1-x^2}\cos \theta \\ \sqrt{1-x^2} \sin \theta \end{array}\right). \nonumber\\
v_{-}(\infty) &=& \left(\begin{array}{cc} y \\ \sqrt{1-y^2}\cos \phi \\ \sqrt{1-y^2} \sin \phi \end{array}\right), \nonumber
\end{eqnarray}
then since the field equations are linear $v_{\rm tot} \rightarrow v_{tot}(\infty) = v_{+}(\infty) + v_{-}(\infty)$. Now in the limit of strong mixing, the $c_{\phi}$ components of $v_{+}(\infty)$, $v_{-}(\infty)$ and $v_{tot}(\infty)$ must all be uniformly distributed random variables on $[-A_{0},A_{0})$.    This imposes a very strong condition on the distributions of $x$ and $y$, in fact one must have  $x = \sqrt{1-X^2} \cos \psi$ and $y = \sqrt{1-X^2} \sin \psi $ where $\psi$ and $X$ are independent uniform random variables: $\psi  \sim U[0, 2\pi)$ and $X \sim U[0,1)$. 

We also know the total flux, $v_{\rm tot}$.  Initially the total flux is $A_{0}^2 = 2$, and finally it is $A_{f}^2 = (x+y)^2 + (\sqrt{1-x^2}\cos \theta + \sqrt{1-y^2}\cos \phi)^2 +(\sqrt{1-x^2}\sin \theta + \sqrt{1-y^2}\sin \phi)^2$. Equating these two gives the consistency condition:
$$
\cos (\theta - \phi) = - \frac{xy}{\sqrt{1-x^2}\sqrt{1-y^2}},
$$
so defining $I_{\gamma}^{+} = 1-x^2$ and $I_{\gamma}^{-} = 1-y^2$ we have
\begin{equation}
\cos^2 (\theta - \phi) = \frac{(1-I_{\gamma}^{+})(1-I_{\gamma}^{-})}{I_{\gamma}^{+}I_{\gamma}^{-}}. \label{coseqn}
\end{equation}

Now the Stokes vectors associated with $v_{\pm}(\infty)$ are
\begin{eqnarray}
S_{+}(\infty) &=& \left(\begin{array}{cc} I_{\gamma}^{+} \\ I_{\gamma}^{+}\cos 2\theta \\ I_{\gamma}^{+}\sin 2\theta  \\ 0\end{array}\right), \\
S_{-}(\infty) &=& \left(\begin{array}{cc} I_{\gamma}^{-} \\ I_{\gamma}^{-}\cos 2\phi\\ I_{\gamma}^{+}\sin 2\phi \\ 0 \end{array}\right),
\end{eqnarray}
so the final Stokes vector of a state with initial Stokes vector $S_{0} = S_{+}(0) + aS_{-}(0)$ is $S_f = S_{+}(\infty) + aS_{-}(\infty)$:
\begin{eqnarray}
S_{f} &=& \left(\begin{array}{cc} I_{\gamma}^{+}+aI_{\gamma}^{-} \\ I_{\gamma}^{+}\cos 2\theta+aI_{\gamma}^{-}\cos 2\phi \\ I_{\gamma}^{+}\sin 2\theta + aI_{\gamma}^{-}\sin 2\phi\end{array}\right).
\end{eqnarray}
Thus the final polarization fraction,  $p_{\infty}$ is:
\begin{eqnarray}
p_{\infty}^2 = \frac{(I_{\gamma}^{+}-aI_{\gamma}^{-})^2 + 4a I_{\gamma}^{+}I_{\gamma}^{-}\cos^2 (\theta -\phi)}{(I_{\gamma}^{+}+aI_{\gamma}^{-})^2}.
\end{eqnarray}
which after some simplification becomes:
\begin{eqnarray}
&p_{\infty} &= F(X^2,\cos 2\psi; p_{0})\\ &=&  \sqrt{1- \frac{4(1-p_{0}^2) X^2}{\left[(1+X^2) -p_{0}(1-X^2)\cos 2\psi \right]^2}}. \nonumber
\end{eqnarray}
where $X \sim U[0,1)$ and $\psi \sim U[0,2\pi)$.   In the simplest case where there is no initial polarization, $p_{0} = 0$, we have 
$$
p_{\infty} = \frac{1-X^2}{1+X^2},
$$
which has mean value
\begin{equation}
\bar{p}_{\infty} = \int_{0}^{1}\,{\rm d}X\,\frac{1-X^2}{1+X^2} = \frac{\pi}{2}-1 \approx 0.57.
\end{equation}
More generally
$$
\bar{p}_{\infty}(p_{0}) = \frac{1}{2\pi}\int_{0}^{2\pi}\,{\rm d}\alpha\,\int_{0}^{1}\,{\rm d}X\, F(X^2, \cos 2\alpha;p_{0}).
$$
$\bar{p}_{\infty}(p_{0}) $ is a monotonically increasing function of
$p_{0}$ and increases from $\pi/2 - 1$ to $1$ as $p_{0}$ goes from $0$
to $1$. 
 
\section{Estimating $BL/2M$ and Confidence Intervals}\label{app:pol}
In this appendix we provide details of how estimates and confidence
intervals for the properties of any chameleon-like field can be
extracted from measurements of the Stokes' parameters, $I_{\gamma}$,
$U$ and $Q$ of a single object.  We suppose that, in the absence of
any chameleon field, the polarization angle of a given object is
roughly independent of wavelength in some interesting part of the
spectrum (e.g. for  UV to visible light).  Since chameleonic effects
die off as $1/\lambda^2$, where $\lambda$ is the wavelength of light,
we can roughly check this assumption by ensuring that the polarization
angle is roughly wavelength independent for the larger wavelengths
that are measured.    We found in \S \ref{sec:Optics} above that the
chameleon induced contributions to the   expected Stokes' vectors oscillate fairly strongly with wavelength up until some critical oscillation wavelength $\lambda_{\rm osc}$.  In all cases, we expect $\lambda_{\rm osc} \lesssim \mathcal{O}$(\AA).  In addition to $\lambda_{\rm osc}$, there is another  critical wavelength $\lambda_{\rm crit}$.  Below $\lambda_{\rm crit}$, the mean magnitude of the chameleonic polarization signal is roughly independent of wavelength, whereas for $\lambda \gtrsim \lambda_{\rm crit} \geq \lambda_{\rm osc}$, the chameleon signal behaves as $1/\lambda^2$.  

We suppose that we have $N_{\rm p}$ measurements of the reduced Stokes' parameter, $Q/I_{\gamma}$ and $U/I _{\gamma}$, for a given object; we denote these measurements $q_{i}$ and $u_{i}$ respectively.   We also require, for this analysis, that $\lambda > \lambda_{\rm osc}$ for all the measurements, and that any intrinsic (i.e. chameleonic) polarization be small i.e. $\ll 100\%$.  We define $\delta \lambda$ to be the spectral resolution of the measurements.   In the weak mixing limit, to leading order, we have that the chameleonic contributions to $Q/I_{\gamma}$ and  $U/I_{\gamma}$ are given by:
\begin{eqnarray}
q_{\rm cham} = &&-P_{0} \frac{\sin^{2} \Delta}{\Delta^2} \sum_{n=1}^{N-1}\sum_{r =0}^{n} \cos(2\Delta(n-r))\\ &&\frac{\sin(\delta \Delta(n-r))}{\delta \Delta(n-r)} \cos(\theta_{n} + \theta_{r}),\nonumber \\
u_{\rm cham} = &&-P_{0} \frac{\sin^{2} \Delta}{\Delta^2} \sum_{n=1}^{N-1}\sum_{r =0}^{n} \cos(2\Delta(n-r))\\ &&\frac{\sin(\delta \Delta(n-r))}{\delta \Delta(n-r)} \sin(\theta_{n} + \theta_{r}),\nonumber
\end{eqnarray}
where $P_{0} = (BL/2M)^2$; $L$ is the coherence length of the magnetic field and $B$ is its strength.  $N$ is the total number of magnetic regions passed through, and $M$ parametrises the strength of the chameleon to photon coupling.  $\Delta = \pi \lambda / 2\lambda_{\rm crit}$ where $\lambda_{\rm crit} = 4\pi^2 \left\vert m^2_{\rm eff} \right\vert L$; $m^2_{\rm eff} = m_{\phi}^2 - \omegap^2$; $\delta \Delta = \pi \delta \lambda / 2 \lambda_{\rm crit}$.    The $\theta_{n}$ define the angle of the magnetic field in the $n^{\rm th}$ region relative to the direction of the light beam.  Without any other prior information, we assume that these angles are essentially random.   Now the total Stokes' parameters are $q = q_0 + q_{\rm cham}$ and $u = u_0 + u_{\rm cham}$.  We assume that the non-chameleonic polarizations $u_0$ and $q_0$ depend on wavelength, but that, compared to the chameleonic contribution, they vary slowly.    This will generally be the case, for instance, if both $u_{0}$, $q_{0}$ have a wavelength dependence similar to the the Serkowski polarization law \cite{Serkowski73} expected for polarization due to interstellar dust, i.e. $u_{0}\,p_{0} \propto \exp(- K \ln^2 (\lambda_{\rm max}/\lambda))$, for some $K$ and $\lambda_{\rm max}$ which we do not require to be the same for both $u_{0}$ and $q_{0}$. Typically $\lambda_{\rm max} \sim 6000$\AA\ and $K \approx 1.15$.    We can then remove much of any intrinsic signal by simply smoothing the data over a scale on which $u_0$ and $q_0$ are expected to be fairly flat, to give $q^{\rm s}$ and $u^{\rm s}$, and then subtracting this smoothed data from the original data. We define $\hat{q} = q - q^{\rm s}$ and $\hat{u} = u - u^{\rm s}$.  

We define $\hat{y}_{i}$ for  $y_{i}$,  with standard error $\sigma_{i}$, made at wavelengths $\lambda_{i}$ as follows:
\begin{itemize}
\item We define some smoothing wavelength scale $\lambda_{\rm smooth}$ and for each $i$ define the $J_{i} = \left\lbrace j\,:\,2\left \vert \lambda_{i}-\lambda_{j})\right\vert < \lambda_{\rm smooth}\right\rbrace$.
\item $N_{i}$ is the number of elements in $J_{i}$.  
\item $S_{i} = \sum_{j \in J_{i}} 1/\sigma_{j}^2$.
\item We define $\hat{y}_{i} = y_{i} - S^{-1}_{i}\sum_{j \in J_{i}} y_{j}/\sigma_{j}^2$. 
\item Assuming that the $y_{i}$ are independent and distributed $N(\mu_{i},\sigma_{i}^2)$ for some $\mu_{i}$, we find the $\delta y_{i}$ have standard error $\hat{\sigma}_{i} =  \sqrt{\sigma_{i}^2 - 1/S_{i}}$.
\end{itemize}

We now have $\hat{q}_{i}$ and $\hat{u}_{i}$ from which the vast majority of any intrinsic signal should have been removed. We assume that any remaining intrinsic signal is sufficiently small compared with the noise as to be negligible. We check the accuracy of this smoothing process by simulations below.  We define $z^{(q)}_{i} = \hat{q}_{i}/\hat{\sigma}_{i}$ and $z^{(u)}_{i} = \hat{u}_{i}/\hat{\sigma}_{i}$.

The chameleonic contributions to $z^{(q)}_{i}$ and $z^{(u)}_{i}$ are predicted be $\beta \mu^{(q)}_{i}$ and  $\beta \mu^{(u)}_{i}$ respectively where:
\begin{eqnarray}
\mu^{(q)}_{i} &=& - \sum_{k = 1}^{N-1} h_{ik}X_{k}, \\
\mu^{(u)}_{i} &=& - \sum_{k = 1}^{N-1} h_{ik}Y_{k},
\end{eqnarray}
where $\beta = P_{0}/2$, $h_{ik} = \hat{H}_{ik}/\hat{\sigma}_{i}$, $\hat{H}_{ik} = H_{ik} - H_{ik}^{\rm s}$ and
\begin{equation}
H_{ik} =  \frac{\sqrt{2(N-k)}\sin^{2} \Delta_{i}}{\Delta^2_{i}}\cos(2k\Delta_{i})\frac{\sin(k \delta \Delta)}{k \delta \Delta}, 
\end{equation}
with $\Delta_{i}  = \pi \lambda_{i} / 2\lambda_{\rm crit}$ and  $\delta \Delta = \pi \delta \lambda / 2 \lambda_{\rm crit}$.
We have also defined
\begin{eqnarray}
X_{k} &=& \frac{\sqrt{2}}{\sqrt{N-k}} \sum_{r = 0}^{N-k-1} \cos(\Theta_{r}^{(k)}), \label{Xkeqn } \\
Y_{k} &=& \frac{\sqrt{2}}{\sqrt{N-k}} \sum_{r = 0}^{N-k-1} \sin(\Theta_{r}^{(k)}),
\end{eqnarray}
where $\Theta_{r}^{(k)} = \theta_{r+k} + \theta_{r}$.  When $N- k \gg
1$, $X_{k}$ and $Y_{k}$ are well approximated as independent
identically distributed $N(0,1)$ random variables. Since we assume that $N \gg 1$ and the largest values of $H_{ik}$ occur for $N-k \gg 1$, we approximate the $X_{k}$ and $Y_{k}$ as being independent and drawn from a $N(0,1)$ distribution. The likelihood of finding $X_{k} = \bar{X}_{k}$ is therefore $\propto \exp(-\bar{X}_{k}^2/2)$.   

Thus the probability density function with measurements $z^{(q)}_{i}$ given $\beta$ is (up to an overall $X_{k}$ and $\beta$ independent number $C_{0}$):
$$
f_{\beta, X_{k}}(z^{(q)}_{i}) = C_{0} e^{-\frac{1}{2}\sum_{i}(z^{(q)}_{i}-\beta \mu^{(q)}_{i})^2 - \frac{1}{2}\sum_{k} X_{k}^2} .
$$
Defining the symmetric matrix $\mathcal{Q}$ thus $\mathcal{Q}_{lk} = \sum_{i} h_{il}h_{ik}$, $M_{\beta} = 1 + \beta^2\mathcal{Q}$, and $v_{k} = \sum_{i} h_{ik} z_{i}^{(q)}$ we have:
$$
f_{\beta, X_{k}}(z^{(q)}_{i}) = C_{0} e^{-\frac{1}{2} \sum_{i} z_{i}^{(q}\, 2}e^{- \beta v^{T} X - \frac{1}{2}X^{T} M_{\beta} X}.
$$
Since the first term is independent of both $\beta$ and $X$ we can incorporate it into a redefinition of the $\beta$ and $X_{k}$ independent number $C_{0}$ i.e. $C_{0} \rightarrow C_{1} = C_{0} \exp(-\sum_{i} z_{i}^{(q),2}/2)$.   By defining $\hat{X}_{k} = \sqrt{M_{\beta}}( X + \beta M_{\beta}^{-1}  v)$, we have the new probability density, $\tilde{f}$, in terms of $\beta$, $z_{i}$ and $\hat{X}_{k}$:
$$
\tilde{f}(\beta) = D_{0}\frac{e^{\frac{\beta^2}{2} v^{T}M_{\beta}^{-1}v}}{\sqrt{{\rm det} M_{\beta}}},
$$
where $D_{0}$ is independent of $\beta$.  The $\beta$ dependent term
is now also independent of the $\hat{X}_{k}$.  We therefore define the
likelihood  of $\beta$ given the $q_{i}$ data, i.e. the $v_{k}$, to be:
\begin{equation}
L_{q}(\beta) = L_{q}(0) \frac{e^{\frac{\beta^2}{2} v^{T}M_{\beta}^{-1}v}}{\sqrt{{\rm det} M_{\beta}}},
\end{equation}
where $L(0)$ is the value of $L$ when $\beta = 0$.  We define $l_{q}(\beta) = \log L_{q}(\beta)/L_{q}(0)$:
\begin{equation}
l_{q}(\beta) = \frac{\beta^2}{2} v^{T} M_{\beta}^{-1} v - \frac{1}{2} \log {\rm det} M_{\beta}.
\end{equation}
Now if the $z_{i} = y_{i}/\sigma_{i}$ are just random noise with mean $0$ and variance $\lambda^2$ then $v = v_{n}$ and:
\begin{eqnarray}
\mathbb{E}\left(\beta^2 v^{T}_{n} M_{\beta}^{-1} v_{n}\right) &=& \beta^2 \sum_{ijkl} \mathbb{E}(z_{i}z_{j}) h_{ik}h_{jl} M_{\beta\, kl}^{-1}, \nonumber \\ 
&=& \beta^2 \lambda^2 \sum_{kl} Q_{kl} M_{\beta\, kl}^{-1}  \\ &=& \lambda^2 {\rm tr} M_{\beta}^{-1}(M_{\beta} - I) \nonumber \\ &=& - \lambda^2{\rm tr} \mathcal{P}(\beta), \nonumber
\end{eqnarray}
where $\mathcal{P}(\beta) = M_{\beta}^{-1} - I$.  Thus if there is only random noise we define $\mathbb{E}(l) = l_{q}^{\rm noise}(\beta)$ and we have:
\begin{eqnarray}
l_{q}^{\rm noise}(\beta)  &=& \frac{1}{2} \left({\rm tr}\log (I + \mathcal{P}(\beta)) - \lambda^2 {\rm tr} \mathcal{P}(\beta)\right) \\ &\leq& \frac{1}{2} (1-\lambda^2)  {\rm tr} \mathcal{P}(\beta).
\end{eqnarray}
with equality when $\beta = 0$ and hence $\mathcal{P}(\beta) = 0$; generally ${\rm tr}\mathcal{P}(\beta) \leq 0$ with equality when $\beta = 0$. Thus if $\lambda =1$, which we should expect if the error estimates for the $y_{i}$ are accurate we have $l_{q}^{\rm noise}(\beta) < 0$ for $\beta > 0$.   A more conservative approach would therefore be to use the data to check whether the scatter in the data points is as one would expect given the quoted errors, and if it is not extend the errors bars.  We outline the method we use to do this in \S \ref{app:Error} below.  Essentially, the highest frequency modes of any chameleonic signal, i.e. those with $k \approx N-1$, also produce the smallest contribution to the overall signal; all other modes are approximately constant over wavelength scales of about $\lambda_{\rm crit}/(N-1)$.  Thus, provided there are enough data points, we can use the variance of the data points on scales $\lesssim \lambda_{\rm crit}/(N-1)$  to estimate their error. 

We make a similar set of definitions for the $\hat{u}_{i}$ data, for
which $l_{u}(\beta)$ is the log-likelihood, and define the total
log-likelihood to be $l(\beta) = l_{u}(\beta)+l_{q}(\beta)$. We define
the maximum likelihood estimate of $\beta$, $\hat{\beta}$, to be the
value of $\beta$ which maximises $l(\beta)$.  There are now a number
of approaches we may take to estimate the confidence intervals. The
simplest approach is to assume that the values of $\beta$ are normally
distributed with some variance $\sigma_{\beta}^2$.  We then estimate
$\sigma_{\beta}^2$ as: 
\begin{equation}
\sigma_{\beta}^2 = -\frac{1}{l_{,\beta\beta}(\hat{\beta})}.
\end{equation}
 A 95\% confidence interval for $\beta$ is then estimated to be $\beta = \hat{\beta} \pm 1.96\sigma_{\beta}$.  Since the quantity we are actually interested in is $x \equiv \vert B\vert L/2M =  \sqrt{2\beta}$, we display all confidence limits as constraints on the value of $x$. We refer to this as the normal approximation and label it (NA). The second approach is to assume that $r(\beta,\hat{\beta}) = 2(l(\hat{\beta}) - l(\beta)) \sim \chi^2_{1}$, which should hold as the number of observations tends to infinity; the approximate 95\% confidence interval for $\beta$ is all $\beta$ for which $r(\beta,\hat{\beta}) < 3.84$. We transform this into an approximate confidence interval for $x$ by taking $\hat{x} = \sqrt{2 \hat{\beta}}$. We refer to this as the $\chi^2$ approximation, labelled ($\chi^2$). A more robust approach to estimating confidence intervals is to bootstrap the $\hat{u}_{i}$ and $\hat{q}_{i}$ data.  We make $B$ bootstrap data sets constructed by resampling  with replacement $N_{p}$ data points for both $\hat{u}_{i}$ and the $\hat{q}_{i}$ from the original data.  For each bootstrap data set we construct the $\hat{\beta}$ and $\sigma_{\beta}^2$ in the same way as was done in the normal approximation, defining them to be: $\hat{\beta}_{\ast}$  and $\sigma_{\beta\,\ast}^2$.  Now there are a number of different bootstrap  methods for estimating confidence intervals. We use the bootstrap-t method which generally has better convergence than the usual bootstrap method. We assume that the distribution of $t= (\hat{\beta} - \beta)/\sigma_{\beta}$ is well approximated by the distribution of the the bootstrap parameter $t_{\ast} = (\hat{\beta}_{\ast} - \hat{\beta})/\sigma_{\beta\,\ast}$.  The lower limit, $\underline{\beta}_{\alpha}$ of the bootstrap-t $100(1-2\alpha)\%$ confidence interval for $\beta$ is therefore given by:
$$
\alpha = P(\frac{\beta-\hat{\beta}}{\sigma_{\beta}} < \frac{\underline{\beta}_{\alpha} - \hat{\beta}}{\sigma_\beta}) = P( t > -\underline{t}) = 1- P(t < -\underline{t}_{\alpha}).
$$
where $\underline{t}_{\alpha} = \frac{\underline{\beta}_{\alpha}-\hat{\beta}}{\sigma_{\beta}}$.  We estimate $P(t < -\underline{t}_{\alpha})$ by $P_{\ast}(t_{\ast} < -\underline{t}_{\alpha}) = G_{\rm Boot}(-\underline{t}_{\alpha})$ where 
$$
P_{\ast}(t_{\ast} < s) \equiv G_{\rm Boot}(\beta) = \frac{\# (t_{\ast} < s)}{B}.
$$
Thus we have:
$$
\underline{\beta}_{\alpha} = \hat{\beta} - \sigma_{\beta} G_{\rm Boot}^{-1}(1-\alpha).
$$
Similarly the upper limit is:
$$
\bar{\beta}_{\alpha} = \hat{\beta} - \sigma_{\beta} G_{\rm Boot}^{-1}(\alpha).
$$
We define the central estimate of $\beta$ to be $\beta_{\rm m} =
\hat{\beta} - \sigma_{\beta} G_{\rm Boot}^{-1}(1/2)$.  If
$\hat{\beta}$ is an unbiased estimator for $\beta$ then $G_{\rm
  Boot}^{-1}(1/2) = 0$ and we have $\beta_{m} =\hat{\beta}$. We label
this approximation (Bt). In all cases given below we have used $B = 5
\times 10^{4}$ bootstrap resamplings.   When the error bars are
rescaled as described above and in \S \ref{app:Error}, we construct
confidence intervals in both the normal approximation, and using the
$\chi^2$ technique; we label these approaches (NA - $\sigma$) and ($\chi-\sigma$) respectively.

Using starlight polarization data for three objects from the WUPPE spectrograph with a spectral resolution of 16\AA\ and a spacing between data points of 2\AA, we are able to find useful constraints.  In all cases we take $\lambda_{\rm crit} = 608$\AA.  A more thorough analysis would attempt to also fix $\lambda_{\rm crit}$, given some reasonable  priors about the electron density, $n_{\rm e}$, the chameleon mass, $m_{\phi}$, and the coherence length.  In all cases we take $\lambda_{\rm smooth} = 100$\AA.  We find that for $75$\AA $\lesssim\ \lambda_{\rm smooth} \lesssim 200$\AA\, our results do not depend greatly on $\lambda_{\rm smooth}$.  For the first star, HD2905 ($d = 880\,{\rm pc}$), we have the following approximate 95\% confidence limits for $x= BL/2M$:
\begin{eqnarray}
&x = \left(6.36_{-1.07}^{+0.92} \right) \times 10^{-2}\quad &{\rm (NA)}, \nonumber \\
&x = \left(6.36^{+1.06}_{-0.91}\right) \times 10^{-2}\quad &{\rm(}\chi^2{\rm)}, \nonumber \\
&x = \left(4.68^{+1.44}_{-1.70}\right) \times 10^{-2} \quad &{\rm (Bt)}, \nonumber \\
&x = \left(4.11_{-0.78}^{+0.66}\right) \times 10^{-2} \qquad &{\rm (NA}-\sigma{\rm)}, \nonumber\\
&x = \left(4.11^{+0.77}_{-0.66}\right) \times 10^{-2} \quad &{\rm(}\chi^2-\sigma{\rm)}.
\end{eqnarray}
Even when the error bars are extended as described above, we have that $r(\hat{x}) = 2l(\hat{x}) = 91.9$; indicating that the maximum likelihood estimate for $x$ deviates from $0$ by more than $9.5\sigma$ in the $\chi^2-\sigma$ approximation. We note that for HD2905, $\lambda_{\rm crit} \approx 610$\AA is a local maximum of the likelihood the MLE estimate for $x$.

The last three techniques are all in rough agreement.  We see the same
behaviour in the analysis of simulated data that we have undertaken;
these simulations also show that the last three techniques are most
accurate, and are robust to the actual errors being larger than the
quoted ones.  From these simulations we also find that by
approximating the $X_{k}$ and $Y_{k}$ as independent identically
distributed $N(0,1)$ random variables, we reduce the likelihood of the MLE for $x$, but do not greatly alter the value of the MLE.  

For two other objects (again assuming $\lambda_{\rm crit} = 608$\AA), we find similar results. For HD39703, at $d=880\,{\rm pc}$, we find:
\begin{eqnarray}
&x = \left(8.69^{+1.36}_{-1.62}\right) \times 10^{-2} \qquad &{\rm (NA)}, \nonumber\\
&x = \left(8.69^{+1.58}_{-1.39}\right) \times 10^{-2} \quad &{\rm(}\chi^2{\rm)}, \nonumber \\
&x = \left(7.59^{+1.63}_{-1.47}\right)\times 10^{-2}\quad &{\rm (Bt)}, \nonumber \\
&x = \left(8.11^{+1.50}_{-1.84} \right) \times 10^{-2} \qquad &{\rm (NA}-\sigma{\rm)}, \nonumber\\
&x = \left(8.11^{+1.70}_{-1.58}\right) \times 10^{-2} \quad &{\rm(}\chi^2-\sigma{\rm)}.
\end{eqnarray}
When the error bars are extended as described above we have $r(\hat{x}) = 2l(\hat{x}) = 74.9$.  This implies that, in the $\chi^2-\sigma$ approximation, $x= 0$ is more than $8.6\sigma$ from the maximum likelihood estimate of $x$.  For HD34078 ($d=610\,{\rm pc}$) we have:
\begin{eqnarray}
&x = \left(9.95^{+1.73}_{-2.11}\right) \times 10^{-2} \qquad &{\rm (NA)}, \nonumber\\
&x = \left(9.95^{+2.09}_{-1.75}\right) \times 10^{-2} \quad &{\rm(}\chi^2{\rm)}, \nonumber \\
&x = \left(8.58^{+2.15}_{-1.85}\right)\times 10^{-2}\quad &{\rm (Bt)}, \nonumber \\
&x = \left(9.41^{+1.93}_{-2.45}\right) \times 10^{-2} \qquad &{\rm (NA}-\sigma{\rm)}, \nonumber\\
&x = \left(9.41^{+2.25}_{-2.03}\right) \times 10^{-2} \quad &{\rm(}\chi^2-\sigma{\rm)}.
\end{eqnarray}
In this case, $r(\hat{x}) = 2l(\hat{x}) = 84.9$ when the error bars are extended.  This again implies that the maximum likelihood estimate for $x$ deviates from $0$ by more than $9\sigma$ in the $\chi^2-\sigma$ approximation.  As expected from simulations, the last three techniques are all in rough agreement.

If we assume that the same value of $BL/2M$ should be appropriate for all three objects (which may not necessarily be the case), then combining all three data sets we find the following 95\% confidence intervals:
\begin{eqnarray}
&x = \left(7.95_{-0.91}^{+0.81}\right) \times 10^{-2} \qquad &{\rm (NA)}, \nonumber\\
&x = \left(7.95^{+0.88}_{-0.83}\right) \times 10^{-2} \quad &{\rm(}\chi^2{\rm)}, \nonumber \\
&x = \left(6.25_{-1.23}^{+1.16} \right)\times 10^{-2}\quad &{\rm (Bt)}, \nonumber \\
&x = \left(6.03^{+0.76}_{-0.87}\right) \times 10^{-2} \qquad &{\rm (NA}-\sigma{\rm)}, \nonumber\\
&x = \left(6.03^{+0.85}_{-0.76}\right) \times 10^{-2} \quad &{\rm(}\chi^2-\sigma{\rm)}.
\end{eqnarray}
The log-likelihood of the MLE of $x$ when the errors have been rescaled is $r = 2\log l(\hat{x}) = 214$; indicating a more than $14.6\sigma$ deviation from $0$ in the $\chi^2$ approximation.  As with all three objects separately, we see that there is rough agreement between the last three approaches, although, as was the case for the three objects separately, the error bars are widest in the bootstrap-t approximation.

Combining all the data in the standard approach (assuming the same value of $x=BL/2M$ is appropriate for all), and using the bootstrap-t method, we find the following 99.9\% confidence intervals:
\begin{equation}
\frac{BL}{2M} = \left(6.25^{2.00}_{-2.19}\right)\times 10^{-2}, \qquad (99.9\%).
\end{equation}
Using the bootstrap-t method, inasmuch as the resolution of the bootstrap distribution allows, we find that, defining $\sigma_{x\,\ast} = \sigma_{\beta\,\ast}/x_{\ast}$, the distribution of $S_{0}(\hat{x}_{\ast} - \hat{x})/\sigma_{x\,\ast} + S_{1}$ is approximately $N(0,1)$ for some $S_{0}$ and $S_{1}$.  If we assume that the distribution of $(\hat{x}_{\ast} - \hat{x})/\sigma_{x\,\ast}$ is a good approximation to that of $(\hat{x} - x)/\sigma_{x}$, we have:
\begin{equation}
x = \frac{BL}{2M} = (6.27 \pm 0.58) \times 10^{-2},
\end{equation}
where this time the quoted error bars are $1\sigma$. This corresponds to a more than $10.7\sigma$ deviation from $0$, and  provides the following 95\% and 99.9\% approximate confidence intervals:
\begin{eqnarray}
\frac{BL}{2M} &=& (6.27 \pm 1.14) \times 10^{-2}, \qquad (95\%), \\
\frac{BL}{2M} &=& (6.27 \pm 1.91) \times 10^{-2}, \qquad (99.9\%).
\end{eqnarray}

Using our estimated values for $B$ and  $L$ we have at 99.9\% confidence:
\begin{equation}
M = \left(1.47_{-0.35}^{+0.64}\right) \times 10^{9} {\rm GeV}, \qquad (99.9\%).
\end{equation}

Although this analysis is only preliminary, it does appear as if there is a reasonably significant, and robust, statistical preference towards the existence of a chameleon-like field in the starlight polarization data of the three objects we have considered here. A fuller analysis would have to take into account more, even all, comparable starlight polarization measurements. Additionally one would also wish to fit for $\lambda_{\rm crit}$.

\subsection{Extending the Estimated Errors}\label{app:Error}
In this subsection we provide further details of how we extend the errors bars on the data to better mask the observed small scale scatter.  We expect any chameleon induced fluctuations of the polarization on wavelength scales smaller than $\lambda_{\rm crit}/(N-1)$ to be small compared with that on larger scales between $\lambda_{\rm crit}/(N-1)$ and  $\lambda_{\rm crit}$.    In all cases we estimate $\lambda_{\rm crit}/(N-1) \gtrsim 16$\AA.   For each smoothed data point $(\lambda_{i}, \hat{u}_{i}, \hat{q}_{i})$, we use the data points labelled $j$ with $2\vert \lambda_{i}-\lambda_{j}\vert < 16$\AA\ to estimate the random (or non-chameleonic) scatter in the data.  The estimated standard errors in the smoothed data points are $\hat{\sigma}_{i}$.   We define, as we did above, $J_{i} = \left\lbrace j\,:\,2\left \vert \lambda_{i}-\lambda_{j}\right\vert < \lambda_{\rm smooth}\right\rbrace$, where this time $\lambda_{\rm smooth}=16$\AA.

We assume that the data points in $J_{i}$ have mean $\mu$ and standard error $\sqrt{\hat{\sigma}^2_{j}+ \delta \sigma^2}$, where $\delta \sigma^2$ is to be found ($\mu$ and $\hat{\sigma}$ will be different for the $\hat{q}_{i}$ and the $\hat{u}_{i}$).  For data points $x_{j}$, with estimated standard error $\hat{\sigma}_{j}$, where $j \in J_{i}$, we estimate $\mu$ by $\bar{\mu}$ its maximum likelihood estimator:
\begin{equation}
\bar{\mu}(\delta \sigma^2) = \frac{\sum_{j \in J_{i}} \frac{x_{j}}{\hat{\sigma}_{j}^2 + \delta\sigma^2}}{\sum_{j \in J_{i}} \frac{1}{{\hat{\sigma}_{j}^2 + \delta\sigma^2}}}.
\end{equation}
Similarly for each $i$, we estimate $\delta \sigma^2$ by its MLE $\delta \bar{\sigma}^2_{i}$ which satisfies:
\begin{equation}
\sum_{j \in J_{i}} \frac{(x_{j}-\bar{\mu})^2}{(\hat{\sigma}_{j}^2 + \delta\sigma^2)^2} = \sum_{j \in J_{i}} \frac{1}{{\hat{\sigma}_{j}^2 + \delta\sigma^2}}.
\end{equation}
If no solutions to this equation exist, then we take $\delta \bar{\sigma}^2_{i} = 0$.  Finally we smooth the $\delta \bar{\sigma}_{i}^2$ over  a $100$\AA smoothing scale, giving $\delta \hat{\sigma}_{i}^2$.  We take the final enhanced error to be $\tilde{\sigma}_{i} = \sqrt{\hat{\sigma}^2_{i} + \delta \hat{\sigma}^2_{i}}$.  Although this procedure is rather ad hoc, by enhancing the error bars, we err on the side of caution and reduce the probability that under-estimated error bars result in a spurious detection of $\beta \neq 0$.

We present estimated confidence intervals where the error bars have been extended using the normal approximation and the $\chi^2$ approximation; we label these two approaches (NA-$\sigma$) and ($\chi^2-\sigma$) respectively.

\subsection{Estimating upper bounds on $BL/2M$}\label{App:BUp}
It is also possible to find upper confidence limits on $BL/2M$, simply from the observation that the component of polarization perpendicular to the mean polarization angle is smaller than some upper bound i.e. $\vert P_{\perp}(\lambda) \vert <  p_{\rm max}$.  Suppose the observations of $P_{\perp}$ are $q_{i}$, and that we have the maximum value of the $q_{i}^2 < p_{\rm max}^2$.  If a chameleon field is present, and assuming that the polarization angle of any intrinsic polarization is roughly constant, we predict $q_{i}^2 = \beta^2 (\sum_{k} X_{k} h_{ki})^2$.  If the intrinsic polarization angle is not constant then we will generally be biased in favour of larger values $\beta$, and so this approach can also be trusted to provide upper bounds on $BL/2M$. Using numerical simulations we can estimate the distribution of $w ={\rm max} (\sum_{k} X_{k} h_{ki})^2$ and for $0 < \alpha < 1$ calculate $w_{\alpha}$ the probability that:
$$
P(w < w_{\alpha}) = \alpha =   P( \beta^2 w < \beta^2 w_{\alpha}).
$$
We defining $q_{\rm max}^2 = \max q_{i}^2$ we then have:
$$
P(\beta^2 < q_{\rm max}^2/w_{\alpha}) = 1-\alpha.
$$
Thus $q_{\rm max}^2 / w_{\alpha} = p_{\rm max}^2/ w_{\alpha} = \bar{\beta}_{\alpha}^2$ is a estimate of the $100(1-\alpha)\%$ upper confidence limit on $\beta^2$.  Generally this is an over-estimate of the true upper confidence limit, and so $\beta < \bar{\beta}_{\alpha}$ with at least $100(1-\alpha)\%$ confidence.

\end{document}